%% file: ms.tex
\documentclass[
journal=apchd5,
manuscript=article,
layout=onecolumn
]{achemso}

\usepackage[utf8]{inputenc}
\usepackage{amsmath,amsfonts,amssymb}
\usepackage{bm}								
\usepackage{xcolor}
\usepackage{float}
\usepackage{graphicx} \graphicspath{{Images/}}
\usepackage{hyperref}
\usepackage[capitalise]{cleveref}
\usepackage{booktabs}
\usepackage[percent]{overpic}
\usepackage{ulem}

\usepackage{nameref} 
\newcommand{\Nnameref}[1]{\textit{\nameref{#1}}}

\usepackage{bbold} 
\usepackage{braket}

\usepackage{xr}
\title{Reduced Density-Matrix Approach to Strong Matter-Photon Interaction}

\author{Florian Buchholz}
\email{florian.buchholz@mpsd.mpg.de}

\author{Iris Theophilou}
\email{iris.theophilou@mpsd.mpg.de}
\affiliation{Theory Department, Max Planck Institute for the Structure and Dynamics of Matter - Luruper Chaussee 149, 22761 Hamburg, Germany}
\author{Soeren E. B. Nielsen}
\affiliation{Theory Department, Max Planck Institute for the Structure and Dynamics of Matter - Luruper Chaussee 149, 22761 Hamburg, Germany}
\author{Michael Ruggenthaler}
\email{michael.ruggenthaler@mpsd.mpg.de}
\affiliation{Theory Department, Max Planck Institute for the Structure and Dynamics of Matter - Luruper Chaussee 149, 22761 Hamburg, Germany}
\author{Angel Rubio}
\email{angel.rubio@mpsd.mpg.de}
\affiliation{Theory Department, Max Planck Institute for the Structure and Dynamics of Matter - Luruper Chaussee 149, 22761 Hamburg, Germany}
\altaffiliation{ Center for Computational Quantum Physics (CCQ), Flatiron Institute, 162 Fifth Avenue, New York NY 10010, USA}


\date{\today}

\begin{document}

\newcommand {\td} {\mathrm{d}}
\newcommand {\br} {\mathbf{r}}
\newcommand {\bz} {\mathbf{z}}
\newcommand {\blambda} {\bm{\lambda}}

\normalem


\begin{abstract}
We present a first-principles approach to electronic many-body systems strongly coupled to cavity modes in terms of matter-photon one-body reduced density matrices. The theory is fundamentally non-perturbative and thus captures not only the effects of correlated electronic systems but accounts also for strong interactions between matter and photon degrees of freedom. We do so by introducing a higher-dimensional auxiliary system that maps the coupled fermion-boson system to a dressed fermionic problem. This reformulation allows us to overcome many fundamental challenges of density-matrix theory in the context of coupled fermion-boson systems and we can employ conventional reduced density-matrix functional theory developed for purely fermionic systems. We provide results for one-dimensional model systems in real space and show that simple density-matrix approximations are accurate from the weak to the deep-strong coupling regime. This justifies the application of our method to systems that are too complex for exact calculations and we present first results, which show that the influence of the photon field depends sensitively on the details of the electronic structure.
\end{abstract}

\input{1introduction}
\input{2rdms}
\input{3fermionization}
\input{4drdmft}

\input{5numerics_results}
\input{6summary}

\makeatletter
\setlength\acs@tocentry@width{4.75cm}
\setlength\acs@tocentry@height{8.5cm}
\makeatother
\begin{tocentry} 
	\centering
	\includegraphics{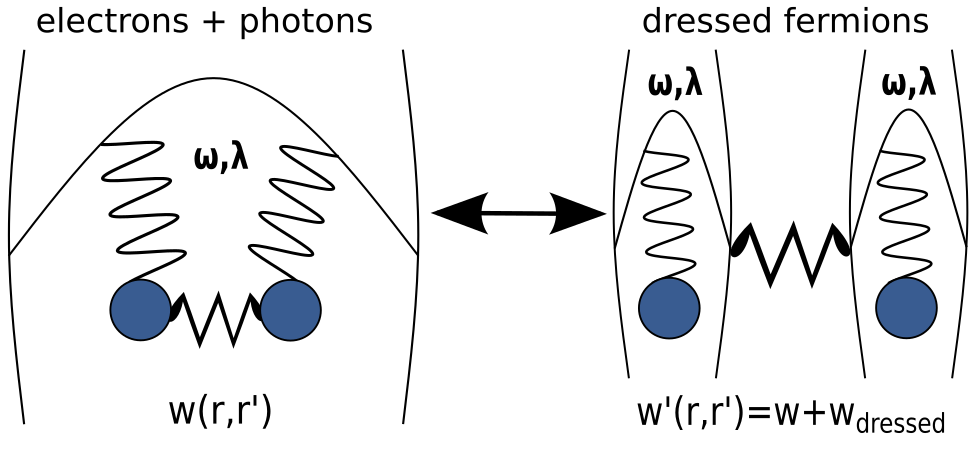}
\end{tocentry}


\bibliography{bibliography}


\end{document}


\newcommand {\td} {\mathrm{d}}
	\newcommand {\br} {\mathbf{r}}
	\newcommand {\bq} {\mathbf{q}}
	\newcommand {\bp} {\mathbf{p}}
	\newcommand {\bz} {\mathbf{z}}
	\newcommand {\blambda} {\bm{\lambda}}

24 pages, 1 figure, 2 tables
	
\newpage
\section*{Short Guide to this supporting information}
In this supporting information, we present in detail how we performed the convergence study of the numerical examples shown in the paper and especially how we validate our implementation in Sec. \ref{APP:ValidationMB} to \ref{APP:Validation_dRDMFT}. We provide a ``HowTo'' for such calculations in Sec. \ref{APP:protocol}. In the last section, we examine the issue of the bosonic symmetry in the photon wave function, which is relevant for the argumentation in the main paper. 

To perform the calculations, we integrated our routine in the Octopus code.\footnote{\url{www.octopus-code.org}} 
To find the accuracy threshold of our implementation we compared the Octopus results to calculations of a private code ``Dynamics'' that was used and validated for Ref. \citenum{Nielsen2018} by S.E.B. Nielsen.\footnote{Contact: \href{mailto:soerenersbak@hotmail.com}{soerenersbak@hotmail.com}} Both codes approximate the polaritonic orbitals on discretized real-space boxes in the x and q coordinate, but they use different boundary conditions, a different level of finite differences for the approximation of the differential operators (fourth order in Octopus vs sixth order in Dynamics), a different method of the grid point evaluation (on-point in Octopus vs. mid-point in Dynamics), and also a different orbital optimization technique. 
Specifically, we present the convergence studies for the example of the one-dimensional Helium atom inside a cavity (He) that we used among others in the main part of this paper. The He atom is described by the nuclear potential $v_{He}(x)=-\frac{2}{\sqrt{x^2+1}}$ (we use atomic units throughout) and coupled to one cavity mode with frequency $\omega$ and coupling parameter $\lambda$. The modified local potential due to the dressed auxiliary construction reads $v_{He}'(x,q)=v_{He}(x) + \tfrac{1}{2} (\lambda x)^2 + \tfrac{\omega^2}{2} q^2 - \tfrac{\omega}{\sqrt{2}}q (\lambda x)$ and the modified interaction kernel is $w'(xq,x'q')= w(x,x')+w_d(xq,x'q')$ with $w(x,x')= \frac{1}{\sqrt{(x-x')^2+1}}$ and
$w_d(xq,x'q')= -\frac{\omega\lambda}{\sqrt{2}}(xq'+x'q) + \lambda^2 xx'$ (see Sec. \Nnameref{sec:fermionization} of the main text.) We approximate all Coulomb terms by a soft-Coulomb potential.
The extension of an electronic structure code to treat such dressed systems requires implementing the extra photon coordinate q that should be treated like the electronic ones (and thus integrated in the calculations of the kinetic energy), the modified local potential $v'(x,q)$, and the modified interaction kernel $w'(xq,x'q')$. Additionally, the many-body wave function that is constructed from dressed orbitals exhibits an additional exchange-symmetry between the q-coordinates. In this work, we do not take this symmetry into account, which is a reasonable approximation for the presented calculations. For further details see Sec. \Nnameref{sec:fermionization} and Ref. \citenum{Nielsen2018}.

The convergence study and validation of our implementation in Octopus requires several steps, which correspond to the sections of this supporting information. We start with converging the exact dressed many-body ground-state and compare it to the results of Dynamics. Although we are limited to a very small Hilbert space for such exact calculations, we only have to solve a linear eigenvalue problem, so it is easy to converge the results to a very high precision within the accessible parameter range. Thus, we obtain an upper bound for the convergence accuracy from these exact calculations. In the next section, we present the convergence of dressed HF (that we abbreviate by dHF in this supporting information) in all its details and compare the converged ground-state to Dynamics. This allows us to determine the accuracy for dHF calculations, that is (because of its non-linear nature) harder to converge than the exact routine. As a side effect, the comparison of the two codes is a good validation of the correct implementation of the dressed modifications. Since these modifications are exactly the same for dressed RDMFT (that we abbreviate by dRDMFT in this supporting information) and dHF, validating the dHF implementation validates also the corresponding changes for dRDMFT. In the last section, we discuss the convergence of dRDMFT, which requires two different minimization procedures that are interdependent, and thus again is harder to converge than dHF.

All the above sections are organized similarly. We start with the separated problem of the electronic system outside the cavity and converge first the electronic counterparts of the routines, i.e. the electronic exact solver, HF or RDMFT. 
The purely photonic problem of the separated problem remains the same at all the discussed levels of theory. We therefore discuss it only once in the first section. Then, we analyze the convergence of the dressed theories in the no-coupling ($\lambda=0$) limit, which theoretically means that the electronic and photonic problems are perfectly separated, but solved in only one large calculation. By comparison of only the electronic (photonic) part of the dressed solutions to the results of the purely electronic (photonic) theories, we can measure if there is a decrease in accuracy due to the simultaneous description of electronic and photonic coordinates. We finish the sections with a convergence study for $\lambda>0$.


\tableofcontents


\section{Validation of the exact dressed many-body ground-state}
\label{APP:ValidationMB}
We start with the validation of the exact many-body ground-state. In both codes, this is calculated by minimizing directly the energy expression of the full many-body wave function (denoted by $\Psi'$ in the main text), discretized on the grid. For the He test system, $\Psi'=\Psi'(x_1q_1,x_2,q_2)$ is four-dimensional. The actual minimization is performed in Octopus by conjugate gradients method, whereas Dynamics makes use of a Lanczos algorithm.

At the beginning of every calculation, we need to find the proper grid, which is defined by the box sizes $L_x,L_q$ and the spacings $dx,dq$ in both dimensions.  For the minimization in Octopus, we used two different convergence criteria, $\epsilon_E=10^{-9}$ and $\epsilon_{\rho}=10^{-8}$. The former tests the energy deviations and the latter the integrated absolute value of the density deviations between subsequent iteration steps.\footnote{Details can be found on \url{http://octopus-code.org/doc/develop/html/vars.php?page=alpha} with the keywords \emph{EigensolverTolerance} and \emph{ConvRelDens}.} These are the criteria already available in Octopus, so we here show that these are sufficient to produce reliable results. Note that we choose $\epsilon_E=\frac{1}{10}\epsilon_{\rho}$, because $\epsilon_{\rho}$ is a much stricter criterion. For the box size and spacing convergence, we perform series $\mathcal{C}=\{ \mathcal{C}^1,\mathcal{C}^2,\dots \}$, where $\mathcal{C}$ can be $L_x,L_q,dx,$ or $dq$ in the following. We perform two types of convergence tests for every $\mathcal{C}$. In the first one, we investigate the deviations in the energy between subsequent elements $\Delta E_{\mathcal{C}^i}=E_{\mathcal{C}^i}-E_{\mathcal{C}^{i-1}}$. We denote the corresponding thresholds with $\epsilon_{E_{\mathcal{C}}}$. The second type of convergence series considers the maximal deviations in the absolute-value of the electronic/photonic part of the polaritonic density $\Delta \rho_{\mathcal{C}^i}=\max_{x/q}|\rho_{\mathcal{C}^i}(x/q)-\rho_{\mathcal{C}^{i-1}}(x/q)|$, with $\rho_{\mathcal{C}^i}(x)=\int\td q\rho_{\mathcal{C}^i}(x,q)$ for the series $L_x,dx$ and $\rho_{\mathcal{C}^i}(q)=\int\td x\rho_{\mathcal{C}^i}(x,q)$ for the series $L_q,dq$.  We denote the corresponding thresholds with $\epsilon_{\rho_{\mathcal{C}}}$.

We start with the electronic part of the example (corresponding to the He atom outside the cavity), choose $dx=0.14$ and vary $L_x=\{8,10,12,\dots\}$. The ground-state energy $E_{L_x}$ drops with increasing $L_x$ (because boundary effects become less important) and we find $\Delta E_{L_x}<10^{-8}$ and $\Delta\rho_{L_x}<10^{-7}$ for $L_x\geq 20$. In the following, we choose $L_x=20$ because the thresholds of $\epsilon_{\rho_{L_x}}=10^{-7}$ and $\epsilon_{E_{L_x}}=10^{-8}$ are already stricter than the maximal accuracy of the later non-linear (dHF,dRDMFT) calculations. Next, we perform  $dx=\{0.2,0.19,0.18,\dots,0.06\}$. We find that $\Delta E_{dx}$ also drops with decreasing $dx$ until $\Delta E_{dx}<10^{-8}$ for $dx = 0.15$ and does not decrease any more (as the convergence criteria of the minimization are no more precise.) 
$\Delta\rho_{dx}$ instead decreases (slowly) until the lowest tested value of $dx=0.06$. We find $\Delta \rho_{dx}<10^{-7}$ already for $dx<0.14$, but to reach $\Delta \rho_{dx}<10^{-8}$, we need to decrease the spacing until $dx=0.07$. Such a small spacing is numerically unfeasible for larger boxes and thus we will not try to go beyond $\epsilon_{\rho_{dx}}=10^{-7}$.
	
We repeat the same series for the harmonic oscillator system of the q-coordinate with frequency $\omega=\omega_{res}\approx0.5535$ (resonance with the transition between ground and first excited state of the He atom outside the cavity) and find that the box size is converged with $\Delta E_{L_q}< 10^{-8}$ and $\Delta \rho_{L_q}<10^{-8}$ for $L_q \geq 14$. 
The spacing is converged in the energy $\Delta E_{dq}<10^{-8}$ and in the density $\Delta \rho_{dq}< 10^{-7}$ for $dq\leq0.20$.

From these preliminary calculations, we can infer the parameters for the dressed calculations: $L_x=20,L_q=14,dx=0.14, dq=0.20$. To be sure that these parameters are still sufficient for non-zero coupling ($\lambda>0$), we perform another box length series in the four-dimensional space of the exact ground-state, e.g. for $\lambda=0.1$ and $\omega=\omega_{res}$. To have a uniform grid distribution, we also set $dx=dq$, although our preliminary calculations suggest that we could choose a larger $dq$. Unfortunately, we cannot explore this space completely, but we are limited with $L_x,L_q\leq 18$.\footnote{The precise reasons is that the memory of one node of the cluster we are using is too small. We would consequently need to distribute the wave function over several nodes, which is possible, but would demand a considerable programming effort.} 
However, we can confirm that the energy is converged in the q-direction for $L_q=14$ and in the x-direction, we have $\Delta E_{L_x}\approx 10^{-7}$ for $L_x=18$.

Finally, we compare the Octopus results with the results from the Dynamics code. For that, we consider the differences in the total energy $E_{OD}=|E_{Dynamics}-E_{Octopus}|$ and the maximum deviations in the polaritonic density $\rho_{OD}\equiv\max_{x,q}|\rho_{Octopus}(x,q)-\rho_{Dynamics}(x,q)|$ between the two codes. Due to the optimization of Dynamics, we are limited to a box length of $L_x=L_q=14$ to have a spacing of $dx=dq=0.14$. 
With these parameters, all the energy and density errors in Octopus increase to $\Delta E_{L_x,L_q,dx,dq}=\Delta\rho_{L_x,L_q,dx,dq}\approx 10^{-5}$. When we compare the ground-state energies of both codes for this maximum possible mesh, we find $E_{OD}\approx 10^{-5}$ and $\rho_{OD}\approx 10^{-5}$.  
So both codes agree on the level of accuracy that we estimated from the Octopus calculations before, although they are quite different from a numerical perspective. We conclude from these calculations that the implementation of the exact many-body routine for dressed two-electron systems\footnote{Larger systems are in principal also possible, but numerically infeasible in real space.} in Octopus is reliable and we can use it as benchmark for the dHF and dRDMFT approximations within Octopus.


\section{Validation of the dHF routine}
\label{APP:dHF}
In this section, we present the validation of dHF, which requires several steps. We start with the electronic HF routine and converge the He atom model (outside the cavity) in box size and spacing, where we proceed like in the previous section. Then, we converge a second HF routine (HF$_{basis}$) that is implemented in Octopus, which makes use of a basis set. The difference of such a basis-set implementation is that the routine calculates all the integral kernels of the total energy as matrix elements of a chosen basis and then searches for the energy minimum by only varying the corresponding coefficients. This routine can be considerably faster than the standard one of Octopus, because the calculation of the exchange-term, which is numerically much more expensive than all the other parts of the Hamiltonian, needs only to be performed once in the beginning for the matrix elements. The conjugate gradient algorithm of the standard HF routine of Octopus instead needs to evaluate the Hamiltonian including the exchange term at every iteration step. Such a basis-set implementation requires a convergence study with respect to the basis size, which we will explain in detail in the second part of this section together with a comparison between HF and HF$_{basis}$.
We conclude the section with the discussion of dHF, which also uses a basis set. However, the basis-set convergence for dHF is more involved than for the purely electronic HF$_{basis}$ and we explain this in detail. Afterwards, we compare dHF in the no-coupling limit to HF/HF$_{basis}$ and we discuss the comparison of Octopus and Dynamics on the level of dHF, where we follow the same strategy like in the previous section.

\subsection{HF minimization on the grid} 
We start with the He atom outside the cavity and calculate the HF ground-state with the standard routine of Octopus, which uses a conjugate gradients algorithm for the orbital optimization. The convergence criteria for the conjugate gradient algorithm are defined exactly as before for the exact many-body ground-state. We set them to $\epsilon_E=10^{-9}$ and $\epsilon_{\rho}=10^{-8}$ and determine $L_x$ and $dx$. We start with a spacing of $dx=0.1$ and find that $\Delta E_{L_x}\approx 10^{-8}$ and $\Delta \rho_{L_x}<10^{-7}$ for all $L_x\geq 20$. 
We choose $L_x=20$ and perform the spacing series, which is converged in energy with $\Delta E_{dx}<10^{-8}$ for $dx\leq 0.15$ and in the density with $\Delta \rho_{dx}< 10^{-7}$ for $dx\leq0.14$. These calculations suggest that all $dx\leq0.14$ are sufficient and we choose $dx=0.1$, which is numerically feasible for the test-system, we consider.

\subsection{HF minimization with a basis set}
\label{APP:HFvsHFbasis}
In this section, we present the convergence of the He test system in the HF$_{basis}$ routine with respect to the basis set. As Octopus is a real-space code, there are no standard quantum-chemistry basis sets implemented. This means that if we wish to express the Hamiltonian and the wave function in a basis set, we need to generate one. For all the calculations of this paper, we do this in the same way by performing a preliminary calculation with the \emph{independent particle} (IP) routine, which solves a simple 1-body Schrödinger equation for every orbital, neglecting interaction. Among the numerically efficient options that Octopus offers, the basis from IP converged fastest with respect to the size of the basis set $M$.\footnote{Note that we used the symbol $\mathcal{M}$ in the main text.}
To allow for enough variational freedom, we calculate besides the occupied orbitals\footnote{Note that in our case, we have always $GS=\frac{N}{2}$, where $N$ is the number of electrons, because we only look at closed-shell systems, which distribute two electrons to every spatial orbital.} (that form the HF ground state and are denoted by $GS$) also excited orbitals, which are called \emph{extra states} ($ES$) in Octopus. The so generated basis set has the size $M=GS+ES$.

The routine performs the minimization by representing the coefficients of the basis-set expansion in a matrix and diagonalizing it repeatedly until self-consistence. The corresponding energy convergence criterion $\epsilon_E$ remains the same in HF$_{basis}$ like in HF.
But the convergence criterion $\epsilon_{\Lambda}$ of the orbital minimization needs to be adapted and there are several possible options. In Octopus, $\epsilon_{\Lambda}$ tests the hermiticity of the Lagrange multiplier matrix $\Lambda$\cite{Andrade2015} that guarantees the orthonormality of the natural orbitals. This is not an obvious choice, but it is more general than typical criteria, and thus allows for using it also for RDMFT minimizations in a basis set. In fact, HF$_{basis}$ and RDMFT in Octopus use exactly the same algorithm for the orbital optimization. This method is explained in detail in Ref. \citenum{Ugalde2008}. As $\epsilon_{\Lambda}$ is a considerably stricter criterion than $\epsilon_E$, it is set as default to $\epsilon_{\Lambda}=10^3\cdot\epsilon_E$.

For the convergence of the size of the basis set M or equivalently the parameter $ES$, we perform a series $ES=\{10,20,\dots,100\}$ and test deviations in energy $\Delta E_{ES, ES_{ref}}=E_{ES}-E_{ES_{ref}}$ and density $\Delta \rho_{ES,ES_{ref}}=\max_x|\rho_{ES}(x)-\rho_{ES_{ref}}(x)|$ from the reference value $ES_{ref}$ that yields the lowest total energy and that typically occurs for the largest basis. We use $L_x$ and $dx$  that we determined before in HF and find that $\Delta E_{ES,100}$ and $\Delta \rho_{ES,100}$ decrease with increasing $ES$ and for $ES\geq40$, $\Delta E_{ES,100}<10^{-8}$ and $\Delta \rho_{ES,100}\approx 10^{-5}$.

Comparing HF with HF$_{basis}$ with these parameters, we find that $|E_{HF}-E_{HF_{basis}}|\approx 10^{-8}$ and $\max_x|\rho_{HF}(x)-\rho_{HF_{basis}}(x)|\approx 10^{-5}$. So both methods are consistent.

\subsection{Validation of dHF}
After having properly understood the convergence of the electronic HF methods, we can now turn to dHF. All the calculations shown in the main part of this paper are done with the basis-set type of implementation,  which is more involved than in the purely electronic HF$_{basis}$ case. Thus, we start in Sec. \ref{APP:dHF_basis_conv} with discussing the new issues that enter when one needs to converge a system in the dressed space with respect to the basis set. In Sec. \ref{APP:dHF_no_coupling}, we present the convergence of dHF with zero-coupling ($\lambda=0$), which means that the electronic and photonic part of the system completely decouple such that we can compare the results of the electronic part to electronic HF. In the last section, \ref{APP:dHF_comp_dynamics}, we compare the dHF results from Octopus with Dynamics.

\subsubsection{Basis-set convergence in the dressed auxiliary space}
\label{APP:dHF_basis_conv}
In the dressed auxiliary space that we explore with dHF and dRDMFT, the basis-set convergence is not as straightforward as in the HF$_{basis}$ case. We illustrate this additional difficulty for dHF in the no-coupling ($\lambda=0$) case. $\lambda=0$ makes the discussion much simpler, because we have two perfectly separated problems, the atomic system and harmonic oscillators that are just calculated at the same time. However, we use a combined basis that consequently needs to include the appropriate degrees of freedom for both systems. But the composition of the basis set is strongly influenced by the parameters of the system, especially $\omega$, because it is generated by a preliminary calculation. This can be illustrated as follows:

For $\lambda=0$, the electronic and photonic part of the system separate. Thus, the coupled (dHF) Hamiltonian is a direct sum of the electronic (photonic) Hamiltonian $\hat{H}^e$ $(\hat{H}^p)$ and the two-dimensional dressed orbitals $\psi_{i\alpha}(x,q)$ can be exactly decomposed in their one-dimensional electronic  $\phi_i(x)$ and photonic $\chi_{\alpha}(q)$ constituents, $\psi_{i\alpha}(x,q)=\phi_i(x)\otimes \chi_{\alpha}(q)$. Here, $\phi_i$ $( \chi_{\alpha})$ is an eigenfunction with eigenvalue $e^e_i$ $(e^p_{\alpha})$ of the electronic (photonic) Hamiltonian $\hat{H}^e$ $(\hat{H}^p)$. Consequently, we know that we can calculate the eigenvalues of the dressed orbitals as sum of the uncoupled ones, i.e. $e^{ep}_{i\alpha}=e^e_i +e^p_{\alpha}$. The basis set for a dHF calculation is then constructed using the ground and the first $ES$ excited orbitals of a preliminary IP calculation. These orbitals are ordered by their eigenvalue $e^{ep}_{i\alpha}$ and for that, the relation between the individual energies of the electron and photon space is crucial. 

Table \ref{tab:basis_contributions} shows the decomposition of such a basis with $ES=6$ (and thus $M=7$.) We see that slightly more electronic states contribute to the basis, but already 3 out of 7 states describe excited photon contributions. So if we wanted to use that basis for an dHF calculation with for example $\lambda=0$, that trivially needs only ground-state contributions for the photonic coordinate, these 3 states would be entirely unnecessary and waste computational resources. Before we discuss this further, we want to explain how $\omega$ influences this distribution between electronic and photonic contributions: If we chose for example a larger $\omega\approx1.3$, the first 4 states of the combined basis would only vary in the electronic contribution. If we instead lowered $\omega$, the opposite would happen and the first states would vary in the photonic contribution. A similar kind of argument could be done for all other ingredients of the Hamiltonian of course, but for $\omega$ this influence is most directly visible and comparatively strong. A detailed investigation of this issue is beyond the scope of this paper. Pragmatically, we sometimes adapt $\omega$ such that the produced basis is reasonable (see the Beryllium test system in Sec. \Nnameref{sec:results} of the main part of the paper.)

At this point, one might ask the question why at all we use the basis-set implementation and the answer remains the same as for the purely electronic case: Although we need large basis sets for the properly converged dHF calculations which makes them numerically very expensive, these computations are still relatively inexpensive compared to calculations with a conjugate gradients algorithm that calculates all integrals on the grid. In the case of the He atom, the dHF calculation with conjugate gradients and $ES=15$ takes still 4 times longer than the same calculation with a basis set and $ES=50$.

Finally, we want to mention that the statements of this subsection carry over straightforwardly to the coupled ($\lambda>0$) case, we just do not have the direct connection to the decoupled spaces and thus cannot visualize this case so clearly. It is clear that we need more variational freedom for the photonic subspace than in the $\lambda=0$ case but the basis will still depend on the system parameters. One can expect that also for $\lambda>0$ many basis states will be unnecessary.
\begin{table}
	\begin{tabular}{@{}ccc|ccc@{}}
		&&&&\multicolumn{2}{c}{contribution} \\
		index & $e^{e}_i$ & $e^{p}_{\alpha}$ & $e^{ep}_{i\alpha}$ & $i$ & $\alpha$\\
		\hline
		1 & -1.483 & 0.277 & -1.207 & 1 & 1 \\
		2 & -0.772 & 0.830 & -0.653 & 1 & 2 \\
		3 & -0.461 & 1.384 & -0.495 & 2 & 1 \\
		4 & -0.263 &       & -0.184 & 3 & 1 \\
		5 &  &  					 & -0.100 & 1 & 3 \\
		6 &  &  					 & 0.014  & 4 & 1 \\
		7 &  &  					 & 0.058  & 2 & 2 \\
	\end{tabular}
	\caption{Basis set structure (IP calculations) with $ES=6$ (thus $M=N/2+ES=7$ basis states) for the dressed He atom with mode frequency $\omega\approx 0.5535$, but no coupling ($\lambda=0$.) On the left side of the table are the first eigenenergies $e^e_i$ ($e^p_{\alpha}$) of the pure electronic (photonic) Hamiltonian $\hat{H}^e$ $(\hat{H}^p)$ shown. On the right side, we see the orbital energies $e^{ep}_{i\alpha}=e^e_i+e^p_{\alpha}$ of the combined Hamiltonian $\hat{H}^{ep}$ and the decomposition of the combined index: The first eigenenergy $e^{ep}_1=e^{ep}_{11}=e^e_1+e^p_1$ is the sum of the two first uncoupled energies. The second energy $e^{ep}_2=e^{ep}_{12}$ instead is formed from the electronic ground but photonic first excited state, etc. In this example, we see that both contributions are similar, although there are slightly more electronic orbitals included. This tendency continues also for increasing $ES$.}
	\label{tab:basis_contributions}
\end{table}

\subsubsection{Validation of dHF for $\lambda=0$}
\label{APP:dHF_no_coupling}
Now we can address the $ES$-convergence of dHF, which we will present for $\lambda=0$, such that we can compare the electronic part of the converged result to HF afterwards. As we set $\lambda=0$, we know that for the photon component, the IP ground-state orbital $\chi_1(q)$ is already the exact solution, because the photons do not interact directly with each other and HF (or any other level of approximation) will not change that. The dHF ground-state orbital thus is $\psi(x,q)=\phi_{HF}(x)\otimes \chi_1(q)$ with the electronic HF ground-state orbital $\phi_{HF}(x)$. As we saw in the last section, when we converged the He test system using the HF$_{basis}$ routine, $\phi_{HF}(x)$ can be approximated by an IP basis with $ES=40$. So we know that in the no-coupling limit of dHF, we do not need a larger basis set, because Kronecker-multiplying the basis-states of the HF$_{basis}$-calculation with $\chi_1(q)$ would be exactly sufficient.
However, here and in the following, we do not want to choose by hand such a well-adapted state space (which for very strong coupling strength is non-trivial), but instead rely on the polaritonic IP-calculations. From our considerations before, we expect to need more basis states to reach the same level of accuracy than using HF$_{basis}$ and the exact number will depend on $\omega$. Indeed, for $\omega=\omega_{res}$, we need $ES=80$ to have $\Delta E_{ES,100}<10^{-7}$ and $\Delta \rho_{ES,100}\approx10^{-5}$, where we calculate the electron density of the dHF-calculation by $\rho(x)=\int\td q\, \rho(x,q)$.
When we instead choose a very high value of $\omega=5.0$, we reach $\Delta E_{ES,100}<10^{-7}$  and $\Delta \rho\approx10^{-5}$ already for $ES\geq 40$. For both calculations, we do not reach $\Delta E_{ES,100}<10^{-8}$ even for higher $ES$. This is due to the many "unnecessary" states that are taken into account in the minimization. These then only introduce numerical noise without providing useful variational information. We find this confirmed by analyzing the ``photonic density'' $\rho(q)=\int\td x\, \rho(x,q)$, which deviates from the correct density $\rho_{ex}(q)=2|\chi_1|^2(q)$ although $\chi_1(q)$ is explicitly part of the basis. Adding basis states consequently cannot improve the photonic orbital. For instance, for $\omega=\omega_{res}$, the density deviations remain at approximately $\Delta \rho_{ES, ES_{ref}}\approx10^{-4}$ for \emph{all} $ES$ \emph{and} $ES_{ref}$. When we instead choose $\omega=5.0$, we increase the accuracy to $\Delta \rho_{ES,100}\approx10^{-5}$ for $ES\geq 40$. Despite being less accurate, we used $\omega=\omega_{res}$ for the results section in the main part of the paper (Sec. \Nnameref{sec:results},) because $\Delta \rho_{ES}\approx10^{-4}$ is still two orders of magnitude smaller than the deviations from the exact or the dRDMFT solution.\footnote{By using harder types of convergence criteria throughout one can likely control $\Delta \rho_{ES}$, but this is beyond our needs.}

Finally, we want to compare the electronic part of dHF to HF. For the energy comparison, we need to subtract the photon part $E_{p}=2\frac{\omega}{2}$ from the total dHF energy $E_{dHF}$, $E_{dHF,e}=E_{dHF}-E_p$. However, this analytical expression is also not exact, because of the just mentioned error in the photonic part of the orbitals. Consequently, we have deviations for $\omega=\omega_{res}$ of $|E_{HF}-E_{dHF,e}^{\omega=\omega_{res}}|\approx 3\cdot10^{-6}$ and $\max_x|\rho_{HF}(x)-\rho_{dHF,e}^{\omega=\omega_{res}}(x)|\approx 10^{-4}$, but for $ |E_{HF}-E_{dHF,e}^{\omega=5.0}|\approx4\cdot10^{-7}$ and $\max_x|\rho_{HF}(x)-\rho_{dHF,e}^{\omega=5.0}(x)|\approx 10^{-5}$.

\subsubsection{Validation of dHF with Dynamics}
\label{APP:dHF_comp_dynamics}
We  conclude this section by comparing the results of our implementation of dHF in Octopus with the Dynamics code, which uses an imaginary time propagation algorithm to calculate the HF ground-state. This comparison allows us to validate also the case of $\lambda>0$. We choose $\omega=\omega_{res}$ and $\lambda=0.1$ for the comparison and perform another $ES$-convergence, confirming that we need $ES=90$ for the same level of accuracy like before in the no-coupling case. For the sake of completeness, we want to mention that for $\omega=5.0$ we need again a significantly smaller basis set for even better converged results, exactly as we found before in the $\lambda=0$ case. However, already for $\omega=\omega_{res}$ the level of convergence is an order of magnitude better than the expected deviations between the codes that we estimated in Sec. \ref{APP:ValidationMB}.  

So we compare the ground-state for these parameters to the result of Dynamics and find for the energy the expected deviations of $E_{OD}\approx 10^{-5}$. For the density, we find instead deviations of $\rho_{OD}\approx 10^{-3}$. This discrepancy is probably due to the different convergence criteria of the two codes. Dynamics tests the eigenvalue equation of the one-body Hamiltonian for a certain subset of all the grid points, which is a much stronger criterion than the one of Octopus, explained in Sec. \ref{APP:ValidationMB}. The influence of these different criteria on a self-consistent calculation are naturally stronger than on the calculation of a linear eigenvalue-problem like the many-body calculation. Still, density errors of the order of $10^{-3}$ are small enough for our purposes.
We find similar errors also for other values of $\lambda$ and conclude that both codes are sufficiently consistent.


\section{Validation of dRDMFT}
\label{APP:Validation_dRDMFT}
In this last section we turn to dRDMFT. This is the first implementation at all of this theory, so we cannot validate it with a reference code any more. However, the difference between HF and RDMFT (using the Müller functional) on the implementation-level essentially is in the treatment of the occupation numbers, which are fixed to 2 and 0 in the former case but are allowed to be non-integer for the latter. The 1-body and 2-body terms, which implementation-wise are the only modifications due to the dressed auxiliary construction ($v(\br)\rightarrow v'(\br,q)$, $w(\br,\br')\rightarrow w'(\br q,\br'q')$) are the same for HF and RDMFT and thus also for dHF and dRDMFT. This means that the validation of dHF that we presented in the previous section at the same time largely validates the implementation of dRDMFT.

Still, we need to analyze and understand the convergence with respect to the basis set in dRDMFT and check for the consistency between the results of RDMFT and dRDMFT in the no-coupling limit.

\subsection{Basis-set convergence of RDMFT}
In RDMFT, we have to perform two minimizations that are interdependent: for the natural orbitals $\phi_i$ \emph{and} the natural occupation numbers $n_i$ (where always $i=1,\dots,M$.) This is done by fixing alternately $\phi_i$ or $n_i$, while optimizing the other until overall convergence is achieved. 

We have the possibility to define different convergence criteria for each minimization, $\epsilon_{E}$ (which is connected to $\epsilon_{\Lambda}$, see Sec. \ref{APP:HFvsHFbasis}) and $\epsilon_{\mu}$. The latter tests the convergence of the Lagrange multiplier $\mu$ that appears in the RDMFT functional to fix the total number of particles in the system\cite{Andrade2015}. The occupation number optimization routine at iteration step m sets $\mu=\mu^m$, minimizes the total energy with respect to the $n_i=n_i^m$ and calculates the particle number $N^m=\sum_{i=1}^{M}n_i$. Based on the deviation to the correct system's particle number, $\mu^{m+1}$ is increased or decreased. The routine exits if $|\mu^{m}-\mu^{m-1}|<\epsilon_{\mu}$. For the remainder of this chapter, we set $\epsilon_E=\epsilon_{\mu}=10^{-8}$.

However, the $ES$-convergence of RDMFT is not as straightforward as in the HF$_{basis}$-case. Contrary to the clear monotonic dependence between $ES$ and the energy that we always found for HF$_{basis}$, the current RDMFT implementation in Octopus is not strictly variational with respect to the number of basis states. For all tested systems, the energy went down with increasing $ES$ until a certain value $ES_{min}$ and then up again. Therefore, it seems that the interplay of the two minimization processes and the relatively soft types of convergence criteria introduces for high $ES$ such large errors that they exceed the gain of accuracy due to more variational freedom.
In Tab. \ref{tab:HF_RDMFT},  we show the variation of the ground-state energy of He (outside the cavity) for a series $ES=\{10,20,\dots,80\}$. The energy first decreases until its lowest value for $ES_{min}\approx 40$ and then increases again. 
\begin{table}[H]
	\centering
	\begin{tabular}{c|c|c}
		$ES$ & Energy & $\Delta E_{ES}=E_{ES}-E_{ES-10}$\\ 
		\hline
		10 & -2.2421837 & -\\ 
		20 & -2.2426908 & $-5.1\cdot10^{-4}$\\ 
		30 & -2.2427080 & $-1.7\cdot10^{-5}$\\ 
		40 & -2.2427085 & $-4.9\cdot10^{-7}$\\ 
		50 & -2.2427049 & $+3.6\cdot10^{-6}$\\ 
		60 & -2.2427035 & $+1.3\cdot10^{-6}$\\ 
		70 & -2.2426928 & $+1.1\cdot10^{-5}$\\ 
		80 & -2.2426937 & $-9.9\cdot10^{-8}$	
	\end{tabular} 
	\caption{Ground state energies of the He atom (outside the cavity) calculated by RDMFT with parameters mentioned in the main text. The energy goes first down with increasing number of ES until it reaches its lowest value at $ES_{min}\approx 40$ and then up again.}
	\label{tab:HF_RDMFT}
\end{table}

This non-strictly variational behavior makes a clear definition of the convergence difficult. However, we find that for every considered system there is an \emph{optimal region} $\mathcal{ES}_{opt}=\{ES\,|\,ES_{min}\leq ES\leq ES_{max}\}$. By optimal region, we mean an interval of $ES$ in which the solutions vary minimally among each other, i.e. their energy and density deviations are minimal. For the energies, we define $\Delta E_{ES,ES'}=|E_{ES}-E_{ES'}|$, the corresponding threshold $\epsilon_{E_{opt}}$, and require $\Delta E_{ES,ES'}<\epsilon_{E_{opt}}$ for \emph{all} $ES,ES'\in\mathcal{ES}_{opt}$. For the densities, we define the point-wise density deviations $\rho_{ES,ES'}(x)=\rho_{ES}(x)-\rho_{ES'}(x)$, their maximal deviations $\Delta \rho_{ES,ES'}=\max_x |\rho_{ES,ES'}(x)|$, and the corresponding threshold $\epsilon_{\rho_{opt}}$. As second condition on  $\mathcal{ES}_{opt}$ we require $\Delta \rho_{ES,ES'}<\epsilon_{\rho_{opt}}$ for \emph{all} $ES,ES'\in\mathcal{ES}_{opt}$.

\begin{figure}[H]
	\centering
	\includegraphics[width=0.48\columnwidth]{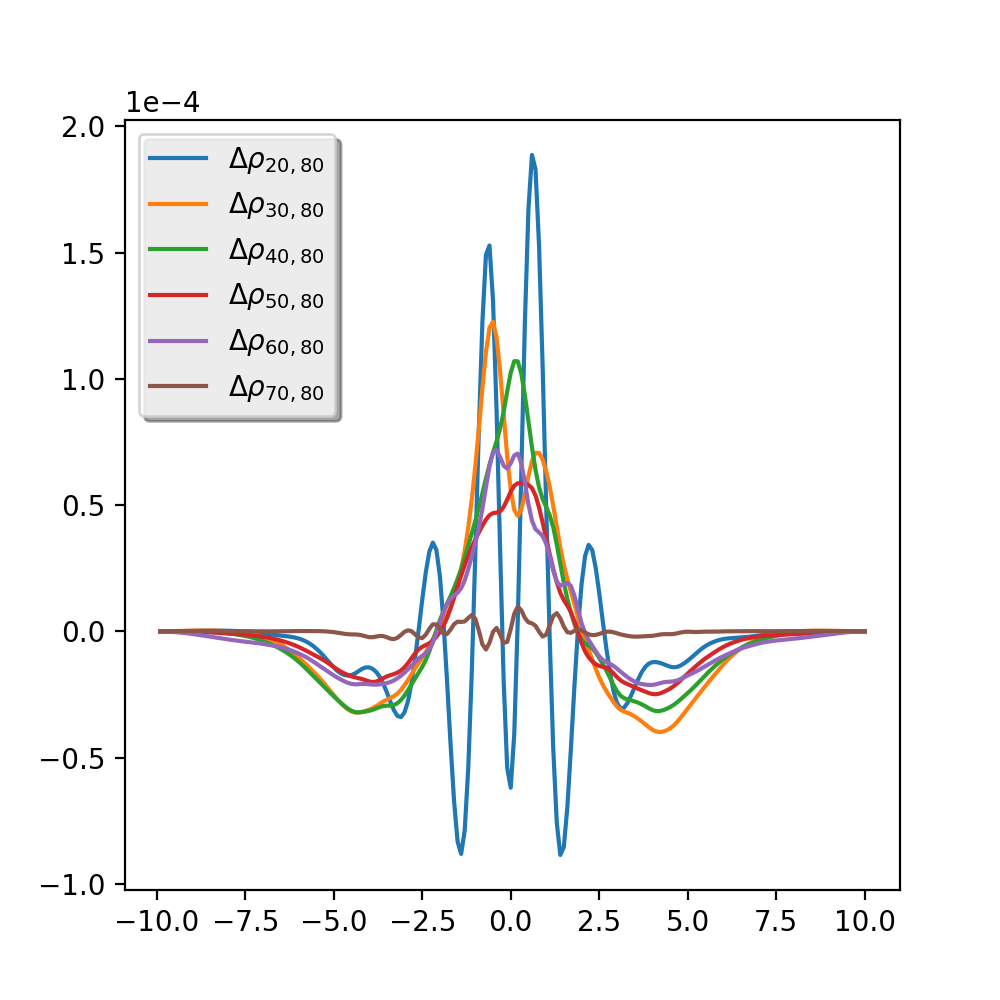} \hfill
	\includegraphics[width=0.48\columnwidth]{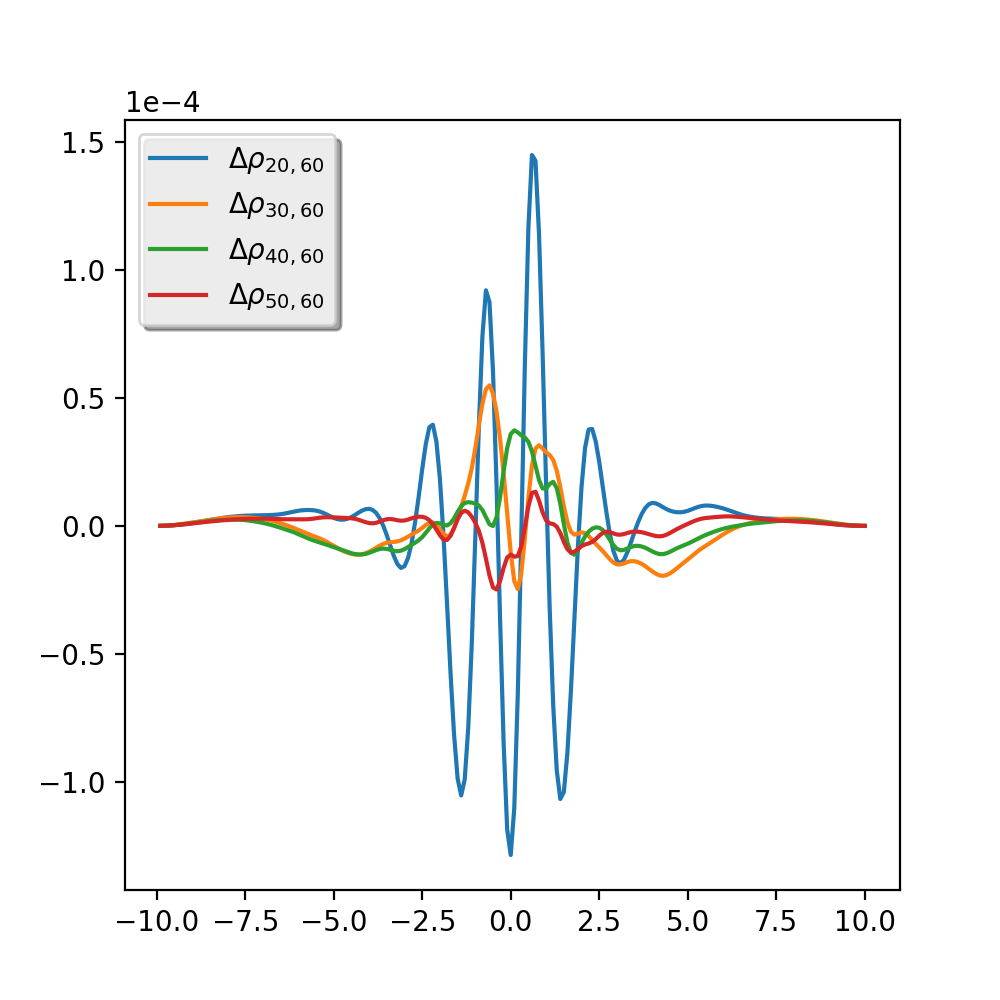}
	\caption{Differences in the ground-state density $\Delta\rho_{ES,ES'}(x)$ for RDMFT calculations of the He atom with $ES'=80$ (left) and $ES'=60$ (right.) $\rho_{ES=20}(x)$ deviates similarly for both reference points, but for $ES_{ref}=80$, we see a systematic (instead of random) $\Delta\rho_{ES}(x)$ for $30\leq ES\leq 60$. The deviations merely drop under $10^{-4}$, except for $ES=70$, which is very close to the reference. When we instead use $ES'=60$, the calculations for $ES=30$ to $ES=60$ only deviate of the order of $10^{-5}$ and the deviations have a random character.}
	\label{fig:HE_RDMFT_DeltaN}
\end{figure}

We start with the investigation of the energies and find that $\Delta E_{ES,ES'}<5 \cdot 10^{-6}$ for all combinations $30 \leq ES,ES'\leq 60$, but $\Delta E_{ES,ES'}>10^{-5}$ when we choose $30 \leq ES\leq 60$ and $30 \leq ES'\leq 60$ or $10 \leq ES'\leq 20$. We conclude that the first condition for $\mathcal{ES}_{opt}$ is met by the interval $30 \leq ES \leq 60$ with threshold $\epsilon_{E_{opt}}= 5 \cdot 10^{-6}$. 
For the investigation of the second condition, we depict in Fig. \ref{fig:HE_RDMFT_DeltaN} the density deviations $\rho_{ES,ES'}(x)$ with $ES'=60$, the upper boundary of the just found interval and $ES'=80$, that corresponds to the largest basis set of this example. For a better visibility, the curve for $ES=10$ is not shown, but it deviates stronger than all the other curves from both $ES'$. We conclude that $\Delta\rho_{ES,ES'}$ goes down until $ES=20$, independently of the reference. However, for $ES\geq 30$ this is not the case any more. We find  $\Delta\rho_{ES,80}\approx10^{-4}$ but $\Delta\rho_{ES,60}< 5\cdot 10^{-5}$. Additionally, we observe two different \emph{types of deviations}: The curves $\rho_{ES,80}(x)$ have a similar form for all $30\leq ES\leq 60$, but when we change the reference point to $ES'=60$, we cannot find pronounced similarities which is what we would expect from fluctuations. When we test also the other possible values for $ES'$, we find $\Delta\rho_{ES,ES'}< 5\cdot 10^{-5}$ for all $30\leq ES,ES'\leq 60$. Thus, the second condition for the optimal region is met by the same interval like the first condition with threshold $\epsilon_{\rho_{opt}}=5\cdot 10^{-5}$. Therefore, we have $\mathcal{ES}_{opt}=\{ES\, |\, 30 \leq ES \leq 60\}$ and conclude that the maximum possible accuracy of RDMFT calculations is already reached for $ES=30$ and it is lower than for HF$_{basis}$.

The strong decrease in accuracy of RDMFT in comparison to HF$_{basis}$ suggests that the occupation number minimization adds a significant error to the entire calculation. The current implementation uses a Broyden–Fletcher–Goldfarb–Shanno algorithm\footnote{A standard routine from the GSL-library. For details, see \url{https://www.gnu.org/software/gsl/doc/html/multimin.html}.} that describes the region around the minimum by a second order Taylor expansion and approximates the corresponding coefficient matrix (Hessian) at every iteration step by a combination of the gradients of the current and the previous iterations, which makes the routine numerically very efficient. It is very difficult to estimate the exact error that is introduced by the method, because of the strong non-linear character of the minimization (especially the interdependence between the $\phi_i$ and $n_i$ optimizations.)
For a thorough understanding of this issue, one needs to implement and test different numerical solvers. However, for the purposes of this paper, we consider the current accuracy as sufficient.

\subsection{Basis-set convergence of dRDMFT and comparison to RDMFT}
For the $ES$-series of dRDMFT, we need to deal with the combination of the inaccuracies introduced by the $n_i$-minimization, explained in the last section and the additional errors due to extra large basis sets that contain many redundant degrees of freedom, that we found for dHF before (Sec. \ref{APP:dHF}.) We know already that a large photon frequency is advantageous in terms of the latter. So let us choose $\omega=5.0$ to make this error as small as possible.  

As expected, it is much harder to define a convergence region in the way we did before for electronic RDMFT. We find the lowest energy at $ES=50$, which deviates from $ES=40$ and $ES=60$ by $|E_{ES=40/60}-E_{ES=50}| > 10^{-5}$. So we cannot find a region, where the energy is converged \emph{better} than $10^{-5}$. Nevertheless, if we accept an accuracy of $\Delta E_{ES}<5\cdot 10^{-5}$ as sufficient, we find a region as large as $20\leq ES\leq 100$ that satisfies the criterion. A look in the density deviations reveals that we can slightly tighten this region and exclude $ES=20$, such that we have deviations $\Delta \rho_{ES,50}\approx10^{-4}$ for all $30\leq ES\leq 100$. We found similar values also for other examples and we can conclude that we lose about one order of magnitude of accuracy for dRDMFT in comparison to RDMFT. Careful fine-tuning of the numerical parameters could improve this, but the current accuracy suffices for our purposes.

Finally, we present the consistency check between RDMFT and dRDMFT in the $\lambda=0$ limit. We find $|E_{RDMFT}-E_{dRDMFT}|\approx 10^{-5}$ and $\max_x|\rho_{RDMFT}(x)-\rho_{dRDMFT}(x)|\approx 10^{-4}$, which means that the deviations between the levels of theory are of the same order as the maximal accuracy that dRDMFT provides. We conclude that both theories are consistent.


\section{Protocol for the convergence of a dHF/dRDMFT calculation}
\label{APP:protocol}
We conclude this supporting information with a step-by-step guide for the proper usage of the dressed orbital implementation in Octopus. All the calculations, presented in the main part of this paper were performed according to this protocol
\begin{enumerate}
	\item Box length $L_x$ and spacing $dx$ convergence for the purely electronic part of the system on the level of IP and electronic HF and for the uncoupled photonic system on the level of IP. 
	\begin{itemize}
		\item Test the deviations in energy $\Delta E_{L_x}<\epsilon_{E_{L_x}}$ and density $\Delta \rho_{L_x}<\epsilon_{\rho_{L_x}}$, as explained in Sec. \ref{APP:ValidationMB}. We chose $\epsilon_{E_{L_x}}=10^{-8}$ and $\epsilon_{\rho_{L_x}}=10^{-5}$ to exclude any numerical artefacts. However, as the dHF and dRDMFT calculations do typically not reach such precisions, one can relax these criteria in general.
		\item For the $dx$-series, test only the deviations in energy $\Delta E_{dx}<\epsilon_{E_{dx}}$ due to the larger density errors. We chose $\epsilon_{E_{dx}}=10^{-8}$, but like for the box length, this criterion can be relaxed.
	\end{itemize}
	\item Basis size convergence for the HF$_{basis}$ routine with the purely electronic part of the system.
	\begin{itemize}
		\item  Perform an $ES$-series and test the deviations in energy $\Delta E_{ES,ES_{ref}}<\epsilon_{E_{ES}}$ and in density $\Delta \rho_{ES,ES_{ref}}<\epsilon_{\rho_{ES}}$ as explained in Sec. \ref{APP:HFvsHFbasis}. Here, we were typically able to reach $\epsilon_{E_{ES}}=10^{-8}$ and $\epsilon_{\rho_{ES}}=10^{-4}$.\footnote{Note that in Sec. \ref{APP:HFvsHFbasis}, we wrote instead $\Delta \rho_{ES,ES_{ref}}\approx 10^{-5}$ because some values were slightly larger than $10^{-5}$. Thus, $\epsilon_{E_{ES}}=10^{-4}$ for sure is satisfied.}
		Again, these criteria can be relaxed.
		\item Compare the converged HF$_{basis}$ and the electronic HF results in energy and density and make sure that both are consistent on their level of accuracy.
	\end{itemize}
	\item Basis size convergence of the dressed theory that is wanted (dHF or dRDMFT) in the no-coupling ($\lambda=0$) limit 
	\begin{itemize}
		\item Perform an $ES$-series like for HF$_{basis}$. Note that one needs to expect considerably larger basis sets for the same level of convergence (see Sec. \ref{APP:dHF_no_coupling} for details.)
		\item Check consistency of the electronic sub-part of the system with HF$_{basis}$ as mentioned in Sec. \ref{APP:dHF_no_coupling}. If this check fails drastically, this is very probably due to the violation of the extra exchange symmetry in the photonic coordinates (see Sec. \Nnameref{sec:fermionization} in the main text.) 
	\end{itemize}
	\item The convergence study is finished with another basis-set convergence for $\lambda>0$. Typically, we also performed another small box length series with the converged basis set to make sure that the coupling does not increase the size of the system crucially such that boundary effects could influence the results.
\end{enumerate}

\section{The bosonic symmetry of the photon wave function}
\label{app:symmetry_photon}
In this appendix we go into a little more detail and show how the mode-representation, which makes the bosonic symmetry explicit, arises and introduce in this setting the usual bosonic density matrices~\cite{Giesbertz2019}. Instead of starting with the displacement representation we start with the definition of the single-particle Hilbert space and its Hamiltonian. We choose the single-particle Hilbert space $\mathcal{H}_1$ to consist of $M$ orthogonal states $\ket{\alpha}$. These states are defined by the eigenstates of the Laplacian with fixed boundary conditions and geometry and correspond to the Fourier modes of the electromagentic field~\cite{greiner2013field, Ruggenthaler2014}. This real-space perspective is a natural choice if one either wants to connect to quantum mechanics and deduce the Maxwell field from gauge independence of the electronic wave function~\cite{greiner2013field}, or when deducing the theory in analogy to the Dirac equation~\cite{Gersten-1999}. It is this analogy of Maxwell's equations as a single-photon wave function with spin 1 that makes the appearance of a bosonic symmetry most explicit when quantizing the theory~\cite{keller2012quantum}. We note, however, that in general the concept of a photon wave function can become highly non-trivial~\cite{Scully1999}. Since we work directly in the dipole approximation we do not go through all the steps of the usual quantization procedure of QED but from the start assume that we have chosen a few of these modes $\ket{\alpha}$ (with a certain frequency and polarization) in Coulomb gauge~\cite{Ruggenthaler2014}. The single-particle Hamiltonian in this representation is then given by
\begin{align*}
	\hat{h}_{ph}' = \sum_{\alpha=1}^{M} \omega_{\alpha} \ket{\alpha} \bra{\alpha}.
\end{align*} 
Since a total shift of energy does not change the physics and for later reference, we can equivalently use $\hat{h}_{ph} = \sum_{\alpha=1}^{M} \left( \omega_{\alpha} + \tfrac{1}{2} \right)\ \ket{\alpha} \bra{\alpha}$. Therefore, the energy of a single-photon wave function $\ket{\phi} = \sum_{\alpha=1}^{M} \phi(\alpha) \ket{\alpha}$ (corresponding to the classical Maxwell field in Coulomb gauge~\cite{keller2012quantum}) is given by
\begin{align*}
	E[\phi] = \sum_{\alpha,\beta =1}^{M} \phi^{*}(\beta) \underbrace{\braket{\beta|\hat{h}_{ph}|\alpha}}_{= \hat{h}_{ph}(\beta,\alpha)} \phi(\alpha) = \sum_{\alpha, \beta}  \hat{h}_{ph}(\beta,\alpha) \gamma_{b}(\alpha, \beta) = \sum_{\alpha} \left(\omega_{\alpha}+ 
	\frac{1}{2}\right)\underbrace{\gamma_b(\alpha,\alpha)}_{|\phi(\alpha)|^2}.
\end{align*}
Here we have introduced the single-particle photonic 1RDM $\gamma_{b}(\alpha, \beta) = \phi^{*}(\beta) \phi(\alpha)$. We can then extend the single-particle space and introduce photonic many-body spaces $\mathcal{H}_{N_{b}}$ which are the span of all \textit{symmetric} tensor products of single-particle states of the form~\cite{Giesbertz2019,Leeuwen2013}
\begin{align*}
	\ket{\alpha_1,...,\alpha_{N_b}} = \frac{1}{\sqrt{N_b!}} \sum_{\wp} \ket{\wp(\alpha_1)}...\ket{\wp(\alpha_{N_b})},
\end{align*}
where $\wp$ goes over all permutations of $\alpha_1,..., \alpha_{N_{b}}$. This construction is completely analogous to the typical construction of the fermionic many-body space with the only difference having minus sings in front of odd permutations. Such a many-body basis is not normalized for bosons, as states can be occupied with more than one particle. Thus, the normalization factor occurs in the corresponding resolution of identity, i.e. $\mathbb{1}=\frac{1}{N_b!}\sum_{\alpha_1,...,\alpha_{N_b}=1}^{M}\ket{\alpha_1,...,\alpha_{N_b}}$ $\bra{\alpha_1,...,\alpha_{N_b}}$. This approach is explained in great detail in Ref. \citenum{stefanucci2013nonequilibrium}. An $N_b$-particle Hamiltonian is then given by a sum of individual single-particle Hamiltonians (interactions among the photons will only come about due to the coupling with the electrons.) Introducing for a general $N_b$ photon state $\ket{\tilde{\phi}} =  \tfrac{1}{\sqrt{N_b!}}\sum_{\alpha_1,...,\alpha_{N_b}=1}^{M} \tilde{\phi}(\alpha_1,...,\alpha_{N_b}) \ket{\alpha_1,...,\alpha_{N_b}}$ with $\tilde{\phi}(\alpha_1,...,\alpha_{N_b})=\frac{1}{\sqrt{N_b!}}\braket{\alpha_1,...,\alpha_{N_b}|\tilde{\phi}}$ the corresponding 1RDM according to Eq.~\eqref{eq:pRDM_boson} as $\gamma_{b}(\alpha,\beta) = N_{b} \sum_{\alpha_2,...,\alpha_{N_{b}}} $ $\tilde{\phi}^*(\beta, \alpha_2,...,\alpha_{N_b})$ $\tilde{ \phi}(\alpha, \alpha_2,...,\alpha_{N_b})$, the energy of that state is given by
\begin{align*}
	E[\tilde{\phi}] = \sum_{\alpha, \beta=1}^{M} \hat{h}_{ph}(\beta,\alpha) \gamma_{b}(\alpha,\beta) = \sum_{\alpha=1}^{M} \left(\omega_{\alpha} + \frac{1}{2} \right) \gamma_{b}(\alpha,\alpha).
\end{align*}
Such a state can be constructed, for instance, as a permanent of $N_b$ single-photon states $\phi(\alpha)$. Note further that the 1RDM of an $N_b$ photon state obeys $N_b = \sum_{\alpha} \gamma_{b}(\alpha,\alpha)$.

Finally, since we want to have a simplified form of a field theory without fixed number of photons, we make a last step and represent the problem on a Hilbert space with indetermined number of particles, i.e., a Fock space. By defining the vacuum  state $\ket{0}$, which spans the one-dimensional zero-photon space, the Fock space is defined by a direct sum of $N_b$-photon spaces $\mathcal{F}=\bigoplus_{N_b=0}^{\infty}\mathcal{H}_{N_b}$. Introducing the ladder operators between the different photon-number sectors of $\mathcal{F}$ by~\cite{Giesbertz2019}
\begin{align*}
	&\hat{a}_{\alpha}^{+} \ket{\alpha_1,...,\alpha_{N_{b}}} = \ket{\alpha_1,...,\alpha_{N_{b}}, \alpha}
	\\
	&\hat{a}_{\alpha} \ket{\alpha_1,...,\alpha_{N_{b}}} = \sum_{k=1}^{N_b} \delta_{\alpha_k,\alpha} \ket{\alpha_1,..., \alpha_{k-1}, \alpha_{k+1},...,\alpha_{N_{b}}}
\end{align*}
with the usual commutation relations, we can lift the single-particle Hamiltonian to the full Fock space and arrive at Eq.~\eqref{eq:PhotonHamiltonian}. The Fock space 1RDM for a general Fock space wave function $\ket{\Phi}$ can then be expressed as
\begin{align*}
	\gamma_{b}(\alpha,\beta) = \braket{\Phi| \hat{a}^{+}_{\beta} \hat{a}_{\alpha} \Phi},
\end{align*}
and $\sum_{\alpha=1}^{M} \gamma_{b}(\alpha,\alpha) = N_b$ now corresponds to the \emph{average} number of photons. And finally, since we know that Eq.~\eqref{eq:PhotonHamiltonian} is equivalent to $\hat{H}_{ph}= \sum_{\alpha=1}^M\left(- \tfrac{1}{2} \tfrac{\partial^2}{\partial p_{\alpha}^2}+ \tfrac{\omega_{\alpha}^2}{2}p_{\alpha}^2\right)$, we also see that the Fock space $\mathcal{F}$ is isomorphic to $L^{2}(\mathbb{R}^{M})$, which closes our small detour.


\bibliography{bibliography}


%% file: 1introduction.tex
Experiments performed in the last decades (see, e.g., Refs.~\citenum{Ebbesen2016, sukharev2017}) have made accessible the strong and ultra-strong\footnote{We follow the definition of the light-matter coupling regimes of Ref.~\citenum{Kockum2018}.} interaction regime between matter degrees of freedom and the quantized modes of optical cavities, which allows for the study of many new phenomena including modification of chemical reaction rates~\cite{Hutchison2012, Thomas2016}, interacting photons in quantum non-linear media~\cite{Firstenberg2013} or super-radiance of atoms in a photonic trap~\cite{Goban2015}. At the same time it creates opportunities such as the modification of energy-transfer pathways within photosynthetic organisms~\cite{Coles2014}, an increase of conductivity in organic semiconductors hybridized with the vacuum field~\cite{Orgiu2015} or the generation of long-distance molecular interactions that, for example, allow for energy transfer way beyond the short-range dipole-dipole mediated transfer (Förster theory)~\cite{Andrew2004}.
All these phenomena are related by the emergence of hybrid light-matter quasi-particle states, called polaritons, that determine the properties of the respective coupled electron-photon system. The physics of these exotic states can be understood impressively well by model systems of Dicke-type~\cite{Dicke1954}, meaning several few-level systems coupled to some photon modes~\cite{Kockum2018}. Many experimentally found features of the (ultra-)strong coupling regime could be described~\cite{Todorov2010,Michetti2015,Cwik2016,Galego2016,Feist2018} and much exciting new physics was predicted~\cite{Feist2015,Herrera2016,Cirio2016,Kockum2017,DeLiberato2014} using such models. This article focuses on the influence of (ultra-)strong coupling on the ground state of light-matter systems, a topic on which considerably less literature exists. Only recently, polaritonic ground states, which are believed to be fundamental for the understanding of polaritonic chemistry~\cite{Ebbesen2016}, have been started to be investigated in more detail~\cite{DeLiberato2017,Kockum2018}. 
However, limits and difficulties of few level-approximations have been pointed out~\cite{Gely2017,Bernardis2018,SanchezMunoz2018,Jaako2016,schafer2018ab,Schafer2018} and recently, new models have been used to investigate polaritonic chemistry~\cite{Galego2015,Galego2016,kowalewski2016non,garcia2017long,zeb2017exact,luk2017multiscale,del2018tensor,Ruggenthaler2018,Schafer2018,groenhof2018coherent,vendrell2018collective,reitz2019langevin,triana2019revealing,Galego2019,csehi2019ultrafast}.
Still, many questions remain open, especially whether the collective (ultra-)strong coupling, predicted by the Dicke-model can actually modify ground state properties of single molecules\cite{Feist2015,Martinez-Martinez2018}. Another example is the ongoing discussion on the theoretical understanding of super-radiance~\cite{Viehmann2011,Bernardis2018}. These debates suggest that there is a need for new theoretical tools that treat matter and photons at the same level of theory~\cite{Ruggenthaler2018,schafer2018ab,Schafer2018}. Up to now, standard theoretical modelling treats in detail either the photons or the matter, which becomes insufficient when the matter and photon degrees of freedom are equally important~\cite{schafer2018ab}. We present in this paper further evidence that the impact of the light-matter interaction on the matter is far from trivial and can change from system to system.

However, the full quantum-mechanical description of just the electronic degrees of freedom is already computationally very challenging due to the exponential scaling of the wave function.\footnote{Imagine for instance a 4-electron system in one spatial dimension, coupled to one cavity mode (which corresponds to the Beryllium example, presented in Sec. \Nnameref{sec:results}.) Every eigenstate of the corresponding Hamiltonian would be a function that depends on 5 variables. Thus, if we wanted to calculate the exact ground state, assuming 100 grid points per coordinate (corresponding to e.g. a box size of 20 with spacing 0.2) and 8 byte per function value (double precision,) we would need $100^5\cdot8 \text{ Byte}\approx 75$ Gigabyte of working memory, which is close to the edge of current high performance technology.} Instead, reformulations of the many-body problem in terms of reduced quantities like the electron density~\cite{Gross1990,dreizler2011density} or the Green's function~\cite{Walecka2003,stefanucci2013nonequilibrium,Onida2002} have been shown to provide accurate results for relatively low computational costs. Thus, working with reduced quantities seems to be a natural choice also for coupled light-matter systems\cite{Theophilou2018}. Recently, quantum-electrodynamical density-functional theory (QEDFT) was introduced as an extension to pure electronic density-functional theory (DFT)~\cite{Ruggenthaler2011, Tokatly2013, Ruggenthaler2014, Ruggenthaler2015}. First calculations showed the feasibility of QEDFT~\cite{Flick2015,Flick2017} leading to the possibility to perform full first-principles calculations of real molecules coupled to cavity modes~\cite{Flick2018a,Flick2018b}. 
However, standard approximations in DFT (usually based on a non-interacting auxiliary Kohn-Sham system) become inaccurate if applied to strongly-correlated systems. This is very well studied in terms of the strong-correlation regime in electronic systems~\cite{Cohen2012} and also observed for the existing QEDFT functionals that approximate the electron-photon interaction~\cite{Flick2018a}. Consequently, to study novel effects arising in the ultra-strong~\cite{Niemczyk2010} or deep-strong coupling regime~\cite{Bayer2017} of light and matter from first principles, i.e., without resorting to simplified few-level systems, one needs to develop new functionals for the combined matter-photon systems or explore alternative many-body methods.

Reduced density-matrix functional theory (RDMFT) is such a method. RDMFT is based on the electronic one-body reduced density matrix (1RDM) instead of the electronic density (as in DFT) as its basic functional variable~\cite{Gilbert1975}. Similarly to DFT, approximations are necessary in RDMFT as it is not known how to express explicitly all expectation values of operators in terms of the 1RDM. Simple approximations within this approach~\cite{Muller1984} have proven very efficient in dealing with difficult electronic structure problems like the correct qualitative description of a dissociating molecule~\cite{Goedecker1998} or the prediction of the Mott-insulating phase of certain strongly-correlated solids~\cite{Sharma2013}. Thus, it seems worth to explore how RDMFT performs in describing also the strong interaction between molecular systems and cavity modes.

\begin{figure}
	\def\svgwidth{0.5\columnwidth}
	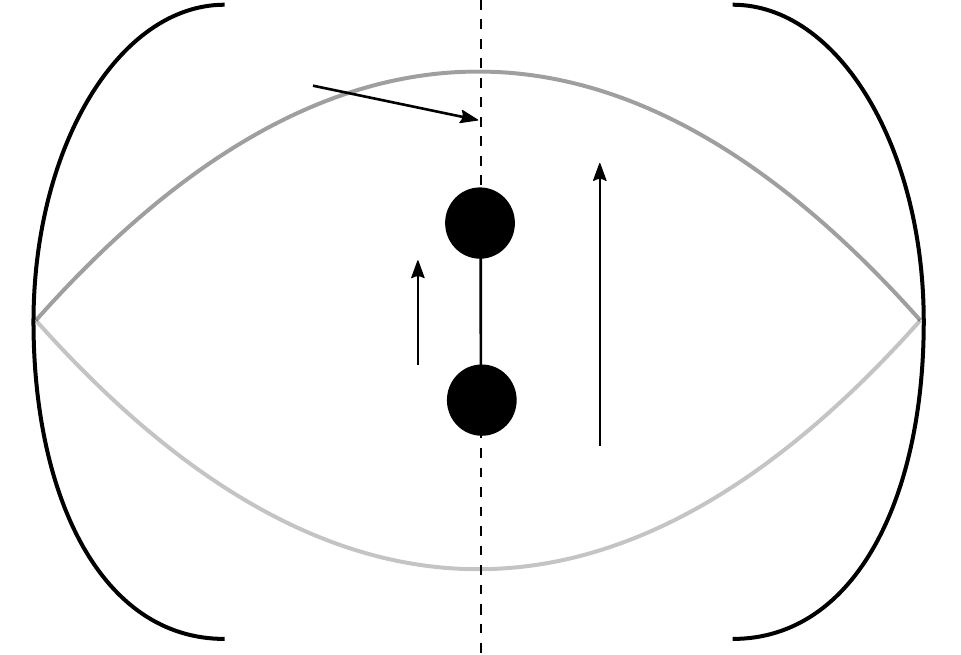
	\caption{Typical setting of a cavity experiment. A matter system (here represented by a diatomic molecule) is put inside an optical cavity that enhances specific modes of the electromagnetic field (here represented by the lowest cavity mode, but in principle many modes can become important.) By that, the coupling between the matter system and the light modes can be considerably enhanced with respect to the free space. The dipole of the molecule should to be aligned with the polarization of the enhanced mode and its position is assumed at the field maximum. Note that in principle also higher multipole moments can become important.}
	\label{fig:cavity_setting}
\end{figure}
Specifically, we will discuss in this work the properties of coupled light-matter systems in a setting that resembles typical cavity experiments (see Fig. \ref{fig:cavity_setting}.) It turns out that transferring RDMFT to such systems involves overcoming additional difficulties in contrast to the DFT framework. The reason behind this are the conditions under which the corresponding reduced density matrices (RDMs) (which will be purely electronic, purely photonic and coupled) connect to the original wave function, which is crucial to construct a well-defined reduced density-matrix framework. Already for the purely electronic system the conditions under which such a connection exists (known as N-representability conditions) are not trivial~\cite{Coleman1963, Klyachko2006, Mazziotti2012}. For RDMs in matter-photon systems, these conditions are entirely unexplored. Nevertheless, we manage to overcome this difficulty by mapping the original system to a higher-dimensional auxiliary system that allows for an effective description of the problem by fermions. Hence, the corresponding RDMs connect to the auxiliary wave functions under N-representability conditions of fermionic systems. Despite being fermions, the newly introduced quasi-particles will depend on electronic and bosonic degrees of freedom. This construction was recently introduced by some of the authors in Ref. \citenum{Nielsen2018} and used to construct a DFT scheme specifically for the strong-coupling regime of light-matter systems. In the auxiliary system, the Hamiltonian consists of only 1- and 2-body terms for the new quasi-particles, thus it has the same structure as the conventional electronic Hamiltonian for molecular systems, which allows us to apply electronic RDMFT theory without major modifications. We will present some results for model systems with the simplest known RDMFT functional, the Müller functional~\cite{Muller1984}, and show that this dressed RDMFT is accurate from the weak to the strong-coupling regime. Then, we will present two examples highlighting that how matter reacts to the interaction with photons depends strongly on the system. For instance, for the same coupling strength, the repulsion between the particles can be locally suppressed in one system but enhanced in another. We finish with commenting on some open issues and challenges for future applications.

This article is structured as follows. First, we explain the physical setting of an electronic system in a cavity in Sec. \Nnameref{sec:phys_setting}.  In Sec. \Nnameref{sec:coupledRDMTheory}, we discuss the RDMs that appear in the ground-state energy expression of the coupled light-matter system. We will explain the difficulties that are introduced by the coupling between fermions and bosons and having non-particle-conserving terms. In Sec. \Nnameref{sec:fermionization}, we introduce the higher-dimensional auxiliary system which will allow us to avoid most of the aforementioned difficulties. We show how to construct a proper RDMFT framework in this system in Sec. \Nnameref{sec:dRDMFT}, explain our numerical implementation in Sec. \Nnameref{sec:implementation}, and present the numerical results in Sec. \Nnameref{sec:results}. We finish with discussing possibilities as well as challenges of dressed RDMFT in Sec. \Nnameref{sec:conclusion}.

\section*{Physical Setting}
\label{sec:phys_setting}
To describe weak and strong matter-photon interaction, it is necessary to go beyond typical quantum-chemistry or solid-state physics theories that describe electrons in a local potential interacting via Coulomb interaction and being perturbed by a classical external electromagnetic field. Instead, we need to treat explicitly the quantum nature of light and the back-reaction between electrons and electromagnetic field excitations (photons)~\cite{Ruggenthaler2018}. Therefore, the framework of quantum electrodynamics (QED) needs to be employed. However, we do not want to treat QED in its full complexity but will apply some well-established approximations (see Ref. \citenum{Ruggenthaler2018} for a detailed discussion.) First, we apply the Born-Oppenheimer approximation and treat the nuclei as fixed classical particles.\footnote{The extension to also include the nuclei as quantum particles is in principle straightforward by following, e.g., similar strategies like discussed in Refs.~\citenum{Flick2017, schafer2018ab}}
Second, we work in the non-relativistic limit, which for the typical energy scales of molecules and their low-energy excitations is usually sufficient. Third, we assume that the wavelength $\lambda$ of the relevant electromagnetic modes is much larger than the spatial extension $d$ of the electronic system ($\lambda \gg d$) such that the dipole approximation (here in the Coulomb gauge) is valid~\cite{Grynberg2010, Rokaj2018}. In the case of the dipole approximation, where every photon mode couples to all Fourier components of the charge current of the electronic subsystem~\cite{Ruggenthaler2014,Ruggenthaler2015}, an effective description with only a few modes is usually sufficient. The continuum of modes is then effectively taken into account by using instead of the bare mass the physical mass of the electrons~\cite{Flick2018b,schafer2018ab}. Since we focus on equilibrium situations, the openness of the cavity can be neglected.

Therefore, the basic Hamiltonian that we use to describe strongly-coupled light-matter systems reads (we use atomic units throughout)
\begin{align}
\label{eq:Hamiltonian}
\hat{H} &= \sum_{k=1}^N \left[ -\tfrac{1}{2} \nabla_{\br_k}^2 + v(\br_{k}) \right] + \tfrac{1}{2}\sum_{k \neq l} w(\br_k,\br_l) + \tfrac{1}{2} \sum_{\alpha=1}^M \left[ -\tfrac{\partial^2}{\partial p_{\alpha}^2} + \left(\omega_{\alpha} p_{\alpha} - \blambda_{\alpha} \cdot \sum_{k=1}^N \br_k \right)^2\, \right].
\end{align}
Here the first two sums on the right-hand side correspond to the usual electronic many-body Hamiltonian $\hat{H}_e=\hat{T} + \hat{V} + \hat{W}$, used to describe the uncoupled matter system consisting of $N$ electrons in an external potential $v(\br)$ interacting via the Coulomb repulsion $w(\br,\br')$. The third sum describes $M$ photon modes, that are characterized by their elongation $p_{\alpha}$, frequency $\omega_\alpha$ and polarization vectors $\blambda_{\alpha}$. The polarization vectors include already the effective coupling strength $g_{\alpha}=|\blambda_{\alpha}|\sqrt{\frac{\omega_{\alpha}}{2}}$\cite{Ruggenthaler2018} and couple to the total dipole  $\hat{\mathbf{D}}=\sum_{k=1}^N \br_k$ of the electronic system. The sum can be decomposed in a purely photonic part $\hat{H}_{ph}= \sum_{\alpha=1}^M\left(- \tfrac{1}{2} \tfrac{\partial^2}{\partial p_{\alpha}^2}+ \tfrac{\omega_{\alpha}^2}{2}p_{\alpha}^2\right)$, the dipole self-interaction\footnote{Note that this term is necessary for the existence of a ground state\cite{Rokaj2018}.} $\hat{H}_d=\sum_{\alpha=1}^M \tfrac{1}{2}\left(\blambda_{\alpha} \cdot \hat{\mathbf{D}}\right)^2$, which we split again for later convenience in its one-body $\hat{H}_d^{(1)}=\sum_{\alpha=1}^{M}\sum_{k=1}^{N}\tfrac{1}{2}(\blambda_{\alpha}\cdot \br_k)^2$ and two-body part $\hat{H}_d^{(2)}=\sum_{\alpha=1}^{M}\sum_{k\neq l } \tfrac{1}{2} (\blambda_{\alpha}\cdot \br_k) (\blambda_{\alpha}\cdot \br_l)$, and the bilinear interaction $\hat{H}_I=-\sum_{\alpha=1}^M\omega_{\alpha} p_{\alpha}\blambda_{\alpha} \cdot \hat{\mathbf{D}}$. Some comments on this Hamiltonian are appropriate. Since we work in Coulomb gauge, the dipole-self energy term arises only for the transversal but not for the longitudinal part of the field\cite{Spohn2004}. However, if we assume a cavity then the Coulomb interaction is modified\cite{Power1982}. We can easily incorporate this into our framework since $w(\br,\br')$ is completely at our disposal and we do not rely on any kind of Coulomb approximation. Thus, we can also treat the influence of e.g., a plasmonic environment\cite{Shahbazyan2013}.

The ground-state wave function of Eq.~\eqref{eq:Hamiltonian} is a function of $4N+M$ coordinates 
\begin{align}
\label{eq:wavefunction}
\Psi(\br_1\sigma_1,...,\br_N\sigma_N;p_1,...,p_M),
\end{align} 
where $\sigma_k$ are the electronic spin degrees of freedom. The wave function $\Psi$ is as usual anti-symmetric with respect to the exchange of any two electron coordinates $\br_j\sigma_j\leftrightarrow\br_k\sigma_k$, and also depends on $M$ photon-mode displacement coordinates $p_{\alpha}$. 

At this point, we want to remind the reader that there is no fundamental symmetry with respect to the exchange of two displacement coordinates $p_{\alpha}$ with $p_{\beta}$. The bosonic symmetry instead refers to the exchange of mode \emph{excitations}, which are interpreted as photons in the number-state representation. To see this, we first use the ladder operators $\hat{a}_{\alpha} =\sqrt{\tfrac{\omega_{\alpha}}{2}}(p_{\alpha}-\tfrac{1}{\omega_{\alpha}}\tfrac{\partial}{\partial p_{\alpha}})$  and $\hat{a}^{+}_{\alpha} =\sqrt{\tfrac{\omega_{\alpha}}{2}}(p_{\alpha}+\tfrac{1}{\omega_{\alpha}}\tfrac{\partial}{\partial p_{\alpha}})$ to represent the sum of harmonic Hamiltonians, i.e., 
\begin{align}
\label{eq:PhotonHamiltonian}
\hat{H}_{ph}=\sum_{\alpha=1}^M\omega_{\alpha}\left(\hat{a}_{\alpha}^{+}\hat{a}_{\alpha}^{\vphantom{+}} + \tfrac{1}{2}\right).
\end{align}
Eigenstates of the individual $\hat{a}_{\alpha}^{+} \hat{a}_{\alpha}$ in the above representation are given by multiple applications of creation operators to the vacuum state $\ket{0}$, i.e., $\ket{\varphi^n_{\alpha}} = (\hat{a}^{+}_{\alpha})^{n} \ket{0}$.\footnote{Note that following Refs. \citenum{Giesbertz2019,stefanucci2013nonequilibrium}, we chose here explicitly a non-normalized basis $\{\ket{\varphi^n_{\alpha}}\}$ of the n-photon sector, with $\braket{\varphi^n_{\alpha}|\varphi^n_{\alpha}}=n!$, which allows for a simpler definition of the bosonic 1RDM. The missing normalization factor is shifted to the resolution of identity in this basis, i.e. $\mathbb{1}=\frac{1}{N_b!}\sum_{\alpha_1,...,\alpha_{N_b}=1}^{M}\ket{\alpha_1,...,\alpha_{N_b}}\bra{\alpha_1,...,\alpha_{N_b}}$, where $\ket{\alpha_1,...,\alpha_{N_b}}=\hat{a}^{+}_{\alpha_1}\cdots \hat{a}^{+}_{\alpha_{N_b}}\ket{0}$ as defined later in the text.} These eigenstates are connected to the displacement representation by $\varphi^n_{\alpha}(p_{\alpha}) = \braket{p_{\alpha}|\varphi^n_{\alpha}} = \braket{p_{\alpha}|(\hat{a}^{+}_{\alpha})^{n}| 0}$. We can then express an $M$-mode eigenfunction as $\phi_{n_1,...,n_M}(p_1,...,p_{M}) = \braket{p_1...p_M|\phi_{n_1,...,n_M}} = \bra{p_1...p_M} \hat{a}^{+}_{1})^{n_1}...$ $(\hat{a}^{+}_{M})^{n_M}\ket{0}$. In this form it becomes clear that a multi-mode eigenstate $\ket{\phi_{n_1,...,n_M}}$ can be considered to consist of $N_b = n_1 + ... + n_M$ photons (mode excitations.) We can associate every such multi-mode eigenstate with a specific photon-number sector, i.e., the zero-photon sector is merely one-dimensional and corresponds to $\ket{\phi_{0_1,...,0_M}} \equiv \ket{0}$, the single-photon sector is $M$-dimensional and corresponds to the span of $\hat{a}^{+}_{\alpha} \ket{0} \equiv \ket{\alpha}$ for all $\alpha$ and so on. For the multi-photon sectors we see due to the commutation relations of the ladder operators the bosonic exchange symmetry appearing, e.g., $ \hat{a}^{+}_{\alpha_1} \hat{a}^{+}_{\alpha_2}\ket{0} = \hat{a}^{+}_{\alpha_2} \hat{a}^{+}_{\alpha_1}\ket{0} \equiv \ket{\alpha_1, \alpha_2}$ for $\alpha_1, \alpha_2 \in \{1,...,M\}$. It is no accident that the bosonic symmetry becomes explicit in this representation since the different modes $\alpha$ determine how the photon wave functions looks like in real space (see also the discussion in App. \ref{app:symmetry_photon}.) A general photon state can therefore be represented by a sum over all photon-number sectors as $\ket{\Phi} = \sum_{n=0}^{\infty} \left(\tfrac{1}{\sqrt{n!}} \sum_{\alpha_1,...,\alpha_n = 1}^{M} \tilde{\Phi}(\alpha_1,...,\alpha_n) \ket{\alpha_1,...,\alpha_n}\right)$.

%% file: typical_cavity_setting.pdf_tex
\begingroup%
  \makeatletter%
  \providecommand\color[2][]{%
    \errmessage{(Inkscape) Color is used for the text in Inkscape, but the package 'color.sty' is not loaded}%
    \renewcommand\color[2][]{}%
  }%
  \providecommand\transparent[1]{%
    \errmessage{(Inkscape) Transparency is used (non-zero) for the text in Inkscape, but the package 'transparent.sty' is not loaded}%
    \renewcommand\transparent[1]{}%
  }%
  \providecommand\rotatebox[2]{#2}%
  \ifx\svgwidth\undefined%
    \setlength{\unitlength}{276.58325436bp}%
    \ifx\svgscale\undefined%
      \relax%
    \else%
      \setlength{\unitlength}{\unitlength * \real{\svgscale}}%
    \fi%
  \else%
    \setlength{\unitlength}{\svgwidth}%
  \fi%
  \global\let\svgwidth\undefined%
  \global\let\svgscale\undefined%
  \makeatother%
  \begin{picture}(1,0.68178897)%
    \put(0,0){\includegraphics[width=\unitlength]{typical_cavity_setting.pdf}}%
    \put(0.15,0.58){\color[rgb]{0,0,0}\makebox(0,0)[lb]{\small\smash{\parbox{4cm}{maximum\\ field}}}}%
    \put(0.63359847,0.33888792){\color[rgb]{0,0,0}\makebox(0,0)[lb]{\footnotesize\smash{polarization}}}%
    \put(0.3067551,0.338393){\color[rgb]{0,0,0}\makebox(0,0)[lb]{\footnotesize\smash{dipole}}}%
    \put(0.28751586,0.21497632){\color[rgb]{0,0,0}\makebox(0,0)[lb]{\footnotesize\smash{molecule}}}%
  \end{picture}%
\endgroup%

%% file: 2rdms.tex
\section*{Reduced Density Matrices for coupled light-matter systems}
\label{sec:coupledRDMTheory}
\begin{table}
	\fbox{
		\begin{tabular}{lcp{9cm}}
			$\Psi(\br_1\sigma_1,...,\br_N\sigma_N;p_1,...,p_M)$ & - & electron-photon many-body state\\
			$\psi_{e}(\br_1\sigma_1,...,\br_N\sigma_N)$ & - & purely electronic many-body state\\
			$\psi_b(\alpha_1,...,\alpha_{N_b})$ & - & photonic many-body state in mode representation with fixed particle number\\
			$\Phi(\alpha_1,\alpha_2,...)$ & - & photonic many-body state in Fock space \\
			$\phi_e^i (\br)/\phi_b^{i}(\alpha)$ & - & electronic/photonic natural orbital
		\end{tabular}	
	}
	\caption{To ease reading, we highlight the different physical wave functions and the corresponding symbols that appear in this section.}
	\label{tab:wf_symbols}
\end{table}
Having introduced our system of interest, we now want to discuss how to find its ground state. A ground state (if it exists) is defined as the state (possibly degenerate) that has the lowest energy expectation value
\begin{align}
	\label{eq:var_principle_psi}
	E_0[\Psi]=\inf_{\Psi} \braket{\Psi|\hat{H}\Psi}.
\end{align}
This is the classical formulation of the variational principle due to Ritz and is well-defined for every Hamiltonian that is bounded from below. While well-known, Eq.~\eqref{eq:var_principle_psi} has the disadvantage that in practice the minimization has to be performed over an enormous configuration space that is spanned by all possible many-body wave functions. A possible reduction of computational complexity presents itself by the fact that the full wave function is usually not necessary to compute the energy expectation value but typically only reduced quantities are sufficient. Varying instead over the space of reduced objects makes the minimization simpler.
For instance, in the case of an electronic many-body state $\psi_e(\br_1\sigma_1,...,\br_N\sigma_N)$ of $N$ electrons (see Tab.~\ref{tab:wf_symbols}), the expectation value of a general (non-local) $q$-body operator $\hat{O}(\br_1,...\br_q; \br_1',...\br_q')$, which is given by $O = \braket{\psi_e|\hat{O}\psi_e}$, can be determined via the electronic (spin-summed)\footnote{We define here only the spin-summed version of the qRDM, because in this work, we do not consider explicitly spin-dependent quantities. For instance, if we included magnetic fields in the Hamiltonian, the situation would change.} $q$-body RDM ($q$RDM)\footnote{Note that there are different conventions in literature for the normalization of the $q$RDM. We followed Ref. \citenum{Bonitz2016}.}
\begin{align}
\label{eq:pRDM_electron}
	\Gamma_e^{(q)}&(\br_1,...,\br_q;\br_1',...,\br_q')= \tfrac{N!}{(N-q)!} \sum_{\sigma_{1},...,\sigma_{N}}\int\td^{3(N-q)}r\\ &\psi_e^*(\br_1'\sigma_1,...,\br_q'\sigma_q,\br_{q+1}\sigma_{q+1},..., \br_N\sigma_N)  \psi_e(\br_1\sigma_1,...,\br_q\sigma_q,\br_{q+1}\sigma_{q+1},...,\br_N\sigma_N).\nonumber
\end{align}
For instance, the well-known electronic 1RDM, that we denote in the following by $\gamma_{e}(\br,\br') = \Gamma_{e}^{(1)}(\br;\br')$, is sufficient to calculate all electronic single-particle observables such as the kinetic energy. A prominent example of a higher-order operator in the electronic case is the two-body Coulomb interaction among the $N$ electrons. To calculate its expectation value, we need to consider the diagonal of the 2RDM $\Gamma^{(2)}(\br_1,\br_2;\br_1,\br_2)$. In the chosen normalization, all RDMs satisfy the sum-rule $\int\td^{3q}r\Gamma^{(q)}(\br_1,...,\br_q;\br_1,...,\br_q)=\tfrac{N!}{(N-q)!}$. So for instance, the 1RDM integrates to the particle number $\int\td^3r\gamma_e(\br,\br)=N$, the 2RDM integrates to two times the number of pairs, etc. Additionally, higher and lower order RDMs are connected via $\Gamma^{(q)}_e(\br_1,...,\br_q;\br_1',...,\br_q')=\tfrac{1}{N-q}\int\td^3r\Gamma^{(q+1)}_e(\br_1,...,\br_{q+1};\br_1',...,\br_{q+1})$. 

For bosons the same construction is possible. In our case, where we have a discrete set of possible single-boson states, a $N_b$ boson state in the (symmetrized) mode-representation $\psi_b(\alpha_1,...,\alpha_{N_{b}})$ (see supporting information, Sec. ~\ref{app:symmetry_photon} for more details) leads to the corresponding bosonic $q$RDM,
\begin{align}
\label{eq:pRDM_boson}
&\Gamma_b^{(q)}(\alpha_1,...,\alpha_q;\alpha_1',...\alpha_q') = \tfrac{N_b!}{(N_b-q)!} 
\sum_{\alpha_{q+1},...,\alpha_{N_b}=1}^{M}\\ 
&\psi_b^*(\alpha_1',...,\alpha_q',\alpha_{q+1},...,\alpha_{N_b})  \psi_b(\alpha_1,...,\alpha_q,\alpha_{q+1},...,\alpha_{N_b}).\nonumber
\end{align}
Accordingly to the electronic case, we denote the 1RDM by $\gamma_{b}(\alpha,\beta) = \Gamma_{b}^{(1)}(\alpha;\beta)$. However, in the specific case of photons, where the number of particles is indetermined and we work with Fock-space wave functions $\ket{\Phi}$, we need to consider a Fock-space 1RDM of the form 
\begin{align*}
\gamma_{b}(\alpha,\beta)  = \braket{\Phi|\hat{a}^{+}_{\beta} \hat{a}_{\alpha} \Phi} = \sum_{N_b=0}^{\infty} N_b \left( \sum_{\alpha_2,...,\alpha_{N_b}=1}^{M} \psi_b^{*}(\beta, \alpha_2,...,\alpha_{N_b})  \psi_b(\alpha, \alpha_2,...,\alpha_{N_b})\right).
\end{align*}
In an according manner one can define a bosonic Fock-space $q$RDM via $\Gamma^{(q)}_{b}(\alpha_1,...,\alpha_q; \alpha_1',...,\alpha_{q}') = \braket{\Phi| \hat{a}^{+}_{\alpha_1'}\cdots\hat{a}^{+}_{\alpha_q'}\hat{a}_{\alpha_q}\cdots\hat{a}_{\alpha_1} \Phi }$.

The fermionic and bosonic RDMs can be extended to the coupled fermion-boson case straightforwardly by just integrating/summing out the other degrees of freedom. That is, if we have a general electron-boson state of the form of Eq.~\eqref{eq:wavefunction} we can accordingly define $\Gamma_{e}^{(q)} \equiv \tfrac{N!}{(N-q)!} \sum_{\sigma_1,...,\sigma_{N}} \int \td^{3(N-q)}r \; \td^{M} p \; \Psi^* \Psi $ as well as $\Gamma^{(q)}_{b} \equiv \braket{\Psi| \hat{a}^{+}_{\alpha_1}\cdots\hat{a}^{+}_{\alpha_q}\hat{a}_{\alpha_q'}\cdots\hat{a}_{\alpha_1'} \Psi}$. 

In a next step, we see whether these standard ingredients of RDMFT are sufficient to express the energy expectation value of the coupled Hamiltonian of Eq.~\eqref{eq:Hamiltonian}. For the purely electronic part, the different contributions can be expressed either explicitly by the electronic 1RDM or by the electronic 2RDM. The single-particle operators of $\hat{H}_e$ and the single-particle part of the dipole self-energy $\hat{H}_d^{(1)}$ are given in terms of the 1RDM by
\begin{align*}
	\left(T+V\right)[\gamma_{e}]&=\int\td^{3}r\left[ -\tfrac{1}{2} \nabla_{\br}^2 + v(\br) \right]\gamma_e(\br;\br')|_{\br'=\br},\\
	H_d^{(1)}[\gamma_{e}]&=\int\td^{3}r\left[\sum_{\alpha=1}^{M}\tfrac{1}{2}(\blambda_{\alpha}\cdot \br)^2\right]\gamma_e(\br;\br).
\end{align*}
Here we have denoted on the left-hand side the explicit dependence of the expectation value on the 1RDM and the subscript $|_{\br'=\br}$ indicates that $\br'$ is set to $\br$ after the application of the semi-local single-particle operator $\left(-\tfrac{1}{2} \nabla_{\br}^2 + v(\br)\right)$. The expectation value of the electronic interaction energy $\hat{W}$ and the two-body part of the dipole self-energy $\hat{H}_d^{(2)}$ are given in terms of the (diagonal) of the 2RDM by
\begin{align*}
W[\Gamma^{(2)}_{e}]=&\tfrac{1}{2}\!\int\td^{3}r\td^3r' w(\br,\br')  \Gamma_e^{(2)}(\br,\br';\br,\br'),\\
H_d^{(2)}[\Gamma^{(2)}_{e}]=&\tfrac{1}{2}\!\int\td^{3}r\td^3r'\left[\sum_{\alpha=1}^M \left(\blambda_{\alpha} \cdot \br\right)\left(\blambda_{\alpha} \cdot \br'\right)\right] \Gamma_e^{(2)}(\br,\br';\br,\br').
\end{align*}
Hence, for the electronic operator expectation values little changes in comparison to a purely fermionic problem, except that we have a coupled electron-boson wave function and the extra contributions of the dipole self-energy. For the purely bosonic part of the coupled Hamiltonian we find
\begin{align*}
 H_{ph}[\gamma_{b}] &=\braket{\Psi | \left\{ \sum_{\alpha=1}^M\left[- \tfrac{1}{2} \tfrac{\partial^2}{\partial p_{\alpha}^2}+ \tfrac{\omega^2_{\alpha}}{2}p_{\alpha}^2\right] \right\} \Psi}
 = \sum_{\alpha=1}^{M} \left(\omega_{\alpha} + \frac{1}{2} \right) \gamma_{b}(\alpha, \alpha).
\end{align*} 
However, the bilinear coupling term is not given in a simple RDM form but becomes
\begin{align*}
H_I[\Gamma^{(3/2)}_{e,b}]=&\braket{\Psi |\left[\sum_{\alpha=1}^M-\omega_{\alpha} p_{\alpha}\blambda_{\alpha} \cdot \hat{\bf D}\right]\Psi}\\
=&\sum_{\alpha=1}^{M} - \omega_{\alpha} \braket{\Psi| \left[\sqrt{\frac{2}{\omega_{\alpha}}} \left( \hat{a}^{+}_{\alpha} + \hat{a}_{\alpha}\right) \blambda_{\alpha} \cdot \hat{\bf D} \right]\Psi}.
\end{align*}
A new reduced quantity appears that mixes light and matter degrees of freedom and can be interpreted as a $3/2$-body operator $\Gamma^{(3/2)}_{e,b}(\alpha; \br, \br')$.\footnote{To see this in a simple manner we also lift the continuous fermionic problem into its own Fock space and introduce genuine field operators $\hat{\psi}_e^{\dagger}(\br \sigma)$ and $\hat{\psi}_e(\br \sigma)$ with the usual anti-commutation relations. Similarly to the discussed bosonic case, the electronic RDMs can then be written in terms of strings of creation and annihilation field operators~\cite{Leeuwen2013}. We re-express $\braket{\Psi| \left[\left( \hat{a}^{+}_{\alpha} + \hat{a}_{\alpha}\right) \blambda_{\alpha} \cdot \hat{\bf D} \right]\Psi} = \sum_{\sigma}\int \td^3 r \braket{\Psi|\left[\left( \hat{a}^{+}_{\alpha} + \hat{a}_{\alpha}\right) \hat{\psi}_e^{\dagger}(\br \sigma) \hat{\psi}_e(\br \sigma) \left(\blambda_{\alpha} \cdot \br\right) \right] \Psi}$. If we then define $\Gamma^{(3/2)}_{e,b}(\alpha; \br, \br') = \sum_{\sigma} \braket{\Psi|\left[\left( \hat{a}^{+}_{\alpha} + \hat{a}_{\alpha}\right) \hat{\psi}_e^{\dagger}(\br \sigma) \hat{\psi}_e(\br' \sigma) \right] \Psi}$ we can re-write $\braket{\Psi| \left[\left( \hat{a}^{+}_{\alpha} + \hat{a}_{\alpha}\right) \blambda_{\alpha} \cdot \hat{\bf D} \right]\Psi} = \int \td^3 r  \left(\blambda_{\alpha} \cdot \br\right) \Gamma^{(3/2)}_{e,b}(\alpha; \br, \br)$.} The bilinear interaction term therefore creates/annihilates bosons by interacting with the electronic subsystem. The $3/2$-body RDM has in general no simple connection to any $q$RDM, even if we extend the definitions to include combined matter-boson $q$RDMs.\footnote{Using the field-operator formulation, the usual $q$RDMs consist of strings of particle-number-conserving combinations of electron and boson operators. Integrating/summing out a number-non-conserving part of it does not lead to a simple relation to half-body RDMs in general.} One obvious reason is that $q$RDMs conserve particle numbers, while half-body RDMs do not. Take, for instance, the Fock-space $\gamma_{b}(\alpha,\beta) = \braket{\Phi|\hat{a}^{+}_{\beta} \hat{a}_{\alpha} \Phi}$. In the special case that $\ket{\Phi}$ consists only of coherent states for each mode (which essentially means that we have treated the photons in mean field) and since the coherent states are eigenfunctions to the annihilation operators, we find $\gamma_{b}(\alpha,\beta) = d_{\beta}^{*}d_{\alpha} $, where $d_{\alpha}$ is the total displacement of the coherent state of mode $\alpha$. In this case we also know $\braket{\Phi| \hat{a}^{+}_{\beta} \Phi} = d^{*}_{\beta}$. If we now assume all but one mode, say mode 1, having zero displacement, then we only know $\gamma_{b}(1,1) = |d_1|^2$ from the bosonic 1RDM. We do, however, in general not know what $d_1^{*}$ is. For other states, such a connection is even less explicit.

Putting the interrelations among the different RDMs aside for the moment, the minimization for the coupled matter-boson problem can be reformulated by
\begin{align}
\label{eq:var_parinciple_RDM}
	E_0 =& \inf_{\Psi} \braket{\Psi|\hat{H}\Psi} \nonumber\\
	=&\inf_{\substack{\{\gamma_e,\Gamma_e^{(2)}, \gamma_b, \Gamma_{e,b}^{(3/2)}}\}\rightarrow \Psi} \left\{\left(T+V\right)[\gamma_e]+\left(W+H_d\right)[\Gamma_e^{(2)}]+H_{ph}[\gamma_{b}] +H_I[\Gamma_{e,b}^{(3/2)}] \right\}.
\end{align}
So in principle, we could replace the variation over all wave functions $\Psi$ by their respective set of RDMs needed to define the energy expectation values. Instead of varying over the full configuration space $(\br_1 \sigma_1,...,p_M)$, the above reformulation seems to indicate that we can replace this by varying over $(\br,\br')$ for the diagonal of $\Gamma_{e}^{(2)}$ and also for the 1RDM $\gamma_{e}$ together with a variation over $(\alpha,\beta)$ for $\gamma_{b}$ and over $(\alpha,\br)$ for $\Gamma^{(3/2)}_{e,b}$. Such a reformulation is the basis of any RDMFT, and for electronic systems the properties of RDMs have been studied for more than 50 years~\cite{Coleman2000}. However, this seeming reduction of complexity is deceptive. In order to find physically sensible results we cannot vary arbitrarily over the above RDMs but need to ensure that they are consistent among each other and that they are all connected to a physical wave function. This is indicated in Eq.~\eqref{eq:var_parinciple_RDM}, where $\{\gamma_e,\Gamma_e^{(2)}, \gamma_b, \Gamma_{e,b}^{(3/2)}\}\rightarrow \Psi$ highlights that the RDMs are contractions of a common wave function. For systems with fixed particle numbers it is in principle known how to restrict the set of trial RDMs to physical ones. The corresponding restrictions are called N-representability conditions~\cite{Coleman1963,Klyachko2006,Mazziotti2012}. However, \emph{only} for the 1RDM of ensembles (fermionic or bosonic) the conditions are simple. In this case, by diagonalizing the 1RDM in its eigenbasis $\gamma_{e/b}=\sum n_i^{e/b} \left(\phi^{i}_{e/b}\right)^{*}\phi^i_{e/b}$, where the $\phi^i_{e/b}$ are called the natural orbitals and the $n^{i}_{e/b}$ the natural occupation numbers, the conditions are
\begin{align}
\label{eq:Nrep_cond}
	0\leq \,&n_i^e \leq 1,\nonumber\\
	0\leq \,&n_i^b,
\end{align}
for fermions and bosons, respectively.
If the particle number $N_{e/b}$ of one species of the system is conserved, the respective sum-rule 
\begin{align}
	\label{eq:Nrep_cond2}
	\sum_{i=1}^{\infty}n_i^{e/b}=N_{e/b}
\end{align}
becomes a second part of the N-representability conditions.
Consequently, to define a proper RDM framework for coupled electron-boson problems, one would need to know the corresponding constraints that connect the wave function with all the necessary RDMs. A glance at the history of an important example, the search for the N-representability conditions of the electronic 2RDM, suggests that finding similar conditions for the novel half-body RDMs together with connections between the fermionic and bosonic $q$RDMs is a very challenging task. The electronic-2RDM problem was proposed in 1960~\cite{Coulson1960} and it took until 2012, to understand how to make the conditions explicit~\cite{Mazziotti2012}.

Although the connection of the different RDMs in coupled fermion-boson systems is a very interesting subject, and recent results for a grand-canonical formulation of fermions or bosons suggest that also a combined formulation is feasible~\cite{Giesbertz2019}, we will follow an alternative route in this work. We reformulate the problem in such a way that we ``fermionize'' the coupled fermion-boson problem, where we can apply the known conditions of the fermionic problem.

%% file: 3fermionization.tex
\section*{The ``fermionization'' of matter-photon systems}
\label{sec:fermionization}
In this section, we explain in detail how a system described by the Hamiltonian \eqref{eq:Hamiltonian}, is mapped to an auxiliary space such that the coupled matter-light degrees of freedom can be modelled with new particles that are fermionic. We call them dressed or polaritonic particles, because they depend on electronic and photonic coordinates.\footnote{Note that the use of a dressed particle picture allows to also describe Landau polaritons, as shown recently in Ref.~\citenum{Rokaj2019}. Thus, also such systems can in principle be considered with the presented approach.} This ``fermionization'' procedure was introduced in a recent work by Nielsen et al.~\cite{Nielsen2018} and can be divided in three steps. First, we introduce for each mode auxiliary extra dimensions $(p_{\alpha,2}, ... , p_{\alpha,N})$, where the number of these extra dimensions depends on the number of electrons $N$. We therefore embed the physical configuration space in a higher-dimensional space, i.e., we now consider wave functions depending on $(\br_1 \sigma_1,...,\br_{N} \sigma_{N}, p_1,..,p_M, p_{1,2},...,p_{1,N},...,p_{M,2}, ... , p_{M,N})$. Second, we add for every photon mode $\alpha$ a linear operator 
\begin{align}
\label{eq:lin_op}
\hat{\Pi}_{\alpha}(p_{\alpha,2}, ... , p_{\alpha,N})=\sum_{i=2}^{N}\left( -\tfrac{1}{2} \tfrac{\partial^2}{\partial p_{\alpha,i}^2} + \tfrac{\omega_{\alpha}^2}{2}p_{\alpha,i}^2\right)
\end{align}
to the physical Hamiltonian of Eq.~\eqref{eq:Hamiltonian}. This auxiliary Hamiltonian is a sum of quantum harmonic oscillators with respect to the auxiliary coordinates. The resulting Hamiltonian in the extended configuration space is $\hat{H}'=\hat{H}+\sum_{\alpha=1}^{M}\hat{\Pi}_{\alpha}$ (we denote all quantities in the auxiliary space with a prime symbol.) Here we see that the auxiliary degrees of freedom do not mix with the physical ones. This will allow in a very simple manner to embed but also to reconstruct the physical wave function. 
In the third step, we perform an orthogonal variable transformation of the photonic plus auxiliary coordinates to new coordinates $(q_{\alpha,1}, ... , q_{\alpha,N})$ such that 
\begin{align}\label{CenterOfMass}
 p_{\alpha} &= \tfrac{1}{\sqrt{N}} \left( q_{\alpha,1} + ... + q_{\alpha,N}  \right) \nonumber\\
 \sum_{i=1}^{N}\left(-\tfrac{1}{2}\tfrac{\partial^2}{\partial q_{\alpha,i}^2} + \tfrac{\omega_{\alpha}^2}{2} q_{\alpha,i}^2\right) &= -\tfrac{1}{2} \tfrac{\partial^2}{\partial p_{\alpha}^2} + \tfrac{\omega_{\alpha}^2}{2} p_{\alpha}^2 + \hat{\Pi}_{\alpha}(p_{\alpha,2}, ... , p_{\alpha,N}).
\end{align}
This whole procedure can be viewed as the inverse of a center-of-mass coordinate transformation~\cite{Watson1968}. In total, we find the auxiliary Hamiltonian in the higher-dimensional configuration space given as
\begin{align}
\label{eq:AuxiliaryHamiltonian1}
\hat{H}' =& \hat{H} + \sum_{\alpha=1}^M \hat{\Pi}_{\alpha}(p_{\alpha,2},...,p_{\alpha,N})\\
 =&\sum_{k=1}^N \left[ -\tfrac{1}{2} \nabla_{\br_k}^2 + v(\br_{k}) \right] + \tfrac{1}{2}\sum_{k \neq l} w(\br_k,\br_l)  -\sum_{\alpha=1}^M\omega_{\alpha} p_{\alpha}\blambda_{\alpha} \cdot \hat{\mathbf{D}}+ \sum_{\alpha=1}^M \tfrac{1}{2}\left(\blambda_{\alpha} \cdot \hat{\mathbf{D}}\right)^2 \nonumber\\
 &+\sum_{\alpha=1}^M\left(- \tfrac{1}{2} \tfrac{\partial^2}{\partial p_{\alpha}^2}+ \tfrac{\omega_{\alpha}^2}{2}p_{\alpha}^2\right) + \sum_{\alpha=1}^M \hat{\Pi}_{\alpha}(p_{\alpha,2},...,p_{\alpha,N})\nonumber\\
 \overset{\eqref{CenterOfMass}}{=}&\sum_{k=1}^{N}  \left\{-\tfrac{1}{2} \nabla_{\br_k}^2 + v(\br_{k}) + \sum_{\alpha=1}^{M}\! \left[ -\tfrac{1}{2}\tfrac{\partial^2}{\partial q_{\alpha,k}^2} 
 + \tfrac{1}{2}\omega_{\alpha}^2q_{\alpha,k}^2  
- \tfrac{\omega_{\alpha}}{\sqrt{N}} q_{\alpha,k}(\blambda_{\alpha}\! \cdot \br_{k})  
 +\tfrac{1}{2}(\blambda_{\alpha}\! \cdot \br_{k})^2 \right]  \right\} \nonumber\\
 &+ \tfrac{1}{2}\sum_{k\neq l} \left[ w(\br_{k}, \br_l) + \sum_{\alpha=1}^{M} \left(  - \tfrac{\omega_{\alpha}}{\sqrt{N}} q_{\alpha,k} \blambda_{\alpha} \cdot \br_{l} - \tfrac{\omega_{\alpha}}{\sqrt{N}} q_{\alpha,l} \blambda_{\alpha} \cdot \br_{k}  +  \blambda_{\alpha}\cdot \br_{k} \blambda_{\alpha} \cdot \br_{l}  \right)\right], \nonumber
\end{align}
where we inserted the definition of the dipole operator, $\hat{\mathbf{D}}=\sum_{k=1}^N \br_k$, and reordered the expressions, such that the terms with only one index and the terms with two different indices are grouped together.
Introducing then a $(3+M)$-dimensional polaritonic vector of space and auxiliary photon coordinates $\bz = (\br, q_1,...,q_{M})$, we can rewrite the above Hamiltonian as
\begin{align}
\label{AuxiliaryHamiltonian2}
\hat{H}' &= \sum_{k=1}^{N}\left[- \tfrac{1}{2} \Delta_k' + v'(\bz_k) \right] + \tfrac{1}{2} \sum_{k \neq l} w'(\bz_k,\bz_l)\\
&= \hat{T}' + \hat{V}' + \hat{W}', \nonumber
\end{align}
where we introduced the dressed Laplacian $\Delta'=\sum_{i=1}^{3}\tfrac{\partial^2}{\partial r_i^2}+\sum_{\alpha=1}^{M}\tfrac{\partial^2}{\partial q_{\alpha}^2}$, the dressed local potential $v'(\bz)=v(\br)+ \sum_{\alpha=1}^{M}\! \left[ \tfrac{1}{2}\omega_{\alpha}^2q_{\alpha}^2  
- \tfrac{\omega_{\alpha}}{\sqrt{N}} q_{\alpha}\blambda_{\alpha}\!\! \cdot\! \br\right.$ $\left.+\tfrac{1}{2}(\blambda_{\alpha}\!\! \cdot\! \br)^2 \right]$ and the dressed interaction kernel $w'(\bz,\bz')= w(\br, \br') + \sum_{\alpha=1}^{M} \left[  - \tfrac{\omega_{\alpha}}{\sqrt{N}} q_{\alpha} \blambda_{\alpha} \!\!\cdot\! \br' \right.$ $\left.- \tfrac{\omega_{\alpha}}{\sqrt{N}} q_{\alpha}' \blambda_{\alpha} \!\!\cdot\! \br  +  \blambda_{\alpha}\!\!\cdot\! \br \blambda_{\alpha} \!\!\cdot\! \br'  \right]$.
Note that the choice of the linear auxiliary operator (Eq.~\eqref{eq:lin_op}) and the coordinate transformation (Eq.~\eqref{CenterOfMass}) have a certain freedom. The operator of Eq.~\eqref{eq:lin_op} must not contain physical coordinates, such that the physical system cannot be influenced by this auxiliary operator. Further, it must together with the transformation give rise to polaritonic 1- and 2-body terms, as shown in Eq.~\eqref{eq:AuxiliaryHamiltonian1}. These requirements are met using an orthonormal transformation\footnote{Note that there are many different orthonormal transformations, but the exact choice is not important for the formalism. It only needs to include the first line of Eq.~\eqref{CenterOfMass}. One specific example of such a transformation is\cite{Nielsen2018} $p_{\alpha,k}=\frac{1}{\sqrt{k^2-k}} (q_1+...+q_{k-1}-(k-1)q_k)$ for $2\leq k\leq N$ alongside the first line of Eq.~\eqref{CenterOfMass}.} together with the harmonic oscillators of Eq.~\eqref{eq:lin_op}. Since the operator of Eq. (10) only acts in the auxiliary coordinates, the normalized physical solution $\Psi$ of the original (time-independent) Schrödinger equation $E_0 \Psi = \hat{H} \Psi$ in the standard configuration space is connected to a new physical solution of the auxiliary Hamiltonian $\hat{H}'$ in a very simple manner, i.e.,
\begin{align*}
 \Psi'(\br_1 \sigma_1, ..., p_{M, N}) \! = \! \Psi(\br_1 \sigma_1, ..., \br_N \sigma_N,p_1,..,p_M) \chi(p_{1,2},..., p_{M, N}).
\end{align*}
Here the normalized solution $\Psi'$ of the auxiliary Schrödinger equation $E_0' \Psi' = \hat{H}' \Psi'$ is found with $\chi$ being the ground state of $\sum_{\alpha} \hat{\Pi}_{\alpha}$. The ground state $\chi$ is merely a tensor product of individual harmonic-oscillator ground states and therefore exchanging $p_{\alpha,i}$ with $p_{\alpha,j}$ does not change the total wave function $\Psi'$. If we rewrite this wave function in the new coordinates
\begin{align*}
\Psi'(\br_1 \sigma_1, ...,\br_N \sigma_N, q_{1,1},..., q_{M, N}) \! = \! \Psi'(\bz_1 \sigma_1,...,\bz_N \sigma_N) ,
\end{align*}
and due to the fact that we constructed $\chi$ to be symmetric with respect to the exchange of $q_{\alpha,k}$ and $q_{\alpha,l}$,\footnote{Note that the auxiliary ground state of mode $\alpha$ is given as a Gaussian with respect to $\sum_{i=2}^{N} p_{\alpha,i}^2 = \sum_{k=1}^{N}q_{\alpha,k}^2 - \tfrac{1}{N}(\sum_{k=1}^{N}q_{\alpha,k})^2$, where we used that $p = \frac{1}{\sqrt{N}} \sum_{k=1}^{N} q_{\alpha,k}$.} we realize that $\Psi'$ is anti-symmetric with respect to the exchange of $(\bz_k \sigma_k)$ and $(\bz_l \sigma_l)$. Thus $\Psi'$ is fermionic with respect to the polaritonic coordinates $(\bz \sigma)$ and can be represented by a sum of Slater determinants of $(3+M)$-dimensional polaritonic orbitals $\varphi(\bz \sigma)$. This makes the application of usual fermionic many-body methods possible and we can rely on fermionic $N$-representability conditions. However, besides the extra dimensions and the new fermionic exchange symmetry, the physical wave functions of the dressed auxiliary space also have a further $q_{\alpha,k} \leftrightarrow q_{\alpha,l}$ exchange symmetry. Simple approximations based on single polaritonic Slater determinants will violate this extra symmetry. We will remark on such violations when we introduce dressed RDMFT in the next section. To further see that indeed the constructed $\Psi'$ is a minimal-energy state in the extended space with the appropriate symmetries, we first point out that for any trial wave function
\begin{align*}
\Upsilon'(\bz_1 \sigma_1,...\bz_N \sigma_N) \equiv \Upsilon'(\br_1 \sigma_1,...,p_M; p_{1,2},...,p_{M,N})	
\end{align*}
the $q$-exchange symmetry implies that the fermionic symmetry is in the $(\br \sigma)$ coordinates. Then it holds since $\hat{H}$ only acts on $(\br_1,...,p_M)$ and $\sum_{\alpha=1}^{M}\hat{\Pi}_{\alpha}$ only acts on $p_{1,2},...,p_{M,N}$ that
\begin{align*}
	\inf_{\Upsilon'}\braket{\Upsilon'|\hat{H}' \Upsilon'} &\geq \inf_{\Upsilon'}\braket{\Upsilon'|\hat{H} \Upsilon'} + \inf_{\Upsilon'}\braket{\Upsilon'|\sum_{\alpha=1}^{M}\hat{\Pi}_{\alpha} \Upsilon'} = \braket{\Psi|\hat{H} \Psi} + \braket{\chi|\sum_{\alpha=1}^{M}\hat{\Pi}_{\alpha} \chi}\\
	&=\braket{\Psi'|\hat{H}' \Psi'}.
\end{align*}	
Although we constructed the auxiliary space explicitly in a way that the physical wave function $\Psi$ can be reconstructed exactly from its dressed counterpart $\Psi'$\footnote{Note that $\hat{H}'$ also has many eigenstates that are not of the form $\Psi' = \Psi\chi$, with $\Psi$ anti-symmetric under exchange of $\br_k\sigma_k \leftrightarrow \br_l\sigma_l$ and $\chi$ symmetric under exchange of $q_{\alpha,k} \leftrightarrow q_{\alpha,l}$. In general, we thus have to enforce these properties to only retain the eigenstates of this form. However, for the ground state it is sufficient to enforce the symmetry under exchange of $q_{\alpha,k} \leftrightarrow q_{\alpha,l}$ together with the anti-symmetry under exchange of $\bz_k\sigma_k \leftrightarrow \bz_l\sigma_l$.} by integration of all auxiliary coordinates, this does not hold for all types of operators. For operators that depend only on electronic coordinates, there is no difference and we have
$\braket{\Psi|\hat{O}\Psi}=\braket{\Psi'|\hat{O}\Psi'}$.
This is not surprising because the coordinate transformation \eqref{CenterOfMass} acts only on the photonic part of the system. For photonic observables instead, the transformation changes the respective operators and thus, the connection between physical and auxiliary space becomes non-trivial in general. However, at least for all observables that depend on photonic $1/2$- or 1-body expressions, there is an analytical connection. For half-body operators, i.e. any operator that depends only on the elongation of $p_{\alpha}=\frac{1}{\sqrt{N}}(q_{\alpha,1}+...+q_{\alpha,N})$ and its conjugate $\frac{\partial}{\partial p_{\alpha}}=\frac{1}{\sqrt{N}}(\frac{\partial}{\partial q_{\alpha,1}}+...+\frac{\partial}{\partial q_{\alpha,N}})$, the coordinate transformation itself provides us with the connection.
For 1-body operators, this becomes slightly more involved. For example, consider the mode energy operator $\hat{H}_{ph}=\sum_{\alpha=1}^{M}\left[-\frac{1}{2}\frac{\partial^2}{\partial p_{\alpha}^2}+\frac{\omega_{\alpha}^2}{2}p_{\alpha}^2\right]\equiv \sum_{\alpha=1}^{M}\hat{h}_{\alpha}$, that we can straightforwardly generalize in the auxiliary space to $\hat{H}_{ph}'=\sum_{\alpha=1}^{M}\sum_{i=1}^{N} \left[-\frac{1}{2}\frac{\partial^2}{\partial q_{\alpha,i}^2}+ \frac{\omega_{\alpha}^2}{2} q_{\alpha,i}^2\right]\equiv\sum_{\alpha=1}^{M} \hat{h}_{\alpha}'$ . The connection between $\hat{H}_{ph}$ and $\hat{H}_{ph}'$ is given by the definition of the coordinate transformation \eqref{CenterOfMass}, 
\begin{align*}
	\hat{H}_{ph} = \hat{H}_{ph}'- \sum_{\alpha=1}^{M}\hat{\Pi}_{\alpha}.
\end{align*}
Since the expectation value of  $\hat{\Pi}_{\alpha}$ is known analytically, $\braket{\Psi'|\hat{\Pi}_{\alpha}(p_{\alpha,2},...,p_{\alpha,N})\Psi'}=\braket{\chi|\hat{\Pi}_{\alpha}\chi}=(N-1)\sum_{\alpha=1}^{M}\frac{\omega_{\alpha}}{2}$, we have $\braket{\Psi|\hat{H}_{ph}\Psi}=\braket{\Psi'|\hat{H}_{ph}'\Psi'}-(N-1)\sum_{\alpha=1}^{M}\frac{\omega_{\alpha}}{2}$.
This can be generalized to any operator that contains terms of the form $p_{\alpha}p_{\beta}, p_{\alpha}\frac{\partial}{\partial p_{\beta}},$ and $\frac{\partial}{\partial p_{\alpha}}\frac{\partial}{\partial p_{\beta}}$, where $\alpha$ and $\beta$ denote any two modes. The reason is that the transformation $\eqref{CenterOfMass}$ preserves the standard inner product of the Euclidean space of the mode plus extra coordinates. This transfers also to their conjugates and combinations of both. From the above, it is straightforward to derive also the expression for the occupation of mode $\alpha$, $\hat{N}^{\alpha}_{ph}=\frac{1}{\omega_{\alpha}}\hat{h}_{\alpha}-\frac{1}{2}$. In the auxiliary system, we have
\begin{align}
\label{eq:mode_occupation_formula}
	\hat{N}^{\alpha}_{ph}=\frac{1}{\omega_{\alpha}}\hat{h'}_{\alpha}-\frac{N}{2}.
\end{align}

%% file: 4drdmft.tex
\section*{Dressed Reduced Density-Matrix Functional Theory}
\label{sec:dRDMFT}

From the observations of the previous chapter, it is clear that a simple dressed Kohn-Sham approximation can capture matter-photon correlations that are hard to capture with standard approximations of Kohn-Sham QEDFT. But from the experience with purely electronic DFT, we expect that for ultra- and deep-strong coupling situations the simple dressed approximations also become less reliable. Instead of developing more advanced approximations for a dressed Kohn-Sham approach, we propose to follow another route in this paper. Similar to electronic-structure theory, where RDMFT becomes a reasonable alternative to DFT methods when strong correlations become important~\cite{Goedecker1998,Sharma2013,Piris2017}, we present a dressed RDMFT approach to capture ultra- and deep-strong electron-photon coupling. 

Let us therefore analyze the structure of $\hat{H}'$, given in Eq.~\eqref{AuxiliaryHamiltonian2}. It consists of only polaritonic one-body terms $\hat{h}^{(1)}(\bz)=- \tfrac{1}{2} \Delta + v'(\bz)$, and two-body terms $ \hat{h}^{(2)}(\bz,\bz')=w'(\bz,\bz')$. It commutes with the polaritonic particle-number operator $\hat{N}'=\int\td^{3+M}z\, \hat{n}(\bz)$, where we used the definition of the polaritonic local density operator $\hat{n}(\bz)=\sum_{i=1}^{N}\delta^{3+M}(\bz-\bz_i)$. This means that the auxiliary system has a constant polaritonic particle number $N$. Additionally, the physical wave function of the dressed system $\Psi'(\bz_1 \sigma_1,\dots,\bz_N \sigma_N)$ is per construction anti-symmetric. This allows for the definition of a dressed (spin-summed) 1RDM
\begin{align}
\label{eq:1RDM_polaritonic}
\gamma(\bz,\bz') &= N \sum_{\sigma_1,...,\sigma_N}\int \td^{(3+M)(N-1)}  z \\
&\hphantom{+}{\Psi'}^*(\bz' \sigma_1,\bz_2 \sigma_2,...,\bz_{N} \sigma_{N}){\Psi'}(\bz \sigma_1,\bz_2 \sigma_2,...,\bz_{N} \sigma_{N}), \nonumber
\end{align}
in accordance to Eq.~\eqref{eq:pRDM_electron}. By construction, this auxiliary density matrix reduces to the physical electron density matrix via $\gamma_{e}(\br,\br') = \int \td^M q \, \gamma(\br,q_1,..,q_M;\br',q_1,..,q_M)$. Furthermore, we introduce the (spin-summed) dressed 2RDM $\Gamma^{(2)}(\bz_1,\bz_2;\bz_1',\bz_2')=N(N-1) \sum_{\sigma_1,...,\sigma_N}\int \td^{(3+M)(N-2)} z$ ${\Psi'}^*(\bz_1'\sigma_{1},$ $\bz_2'\sigma_2, \bz_3 \sigma_3,..,\bz_{N} \sigma_{N})$ ${\Psi'} (\bz_1\sigma_{1},\bz_2\sigma_2, $ $\bz_3 \sigma_3,..,\bz_{N} \sigma_{N})$. These dressed RDMs allow for expressing the energy expectation value of the dressed system by
{\small
	\begin{align*}
	E_0'&=\braket{\Psi'|\hat{H}'|\Psi'}=\braket{\Psi'|\sum_{k=1}^{N}\hat{h}^{(1)}(\bz_k) + \tfrac{1}{2} \sum_{k \neq l} \hat{h}^{(2)}(\bz_k,\bz_l)|\Psi'}\\
	=& \!\int\!\td^{3+M}z \hat{h}^{(1)}(\bz) \gamma(\bz,\bz')|_{\bz'=\bz}+
	\!\tfrac{1}{2}\!\int\!\td^{3+M}z\td^{3+M}z' \hat{h}^{(2)}(\bz,\bz') \Gamma^{(2)}(\bz,\bz',\bz,\bz').
	\end{align*}}
Thus, we can define the variational principle for the ground state only with respect to \emph{well-defined} reduced quantities,
\begin{align}
\label{eq:min_principle}
E_0'=\inf_{\{\gamma,\Gamma^{(2)}\}\rightarrow \Psi'} E[\gamma,\Gamma^{(2)}].
\end{align}
To perform this minimization, we need to constrain the configuration space to the physical dressed RDMs that connect to an anti-symmetric wave function with the extra $q$-exchange symmetry by testing the appropriate N-representability conditions of the dressed 2RDM and the  dressed 1RDM. Besides the by now well-known conditions for the fermionic 2RDM~\cite{Mazziotti2012} and the fermionic 1RDM~\cite{Coleman1963} we would in principle get further conditions to ensure the extra exchange symmetry. However, already for the usual electronic 2RDM the number of conditions grows exponentially with the number of particles, and it is out of the scope of this work to discuss possible approximations. The interested reader is referred to, e.g., Ref.~\citenum{Mazziotti2012rev}. Instead, we want to stick to the  dressed 1RDM $\gamma$ and approximate the 2-body part as a functional of the $\gamma$. The mathematical justification of RDMFT is given by Gilbert's theorem~\cite{Gilbert1975}, which is a generalization of the Hohenberg-Kohn theorem of DFT~\cite{Hohenberg1964}. More specifically, Gilbert proves that the ground state energy of any Hamiltonian with only 1-body and 2-body terms is a unique functional of its 1RDM. Following this idea, we will express the ground-state energy of the dressed system as a partly unknown functional $F'$ of only the system's dressed 1RDM  
\begin{align}
E_0'=\inf_{\gamma}F'[\gamma] = \inf_{\gamma}\left\{\!\int\!\td^{3+M}z \hat{h}^{(1)}(\bz) \gamma(\bz,\bz')|_{\bz'=\bz} + \underbrace{\tfrac{1}{2}
\!\int\!\td^{3+M}z\td^{3+M}z' \hat{h}^{(2)}(\bz,\bz') \Gamma^{(2)}([\gamma];\bz,\bz',\bz,\bz')}_{=W'[\gamma]}\right\}.
\end{align}
For this minimization, we need a functional of the diagonal of the dressed 2RDM in terms of the dressed 1RDM as well as adhering to the corresponding $N$-representability conditions when varying over $\gamma$. We also see now the advantage of the dressed RDMFT approach, which avoids the original variation over all wave functions as well as a variation over many different RDMs as shown in Eq.~$\eqref{eq:var_parinciple_RDM}$. Instead, we only need the dressed 1RDM, which has a comparatively simple connection to fermionic wave functions (at least when we vary over ensembles, i.e., Eq.s~(\ref{eq:Nrep_cond},\ref{eq:Nrep_cond2}).) The price we pay for this is twofold: First, we have a new symmetry that will most likely lead to extra $N$-representability conditions. Second, we need to increase the dimension of the natural orbitals by one for every photon mode. However, to capture the main physics of usual cavity experiments, often one effective mode is enough. Computations with four-dimensional dressed orbitals are numerically feasible. It is specifically such settings, where we envision a dressed RDMFT to be a reasonable alternative to other ab-initio approaches to cavity QED\cite{Nielsen2018,Ruggenthaler2014,Flick2017,Galego2019,kowalewski2016non,csehi2019ultrafast}.

Another advantage of RDMFT in general is the direct access to all one-body observables. This transfers also to dressed RDMFT. The calculation of expectation values of purely electronic one-body observables is trivial with the knowledge of the dressed 1RDM,  but also photonic one-body (and half-body) observables can be calculated, using the connection formula shown in the end of Sec. \Nnameref{sec:fermionization}. Thus, we are able to calculate very interesting properties of the cavity photons like the mode occupation or quantum fluctuations of the electric and magnetic field.

To see whether our approach is practical and accurate, we perform first simple calculations for coupled matter-photon systems. We will make the following pragmatic approximations: We only enforce the fermionic ensemble $N$-representability conditions\footnote{This approximation is similarly employed in electronic RDMFT. For details we refer to, e.g., Ref.~\citenum{Theophilou2015}.} and ignore presently the extra exchange symmetry between the q-coordinates,\footnote{In all the numerical examples that we studied, this ``fermion polariton approximation'' was very accurate. One can find more details in the supporting information.} and we employ simple approximations to the unknown part $W'[\gamma]$ that have been developed for the electronic case. To do so, we further, similarly to the electronic case, decompose $W'[\gamma]=E_H[\gamma]+E_{xc}[\gamma]$ into a classical Hartree part $E_{H}[\gamma]=\frac{1}{2}\int\int\td^{3+M}z\td^{3+M}z' \gamma(\bz,\bz)\gamma(\bz',\bz')w'(\bz,\bz')$ and an unknown exchange-correlation part $E_{xc}[\gamma]$. Almost all known functionals $E_{xc}[\gamma]$ are expressed in terms of the eigenbasis and eigenvalues of the 1RDM. In our case the dressed natural orbitals $\phi_i(\bz)$ and occupation numbers $n_i$ are found by solving $\int \gamma(\bz,\bz') \phi_i (\bz') \td^{3+M} z' = n_i \phi_i (\bz)$. The simplest approximation is to only retain the fermionic exchange symmetry and employ the Hartree-Fock (HF) functional
\begin{align}
\label{eq:HF}
E_{xc}[\gamma]=E_{\text{HF}}[\gamma]=-\tfrac{1}{2} \sum_{i,j}n_in_j&\int\!\!\int\td^{3+M}z\td^{3+M}z'\\ &\phi^*_i(\bz)\phi^*_j(\bz')\, w'(\bz,\bz') \phi_i(\bz')\phi_j(\bz).\nonumber
\end{align}
As the HF functional depends linearly on the natural occupation numbers, any kind of minimization will lead to the single-Slater-determinant HF ground state (which corresponds to occupations of 1 and 0.)~\cite{Lieb1981} We call this approximation dressed Hartree-Fock (dressed HF.) We can go beyond the single Slater determinant in dressed RDMFT, if we employ a non-linear occupation-number dependence in the exchange-correlation functional. We here employ the M\"{u}ller functional~\cite{Muller1984}
\begin{align}
\label{eq:Mueller}
E_{xc}[\gamma]=E_{\text{M}}[\gamma]=-\tfrac{1}{2} \sum_{i,j}\sqrt{n_in_j}&\int\!\!\int\td^{3+M}z\td^{3+M}z'\\ &\phi^*_i(\bz)\phi^*_j(\bz')\, w'(\bz,\bz') \phi_i(\bz')\phi_j(\bz),\nonumber
\end{align}
which has been re-derived by Bjuise and Baerends~\cite{Buijse2002}. The Müller functional has been studied for many physical systems~\cite{Goedecker1998,Buijse2002} and gives a qualitatively reasonable description of electronic ground states. Additionally, it has many advantageous mathematical properties~\cite{Muller1984,Frank2007}. A thorough discussion of different functionals goes beyond the scope of this work, and we only want to remark that a variety of functionals were proposed after $E_{M}[\gamma]$ and it is likely to have even better agreement with the exact solution by choosing more elaborate functionals.

%% file: 5numerics_results.tex
\section*{Numerical Implementation}
\label{sec:implementation}
Besides the fact that we can re-use many ideas from electronic RDMFT, a further advantage of the dressed reformulation is that we can also re-use most of the numerical techniques developed for quantum chemistry and materials science.
For instance, to determine the dressed orbitals we merely need to be able to solve higher-dimensional static Schr\"odinger-type equations. We only have to change the usual electronic potential $v$ to its dressed counter part $v'$. This, together with a change of the electronic Coulomb interaction $w$ to its dressed counter part $w'$, already allows to perform dressed HF calculations, at least under the approximation of violating the additional symmetry constraint, discussed in the previous section. If the code one uses is furthermore able to perform RDMFT minimizations, it is straightforward to extend the implementation to also solve coupled electron-photon problems via dressed RDMFT from first principles. We have done so with the electronic-structure code Octopus~\cite{Andrade2015} and the implementation will be made available with the upcoming release. 

Specifically, we rewrite the approximated energy functional in the natural orbital basis as
\begin{align*}
E[\gamma]&=\sum_{i=0}^{\infty} n_i \int \td^{3+M}z \, \phi^*_i (\bz) \left[-\tfrac{1}{2} \Delta + v'(\bz) \right] \phi_i (\bz) + \\
&\tfrac{1}{2} \sum_{i,j}n_in_j\int\!\! \int\td^{3+M}z\td^{3+M}z' \left|\phi_i(\bz)\right|^2 \left|\phi^*_j(\bz')\right|^2 \, w'(\bz,\bz') + E_{\text{M}}[\gamma].
\end{align*}
We use this form to minimize the energy functional by varying the natural orbitals as well as the natural occupation numbers.  To impose fermionic ensemble $N$-representability, we first represent the occupation numbers as the squared sine of auxiliary angles, i.e. $0 \leq n_i=2\sin^2(\alpha_i)\leq 2$, to satisfy Eq.~\eqref{eq:Nrep_cond}.\footnote{Note that the $n_i$ are bounded by 2 because we employed a spin-summed formulation. If we considered natural \emph{spin-orbitals} instead, the upper bound would be 1.}  
The second part of the conditions (Eq.~\eqref{eq:Nrep_cond2}), i.e., $\sum_{i=1}n_i=N$, as well as the orthonormality of the dressed natural orbitals, i.e., $\int\td^{3+M}z\phi^*_i(\bz)\phi_j(\bz)=\delta_{ij}$, are imposed via Lagrange multipliers as, e.g., explained in Ref.~\citenum{Andrade2015}. We have available two different orbital-optimization methods, a conjugate-gradient algorithm\footnote{We used this method only for some benchmark calcualtions as it is not yet optimized for the needs of RDMFT.} and an alternative method that was introduced by Piris \textit{et al.} in Ref.~\citenum{Ugalde2008}. The latter expresses the $\phi_i$ in a basis set and can use this representation to considerably speed up calculations in comparison to the conjugate-gradient algorithm. It was used for all results presented in this paper. However, it is not trivial to converge such calculations in practice and we developed a protocol to obtain properly converged results. The interested reader is referred to Sec.~\ref{APP:protocol} of the supporting information.

\section*{Numerical Results}
\label{sec:results}
In the following, we present some examples of few electron systems in one spatial dimension. In the first part of this section, we validate our method by comparing to exact solutions of simple atomic and molecular systems. Then we show that our method also provides reasonable results for a more complex system. We finish the section with two examples that illustrate how our method can describe and uncover non-trivial changes of the matter due to its coupling to photons. We first present the dissociation of a molecule as an example for a chemical reaction. Despite the dipole approximated coupling the cavity photons affect the ground state locally differently. The changes also have a non-trivial dependence on the inter-atomic distance. As this system can be solved exactly, we can again validate that dressed RDMFT reproduces these intricate effects accurately. Finally, we show also that the ground state modifications of atomic systems, that have a very similar density profile outside of the cavity, are localized and depend strongly on the detailed electronic structure. These results highlight how cavity photons can at once locally enhance and suppress electronic repulsion and modify the electronic structure considerably.

The different systems are described by a local potential $v(x)$ and coupled to one photon mode. We transfer the systems in the dressed basis, that leads to a dressed local potential $v'(x,q) = v(x) + \tfrac{1}{2} (\lambda x)^2 + \tfrac{\omega^2}{2} q^2 - \tfrac{\omega}{\sqrt{2}}q (\lambda x)$. 
Specifically, we consider a one-dimensional model of a helium atom ($He$), i.e., $v_{He}(x) = -\frac{2}{\sqrt{x^2+1}}$, a one-dimensional model of a hydrogen molecule ($H_2$), i.e., $v_{H_2}(x) =-\frac{1}{\sqrt{(x-d)^2+1}} -\frac{1}{\sqrt{(x+d)^2+1}}$, first at its equilibrium position $d= d_{eq}=1.628$ a.u., later with varying $d$, and a one-dimensional model of a Beryllium atom $Be$, i.e., $v_{Be}(x) = -\frac{4}{\sqrt{x^2+\epsilon^2}}$. For the latter, we consider a smaller softening parameter $\epsilon=0.5$ to make sure that all electrons are properly bound. We use the soft Coulomb interaction $w(x,x') = 1/\sqrt{|x-x'|^2+1}$~\cite{Ruggenthaler2009, Fuks2011} for all test-systems. For the two-electron examples, we set the photon frequency in resonance with the lowest excitations of the respective ``bare'' systems, so outside of the cavity. For that we calculate the ground and first excited state of each system with the exact solver and find the corresponding excitation frequencies $\omega_{He}=0.5535$ a.u. and $\omega_{H2}=0.4194$ a.u.
For $Be$ instead, we choose $\omega_{Be}=3.0$, which is not a resonance of the $Be$ atom, for rather numerical than physical reasons. As resonance is not an important feature for ground state calculations, different choices of $\omega$ do not crucially change the physics of the investigated system. This is in contrast to the excited states and the ensuing Rabi splitting~\cite{schafer2018ab}. Instead, the chosen $\omega_{Be}$ considerably enhances the numerical stability of the calculations, which has the following reason: At the current state of our implementation, we need to make use of a basis set that we generate by a preliminary calculation. To generate a basis that captures electronic and photonic parts of the system equally well, we need to make sure that the energy scales of both degrees of freedom are similar. This can be controlled easiest by varying $\omega$. We want to stress that this basis-set issue is not a fundamental problem of the dressed orbital approach. On the one hand, we plan to control the photonic basis directly and on the other hand, we are working on optimizing an alternative conjugate-gradient routine that does not use a specific basis. Details can be found in Sec. \ref{APP:dHF_basis_conv} of the supporting information. 

We start with the discussion of the $(N\!=\!2)$-particles examples. In this setting the dressed auxiliary system is 4-dimensional (2 particles with 2 coordinates each), which is still small enough to be solved exactly in a 4-dimensional discretized simulation box, so that we can compare  dressed RDMFT (with the M\"uller functional of Eq.~\eqref{eq:Mueller}), dressed HF (see Eq.~\eqref{eq:HF}) and the exact solutions. We used the box lengths of $L_x=L_q=16$ a.u. and spacings of $dx=dq=0.14$ a.u. to model the electronic x and photonic coordinates q of the two dressed particles in the exact routine.\footnote{We want to mention that the box length is not entirely converged with these parameters. In a (numerically very expensive) benchmark calculation, we observed a further decrease of energy with larger boxes (the calculations with respect to the spacing are converged,) but the changes in energy and density are only of the order of $10^{-5}$ or less. All the following results require a maximal precision of the order of $10^{-2}$ in energy as well as in the density and thus we can safely use the given parameters. Details can be found in Sec. \ref{APP:ValidationMB} of the supporting information.} For dressed RDMFT and dressed HF instead, we needed to consider 2-dimensional simulation boxes for every dressed orbital and we set $L_x=L_q=20$ a.u. and $dx=dq=0.1$ a.u. We obtained converged results for $\mathcal{M}=41$ ($\mathcal{M}=71$) natural orbitals for $He$($H_2$). Details about how we determined these parameters can be found in the supporting information.
\begin{figure}[H]
		\begin{overpic}[width=0.49\columnwidth]{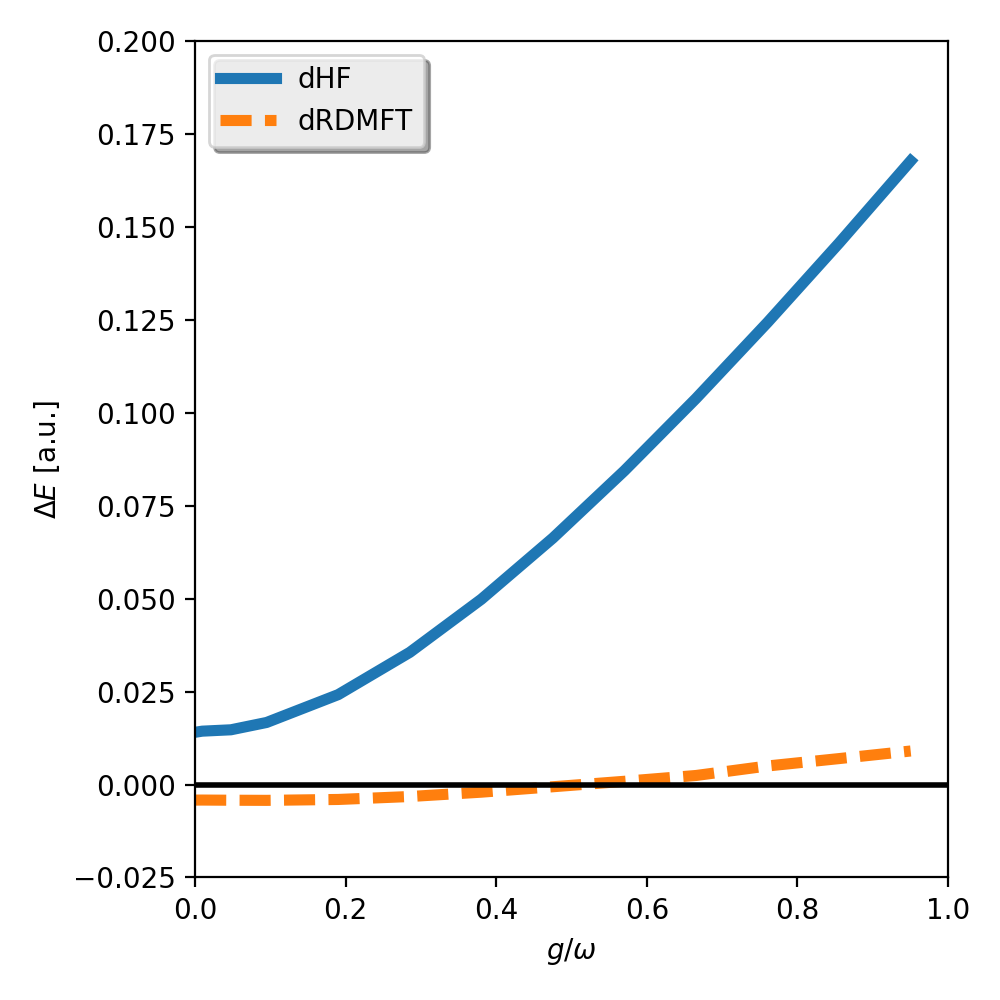}
		\put (50,90) {\textcolor{black}{He}}\hfill
	\end{overpic}
	\begin{overpic}[width=0.49\columnwidth]{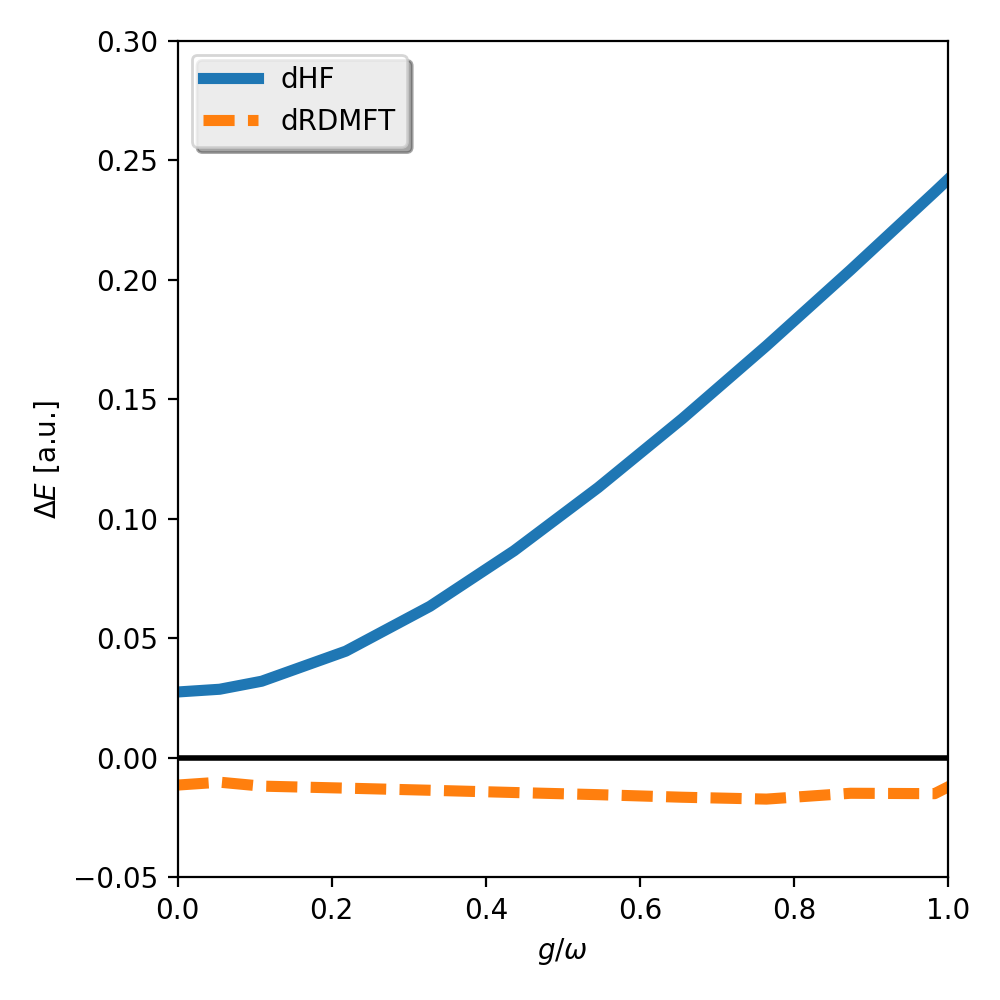}
		\put (50,90) {\textcolor{black}{$H_2$}}
	\end{overpic}
	\caption{Differences of dressed HF (dHF) and dressed RDMFT (dRDMFT) from the exact ground state energies (in Hartree) as a function of the coupling $g/\omega$ for the (one-dimensional) $He$ atom (left) and (one-dimensional) $H_2$ molecule (right) in the dressed orbital description. Dressed RDMFT improves considerably upon dressed HF. For both systems, the energy of dressed RDMFT remains close to the exact one, the error of dressed HF instead increases with the coupling strength.}
	\label{fig:H2_He_lambda}
\end{figure}
We first show (see Fig.~\ref{fig:H2_He_lambda}) the deviations of the ground state energies for dressed RMDFT and dressed HF from the exact dressed calculation as a function of the dimensionless relation between effective coupling strength and photon frequency $g/\omega$\footnote{This quantity is typically used as a measure for the strength of the light-matter interaction, see e.g. Ref. \citenum{Schafer2018}.} for $He$ and $H_2$, respectively. We thereby go from weak to deep-strong coupling with $g/ \omega = 1$\cite{Kockum2018}. The deep-strong coupling regime has been reached in different systems like for instance for Landau polaritons\cite{Bayer2017}. For (organic) molecules the highest reported coupling strengths are in the ultra-strong regime of $g/\omega\approx 0.4$\cite{Kockum2018}. We see that while dressed HF deviates strongly for large couplings, dressed RDMFT remains very accurate over the whole range of coupling strength. 
Still, a more severe test of the accuracy of our method is if instead of merely energies, we compare spatially resolved quantities like the ground-state density $\rho(x,q)\equiv\gamma(x,q;x,q)$. To simplify this discussion, we separate the electronic and photonic parts of the two-dimensional density by integration, i.e., $\rho(x)=\int\td q\, \rho(x,q)$ and $\rho(q)=\int\td x\, \rho(x,q)$. The exact reference solutions show that with increasing $g/\omega$ the electronic part of the density becomes more localized, while the photonic part becomes broadened (for details of the effects of matter-photon coupling see the discussion following Fig.~\ref{fig:H2_diss}.)
This behavior is captured qualitatively with dressed HF as well as with  dressed RDMFT. The latter performs for the electronic density considerably better over the whole range of coupling strength, whereas for the photonic densities both levels of theory deviate in a similar way from the exact result. This is shown for $g/\omega=0.1$ in Fig.~\ref{fig:H2_He_dens} for both test systems. Looking at the electronic densities, we can observe a feature that the ground state energy does not reveal. In some cases the effects of the two approximations are contrary to each other as we can see in the $He$ case. Here, the dressed RDMFT electronic density is more localized around the center of charge than the exact reference and the electronic density of dressed HF less. In other cases instead, both theories over-localize $\rho(x)$ (here visible for $H_2$.)

\begin{figure}[H]
		\begin{overpic}[width=0.49\columnwidth]{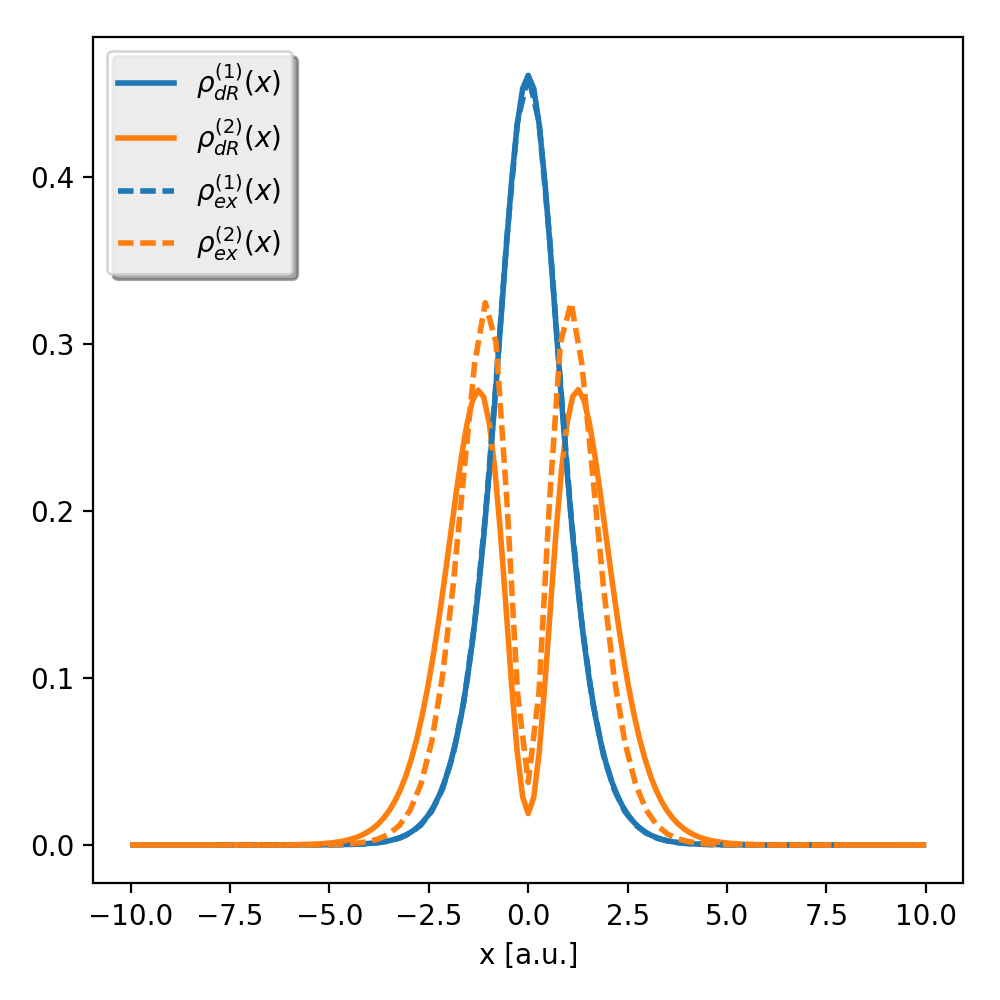}
		\put (85,85) {\textcolor{black}{He}}\hfill
	\end{overpic}
	\begin{overpic}[width=0.49\columnwidth]{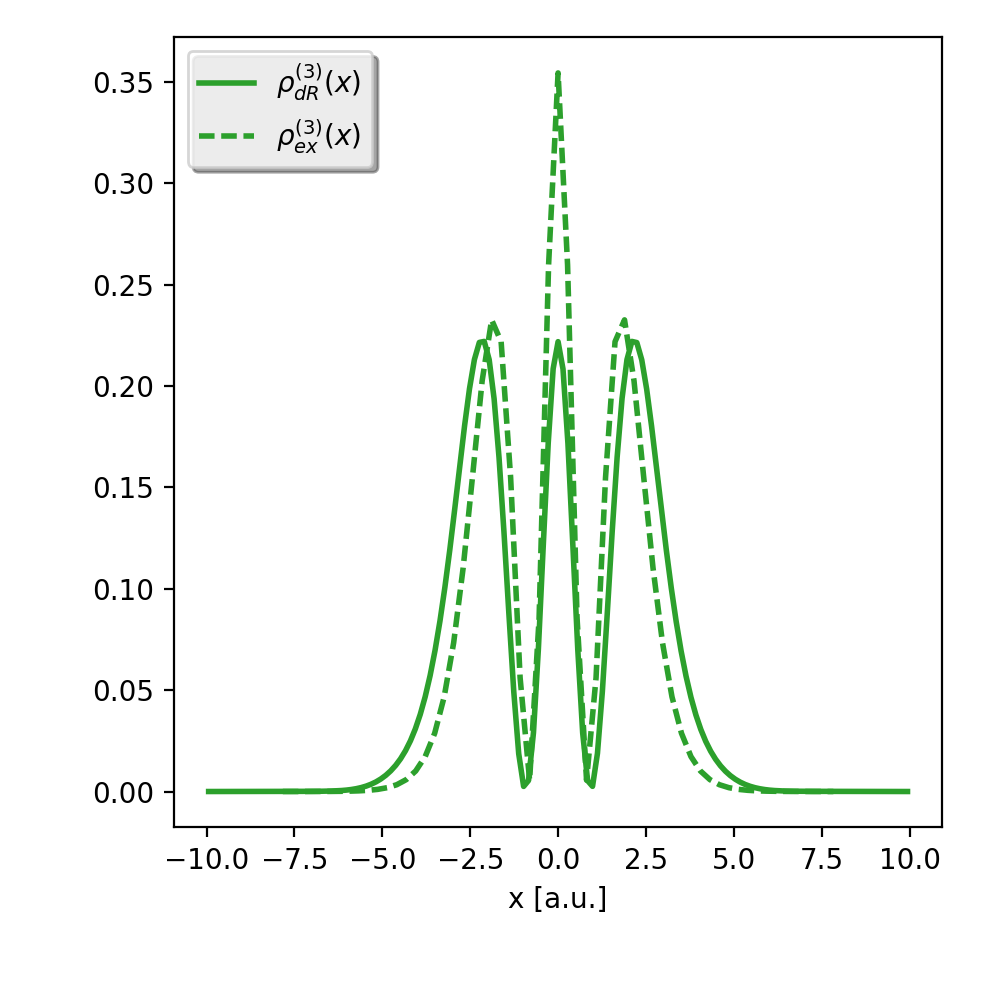}
		\put (85,85) {\textcolor{black}{He}}
	\end{overpic}\\
	\begin{overpic}[width=0.49\columnwidth]{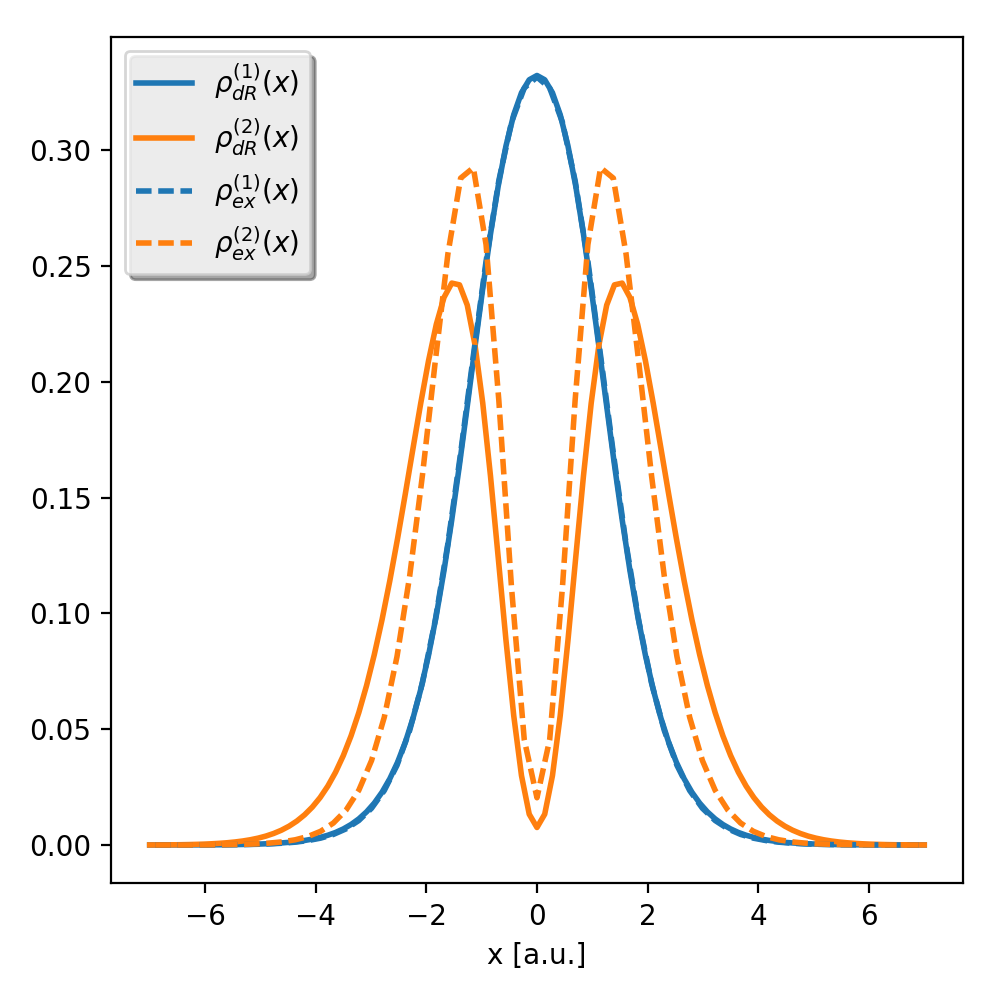}
		\put (85,85) {\textcolor{black}{$H_2$}}\hfill
	\end{overpic}
	\begin{overpic}[width=0.49\columnwidth]{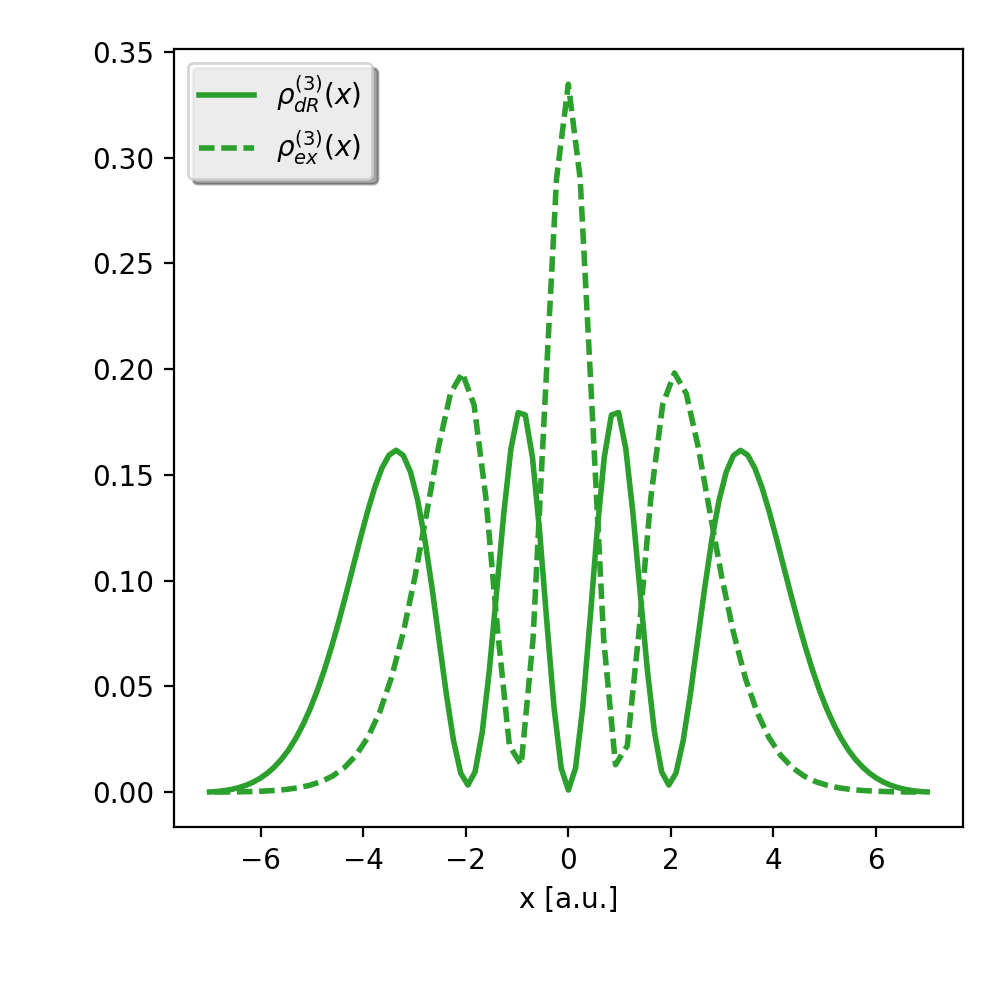}
		\put (85,85) {\textcolor{black}{$H_2$}}
	\end{overpic}
	\caption{The first three natural orbital densities $\rho^{(i)}_{ex/dR}(x)$ of the exact (ex) and dressed RDMFT (dR) calculations are depicted for the $He$ atom (top) and the $H_2$ molecule (bottom) with coupling $g/\omega =0.1$. We see in both cases that $\rho^{(1)}_{ex}(x)$ is almost exactly reproduced by dressed RDMFT, but $\rho^{(2)}_{dR}(x)$ deviates already visibly from $\rho^{(2)}_{ex}(x)$ (left.) However, it is in both cases \emph{qualitatively} correct. This changes for $\rho^{(3)}_{dR}(x)$ of $H_2$, which has one node more than $\rho^{(3)}_{ex}(x)$. Nevertheless, $\rho^{(3)}_{dR}(x)$ of $He$, is reproduced correctly (right.)}
	\label{fig:H2_NO}
\end{figure}

An even more stringent test of the accuracy of the dressed RDMFT approach is to compare the dressed 1RDMs. The essential ingredients of the dressed 1RDMs are their natural orbitals $\phi_{i}(x,q)$. Again, we separate electronic and photonic contributions and show their reduced electronic density $\rho_i(x)=\int\td q\, |\phi_i(x,q)|^2$. Fig.~\ref{fig:H2_NO} depicts the first three dressed natural orbital densities of dressed RDMFT in comparison with the exact ones for both test systems. While it holds that for both systems, the lowest natural orbital density of the dressed RDMFT approximation is almost the same as the exact one, and the second natural orbital densities are only slightly different, the third natural orbital densities of $H_2$ differ even qualitatively. For $He$, similar strong deviations are visible for the fourth natural orbital. However, as long as such strong deviations only occur for natural orbitals with small natural occupation numbers, like in these cases ($H_2: n_1=1.878, n_2=0.102, n_3=0.015$, He: $n_1=1.978, n_2=0.020, n_3=0.001$), their (inaccurate) contribution to the density and total energy remains small.
\begin{figure}[H]
	\begin{overpic}[width=0.49\columnwidth]{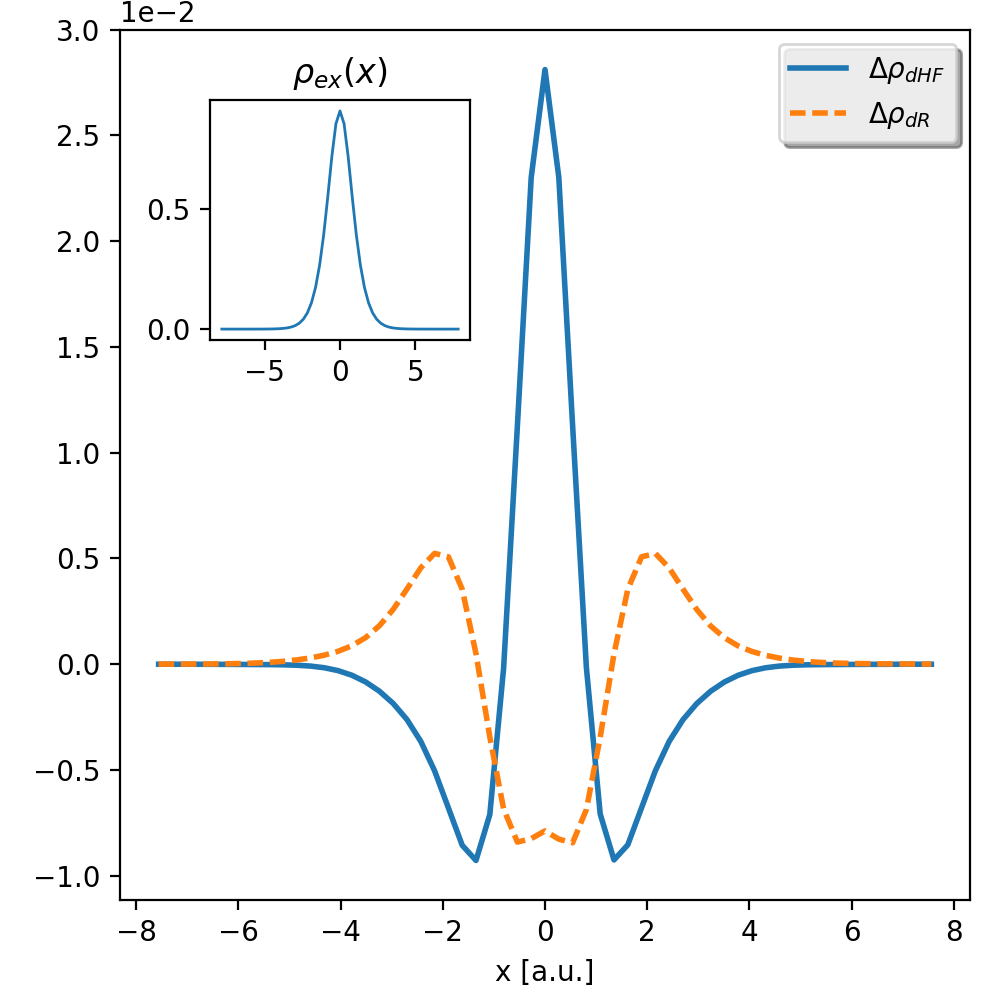}
		\put (85,15) {\textcolor{black}{He}}\hfill
	\end{overpic}
  \begin{overpic}[width=0.49\columnwidth]{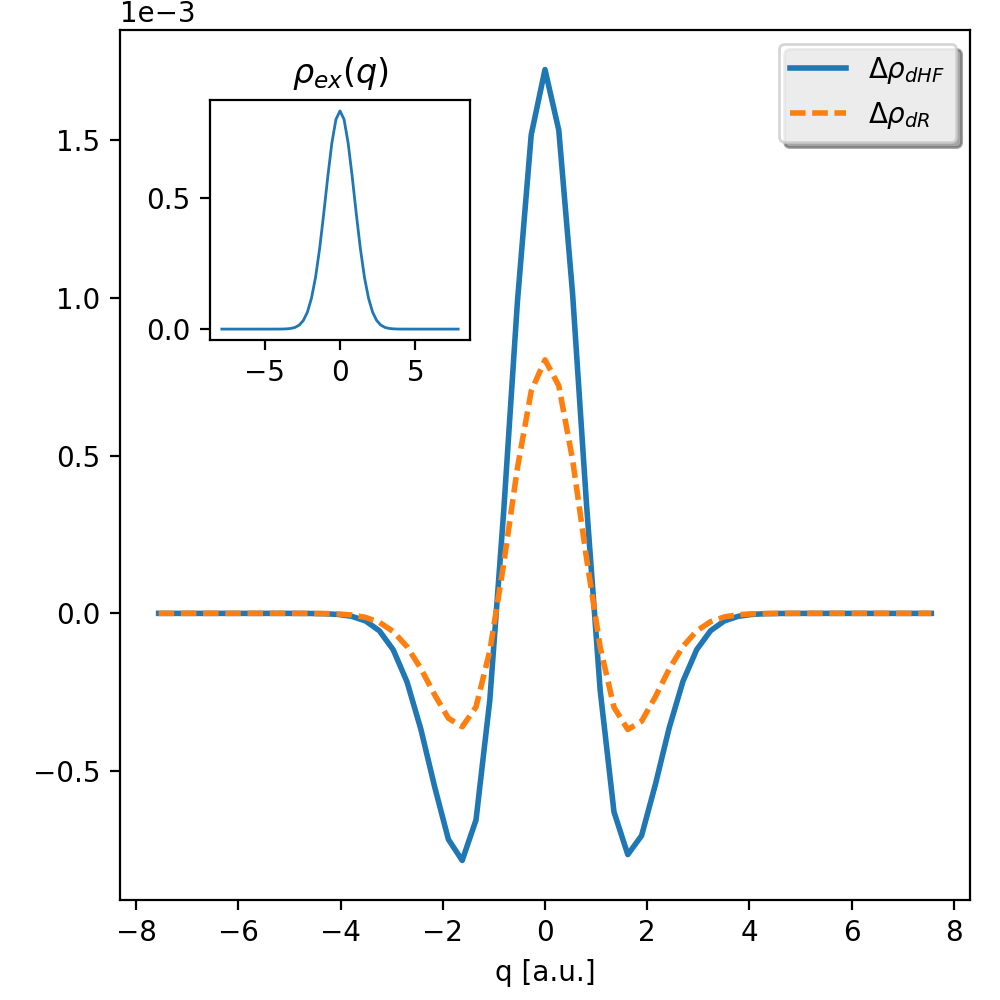}
  	\put (85,15) {\textcolor{black}{He}}
  \end{overpic}\\
  \begin{overpic}[width=0.49\columnwidth]{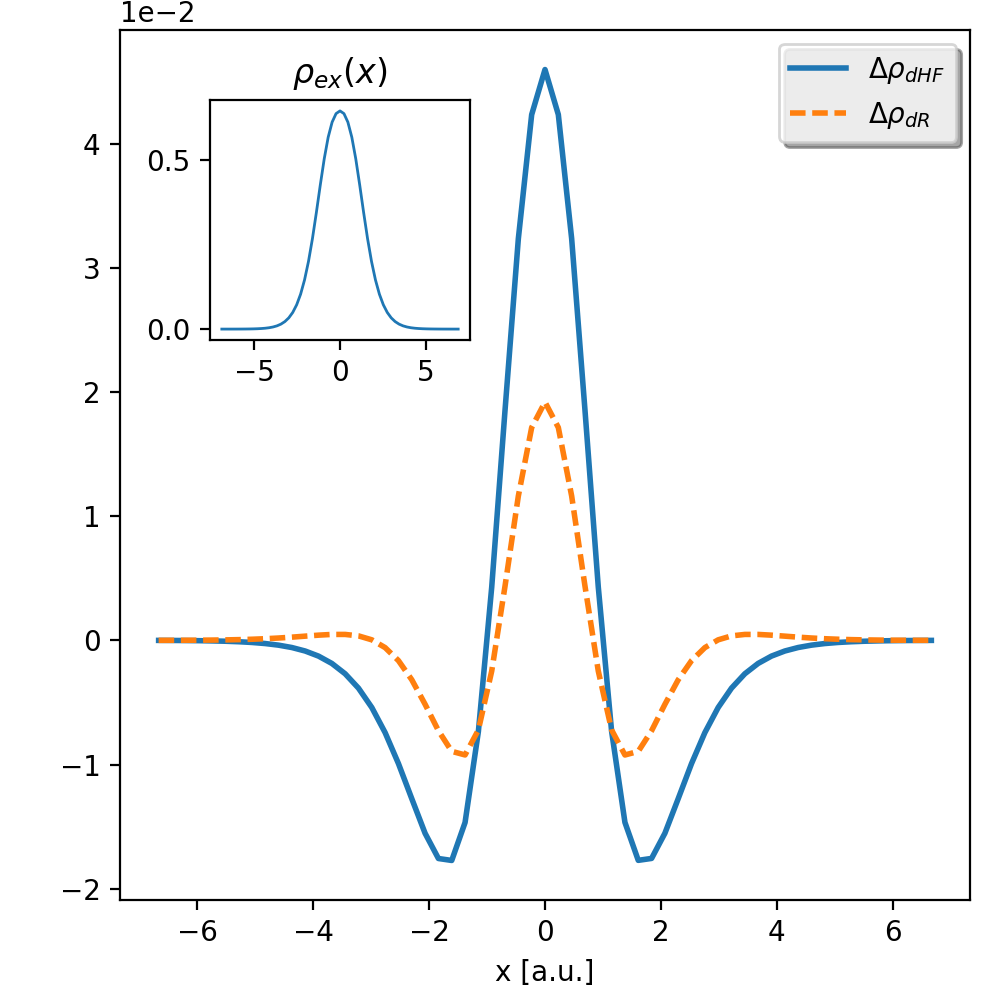}
  	\put (85,15) {\textcolor{black}{$H_2$}}\hfill
  \end{overpic}
  \begin{overpic}[width=0.49\columnwidth]{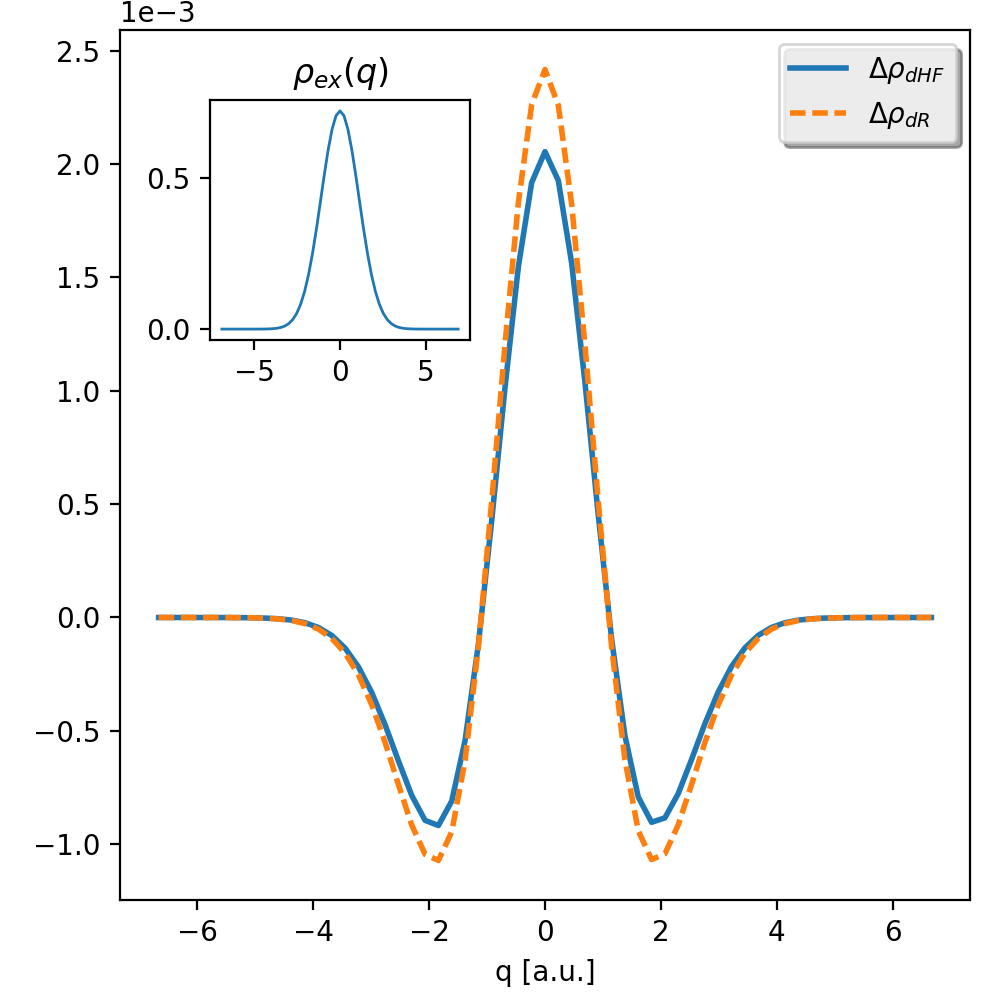}
  	\put (85,15) {\textcolor{black}{$H_2$}}
  \end{overpic}
\caption{Deviations of dressed HF (dHF) and dressed RDMFT (dR) ground state densities from the exact solution (depicted in the insets) for the $He$ atom (top) and the $H_2$ molecule (bottom) with coupling $g/\omega=0.1$. We separate the electronic (x, left) and photonic (q, right) coordinates as explained in the text. For both systems, dressed RDMFT finds a considerably better electronic density than dressed HF, which is consistent with the better result in energy (see Fig.~\ref{fig:H2_He_lambda}.) The photonic densities are reproduced almost exactly for both levels of theory.}
\label{fig:H2_He_dens}
\end{figure}

\begin{figure}[H]
	\begin{overpic}[width=0.49\columnwidth]{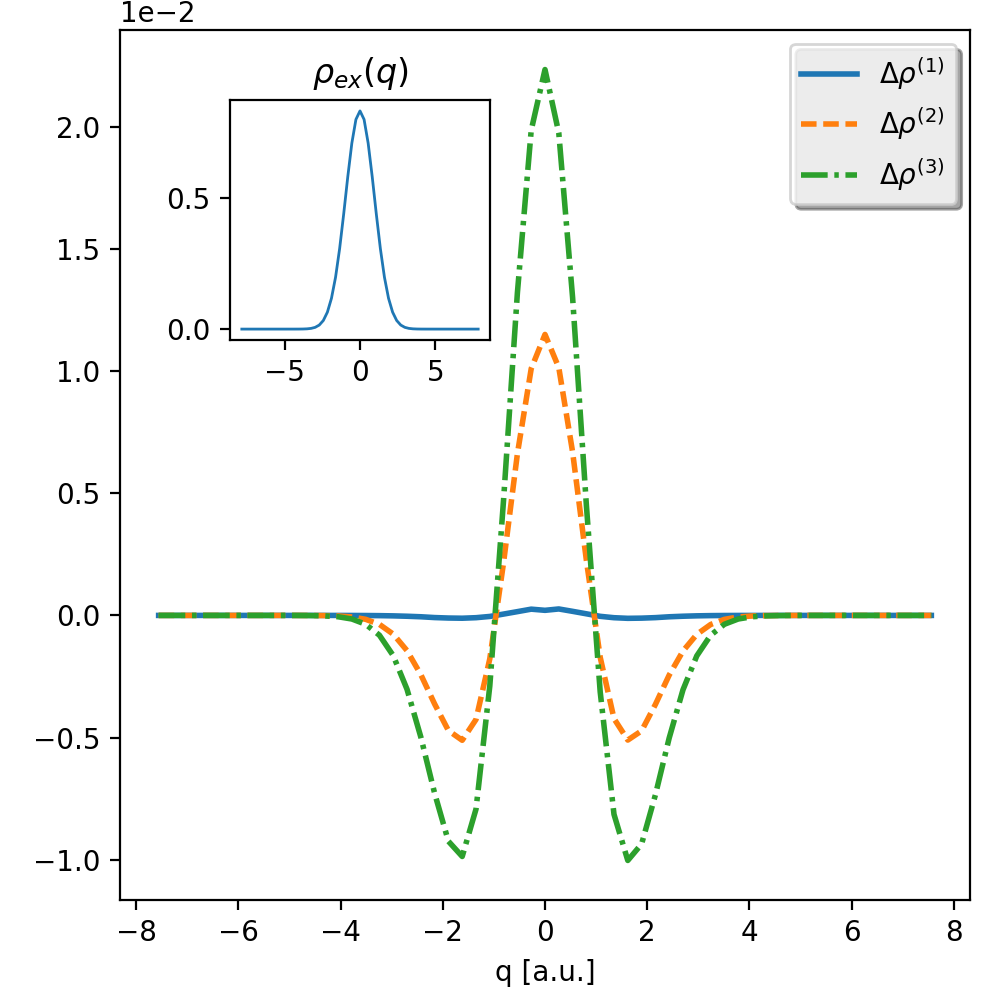}
		\put (80,20) {\textcolor{black}{$He$}}\hfill
	\end{overpic}
	\begin{overpic}[width=0.49\columnwidth]{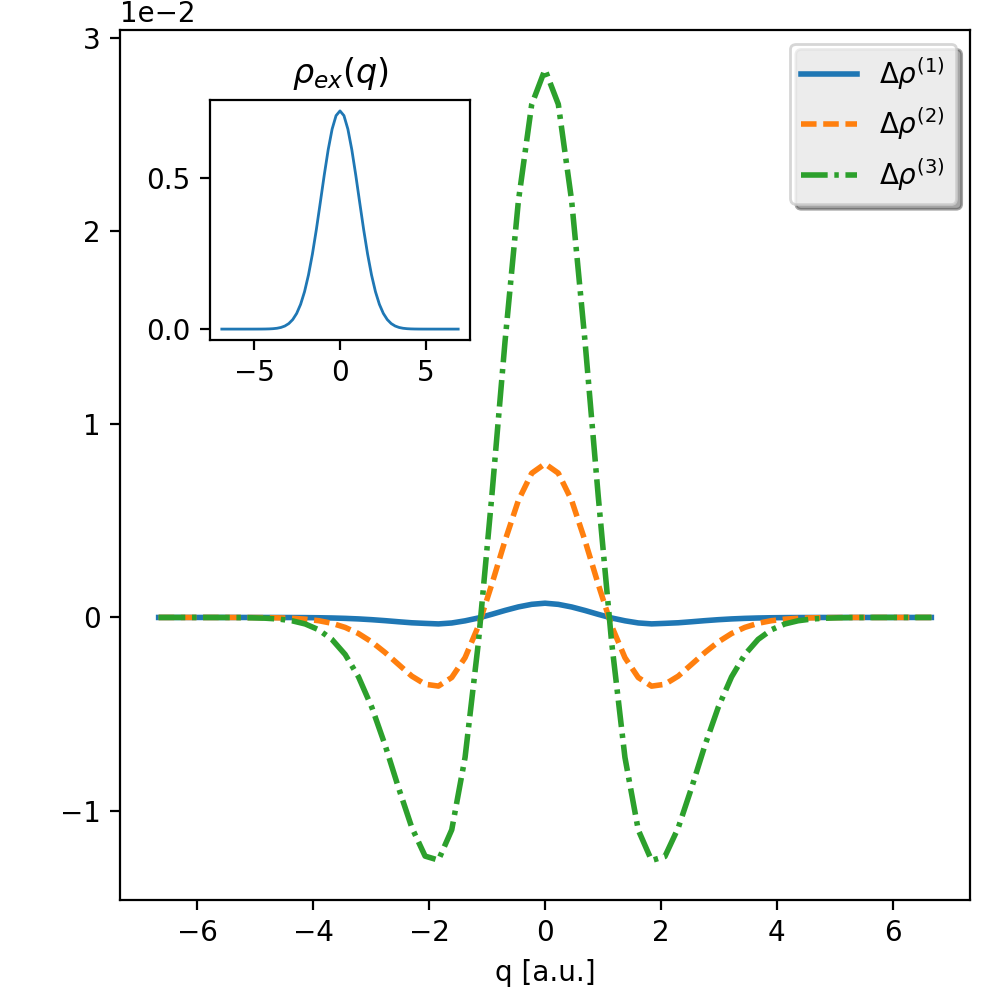}
		\put (80,20) {\textcolor{black}{$H_2$}}
	\end{overpic}
	\caption{We show the differences $\Delta \rho^{(i)}=\rho^{(i)}_{dR}(q)-\rho^{(i)}_{ex}(q)$ between the dressed RDMFT (dR) and the exact (ex) photonic natural orbital densities $\rho^i_{ex/dR}(q)$ for the 3 highest occupied natural orbitals for the $He$ atom (left) and the $H_2$ molecule (right) for coupling strength $g/\omega =0.1$. For both systems, the exact $\rho^{(i)}_{ex}(q)$ have a similar shape as the density (see inset.)  We see in both cases that dressed RDMFT captures the exact solution very well.}
	\label{fig:H2_He_q_orbs}
\end{figure}

To complete the picture, we also look at the photonic natural orbital densities, $\rho_i(q)=\int\td x\, |\phi_i(x,q)|^2$, the first 3 of which are plotted in Fig. \ref{fig:H2_He_q_orbs}, for $He$ and $H_2$. Here, the dressed RDMFT results even agree better with the exact solution than their electronic counterparts. Apparently, dressed RDMFT captures the photonic properties of the tested systems very accurately for the ultra-strong coupling regime. The accuracy drops with increasing $g/\omega $.

\begin{figure}[H]
	\begin{overpic}[width=0.49\columnwidth]{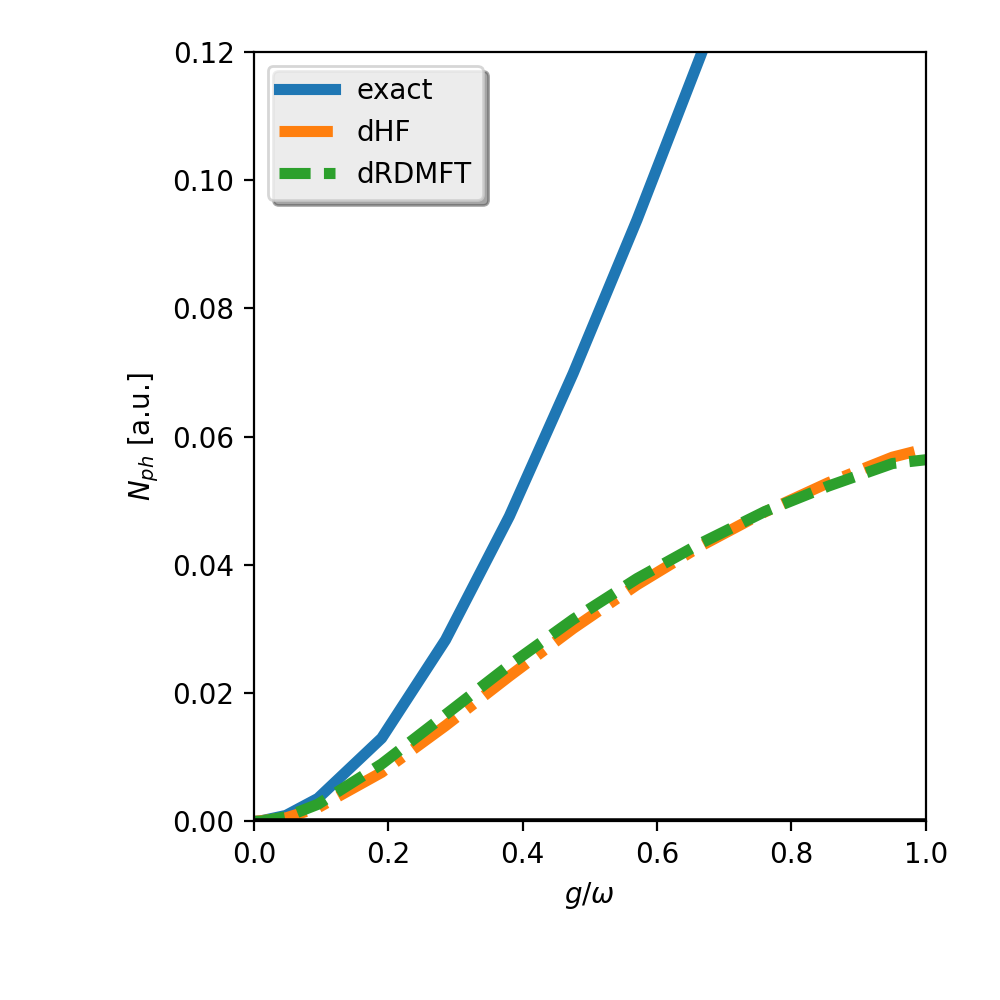}
		\put (80,25) {\textcolor{black}{He}}\hfill
	\end{overpic}
	\begin{overpic}[width=0.49\columnwidth]{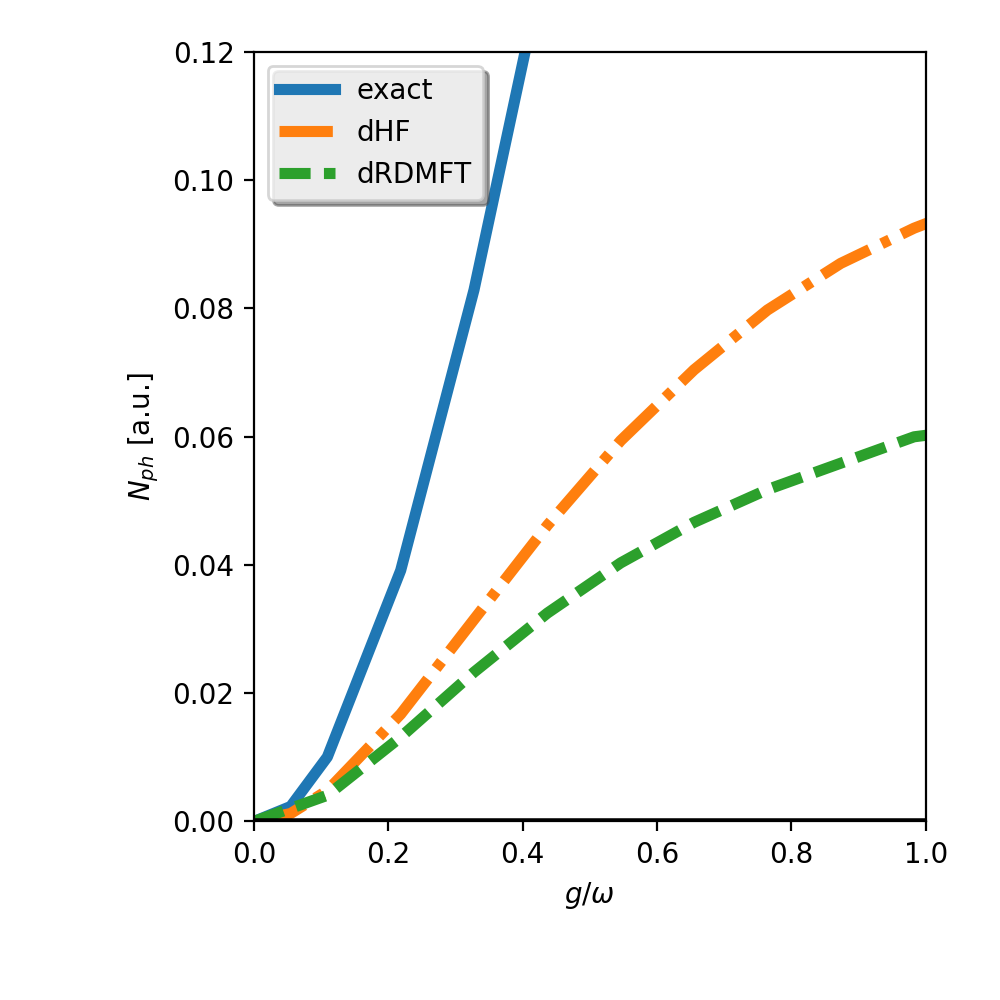}
		\put (80,25) {\textcolor{black}{$H_2$}}
	\end{overpic}
	\caption{The total mode occupation $N_{ph}$, calculated from the exact, dressed HF and dressed RDMFT solutions is shown for $He$ (left) and $H_2$ (right.) We see that both dressed RDMFT and dressed HF underestimate $N_{ph}$. In the ultra-strong coupling regime for $g/\omega >0.3$ both dressed HF and dressed RDMFT (with the Müller functional) deviate strongly from the exact solution.}
	\label{fig:H2_He_nphot}
\end{figure}

As an example for a photonic observable, we show in Fig. \ref{fig:H2_He_nphot} the mode occupation $N_{ph}$ as a function of the coupling strength $g/\omega $ that we calculated by using Eq. \eqref{eq:mode_occupation_formula}, i.e. $N_{ph}=\frac{E_{ph}}{\omega}-\frac{N}{2}$, with the photon mode energy $E_{ph}=\sum_{i=1}^{\mathcal{M}} n_i \int \td x \td q \, \phi^*_i (x,q)  \left(-\frac{1}{2}\frac{\td^2}{\td q^2} + \frac{w^2}{2}q^2\right) \phi_i (x,q)$. From weak to the beginning of the ultra-strong coupling regime ($g/\omega\approx 0.1$,) both dressed HF and dressed RDMFT capture $N_{ph}$ well. For very large coupling strengths, the deviations to the exact mode occupation becomes sizeable. This might sound counter-intuitive, as the photonic density is described comparatively well. The reason is that the photon occupation, in contrast to the density, is mainly determined by the second and third natural orbital, because the first natural orbital resembles a photonic ground state with occupation number zero in the studied cases. Dressed HF does not consider a second orbital (the first instead is doubly occupied) and thus cannot capture the effect. And for dressed RDMFT, the error in the second and third natural orbital is much larger than in the first (see Fig. \ref{fig:H2_He_q_orbs}.) However, it is probable that this can be improved by better functionals.

\begin{figure}[H]
	\includegraphics[width=0.49\columnwidth]{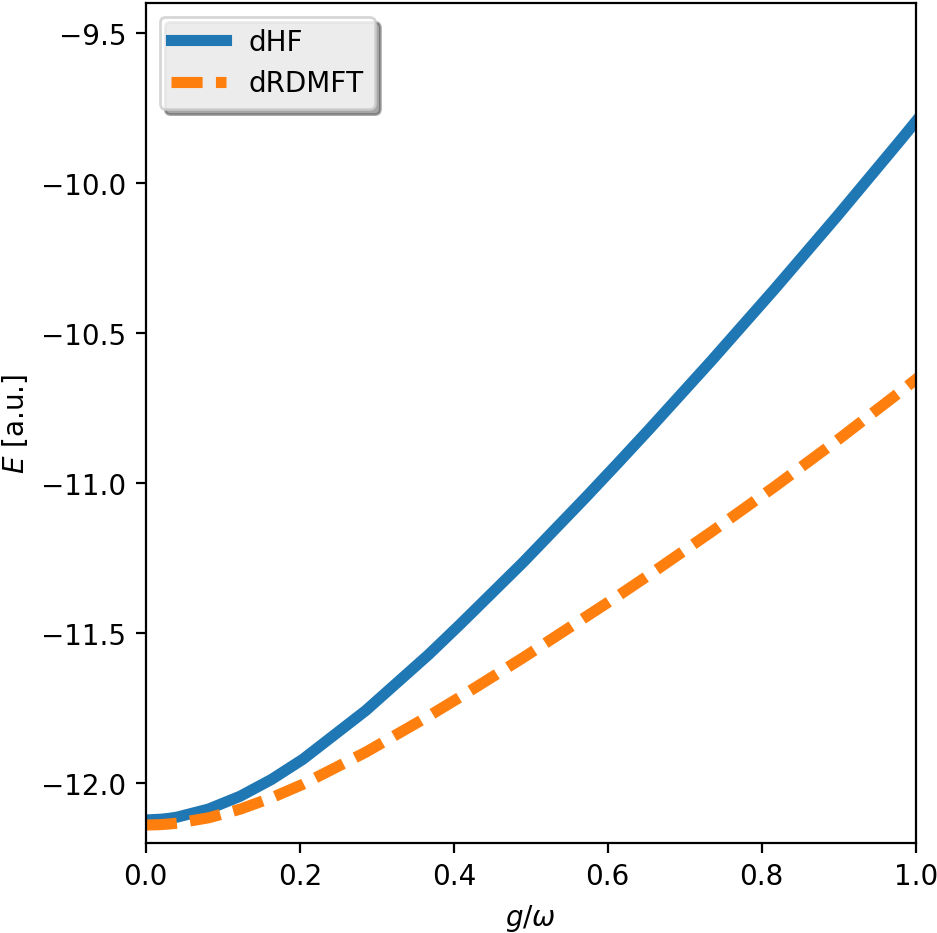}
	\caption{The plot shows the total energy of the dressed HF and dressed RDMFT calculations of $Be$ for increasing $g/\omega $. We observe the same trend as for the two-electron systems: for both levels of theory, the energy grows with increasing $g/\omega $, though for dressed HF faster than for dressed RDMFT.} 
	\label{fig:Be_Etot}
\end{figure}

\begin{figure}[H]
	\includegraphics[width=0.49\columnwidth]{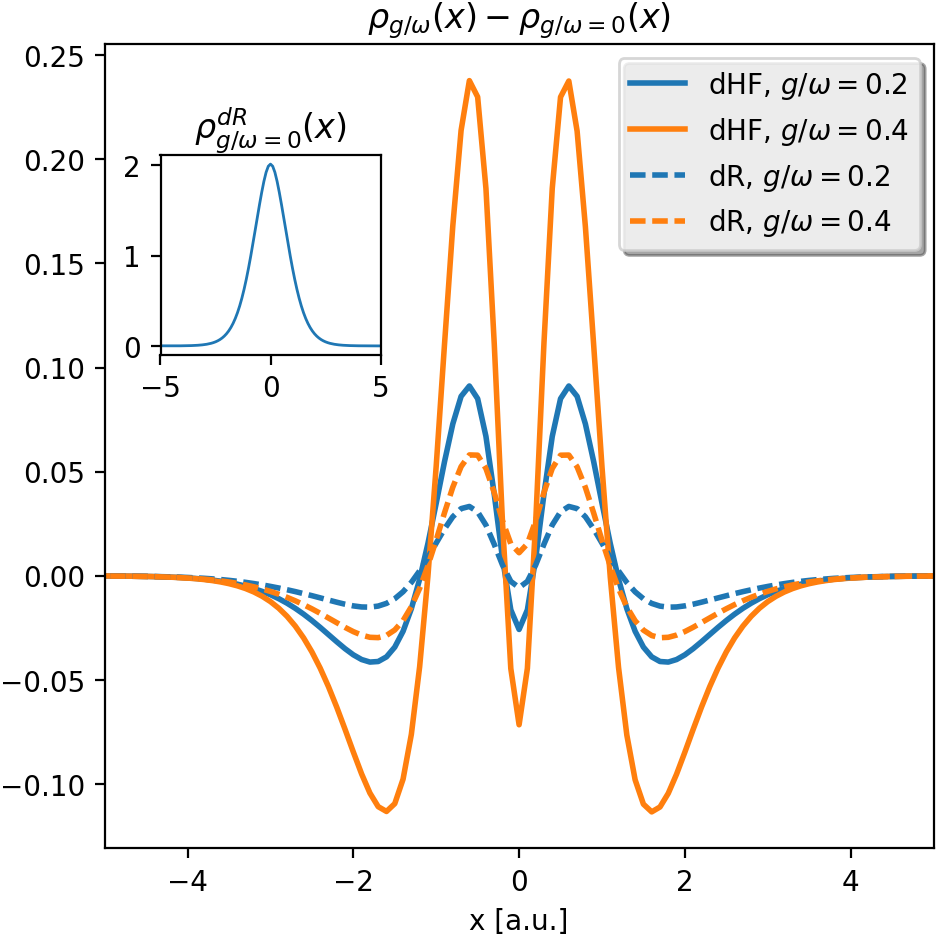}
	\includegraphics[width=0.49\columnwidth]{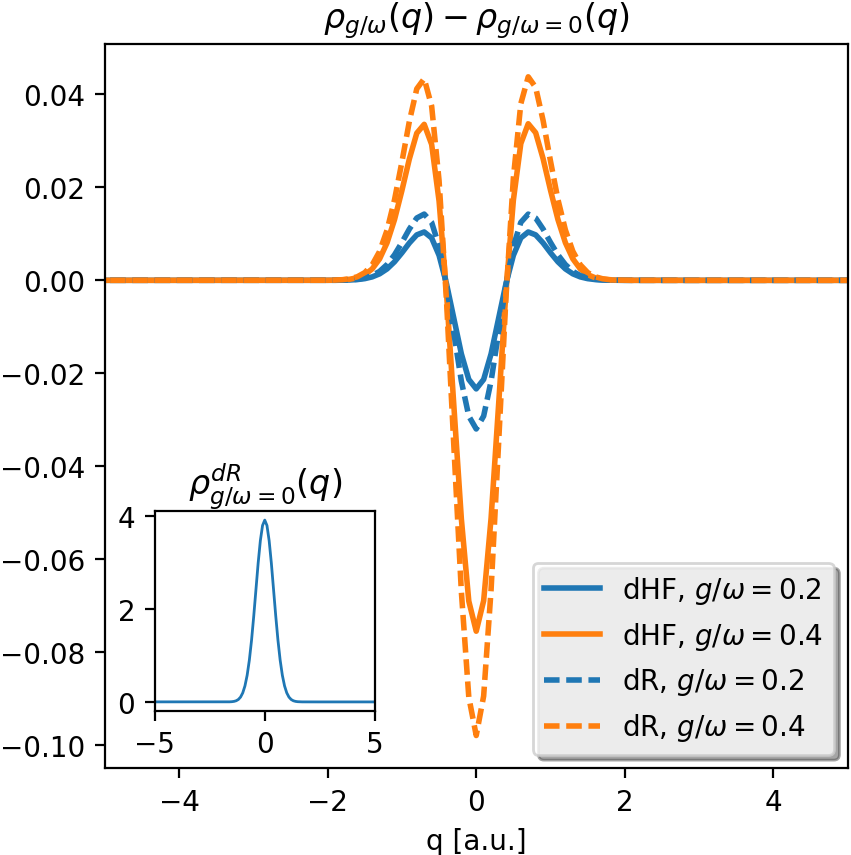}
	\caption{Shown are the electronic ($\rho^{dHF/dR}_{g/\omega}(x)$, left) and photonic ($\rho^{dHF/dR}_{g/\omega}(q)$, right) densities of $Be$ for dressed HF (dHF) and dressed RDMFT (dR) for 2 different coupling strengths subtracted from their counterparts in the no-coupling limit ($\rho^{dHF/dR}_{g/\omega=0}(x/q)$.) We see in the electronic (photonic) case that the dressed RDMFT deviations are less (more) pronounced than for dressed HF.} 
	\label{fig:Be_dens}
\end{figure}

\begin{figure}[H]
	\includegraphics[width=0.49\columnwidth]{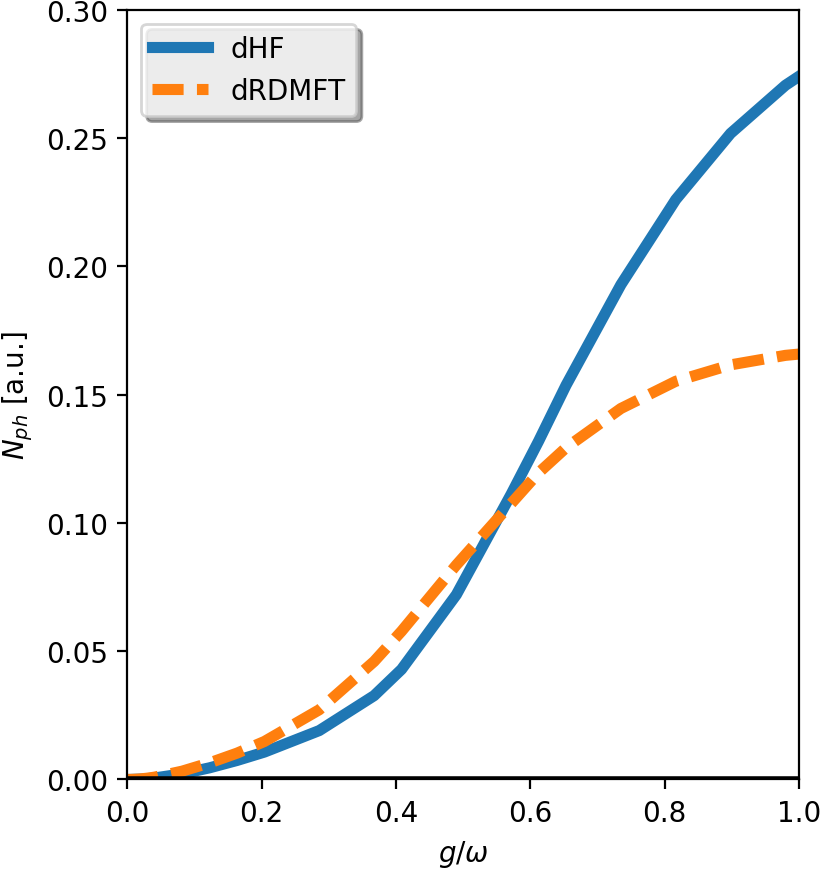}
	\caption{The total mode occupation $N_{ph}$ for dressed HF and dressed RDMFT is shown for $Be$. We see that dressed RDMFT exhibits larger $N_{ph}$ until a coupling strength of $g/\omega\approx 0.5$. For larger coupling the dressed HF mode occupation becomes higher.} 
	\label{fig:Be_nphot}
\end{figure}

\begin{figure}[H]
	\begin{overpic}[width=0.49\columnwidth]{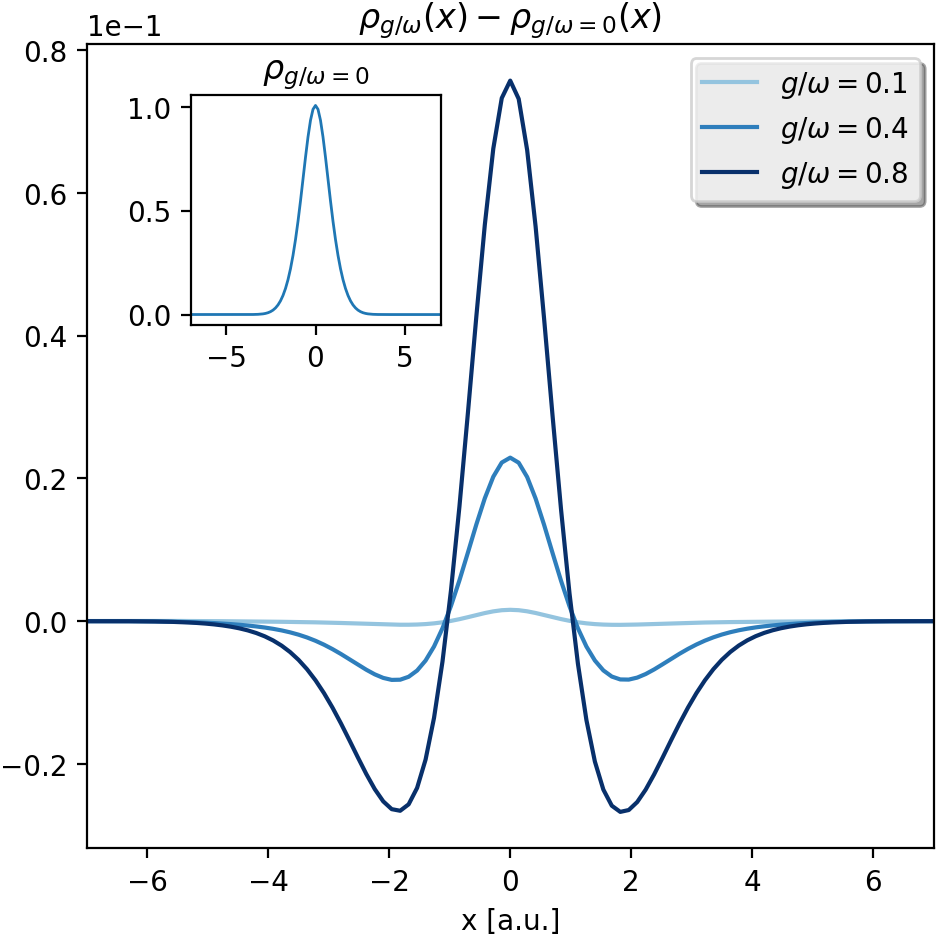}
		\put (85,20) {\textcolor{black}{He}}\hfill
	\end{overpic}
	\begin{overpic}[width=0.49\columnwidth]{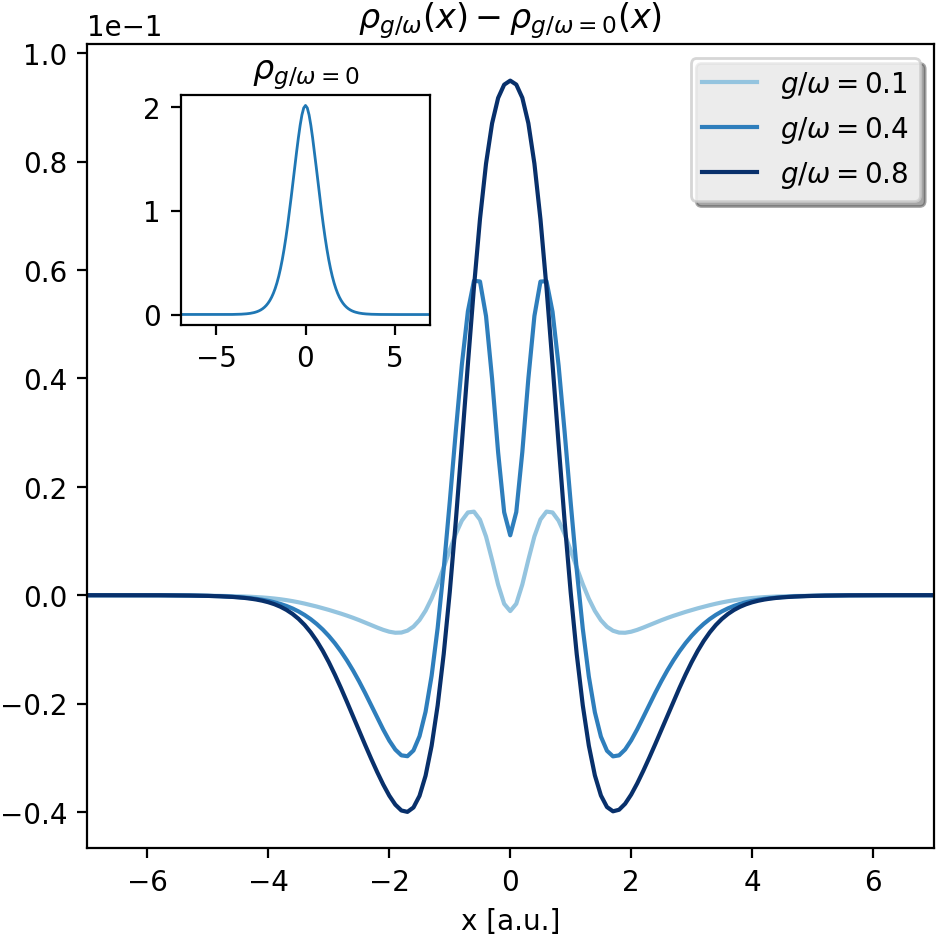}
		\put (85,20) {\textcolor{black}{Be}}\hfill
	\end{overpic}
	\caption{We show the differences in the electronic density ($\rho_{g/\omega}(x)$) of $He$ (left) and $Be$ (right) for 3 different coupling strengths compared to the atoms outside the cavity (insets,) calculated with dressed RDMFT. We see that the effect of the cavity is very different for both systems: The strong localization of the electronic density for $He$ indicates the suppression of electronic repulsion for all coupling strengths. For $Be$ instead, we see additionally local enhancement of the repulsion. The interplay of enhancement and suppression changes with increasing coupling strength.}
	\label{fig:He_Be_comp}
\end{figure}

By comparing to the exact solution, we showed that the dressed-orbital construction seems to be a reasonable starting point for an approximate description of both the electronic and the photonic part of coupled matter-photon systems. Thus, we can now go one step further and present results for a many-body systems that cannot easily be solved exactly: the one-dimensional $Be$-atom in a cavity. In Fig. \ref{fig:Be_Etot}, we see the total energy as a functional of the coupling strength $g/\omega $ for dressed HF and dressed RDMFT, respectively. Like in the two-electron systems, the deviation between both curves increases for larger $g/\omega $ and as expected the dressed RDMFT energies are lower than the dressed HF results. 
Analyzing the ground-state densities, we see a similar trend as in the 2-particle systems. With increasing $g/\omega$, the electronic (photonic) part of the density becomes more (less) localized, though the details differ as we show in the last part of this section (see Fig. \ref{fig:He_Be_comp} and the corresponding part in the main text.)  Comparing dressed RDMFT with dressed HF, we observe that the variation of the electronic (photonic) density with increasing coupling strength is less (more) prounounced for dressed RDMFT, as Fig. \ref{fig:Be_dens} shows.  We conclude the survey of $Be$ with the mode occupation under variation of the coupling strength (see Fig. \ref{fig:Be_nphot}.) We see that the value of $g/\omega\approx 0.5$ separates two regions. For $g/\omega< 0.5$ dressed RDMFT finds a larger mode occupation than dressed HF and for $g/\omega> 0.5$ instead the dressed HF mode is stronger occupied. We found similar behavior also for the 2-particle systems, although the boundary between the 2 regions was considerably different there ($He: g/\omega\approx0.8, H_2:g/\omega\approx0.1$, see Fig. \ref{fig:H2_He_nphot}.)

\clearpage
\begin{figure}[H]
	\includegraphics[width=0.49\columnwidth]{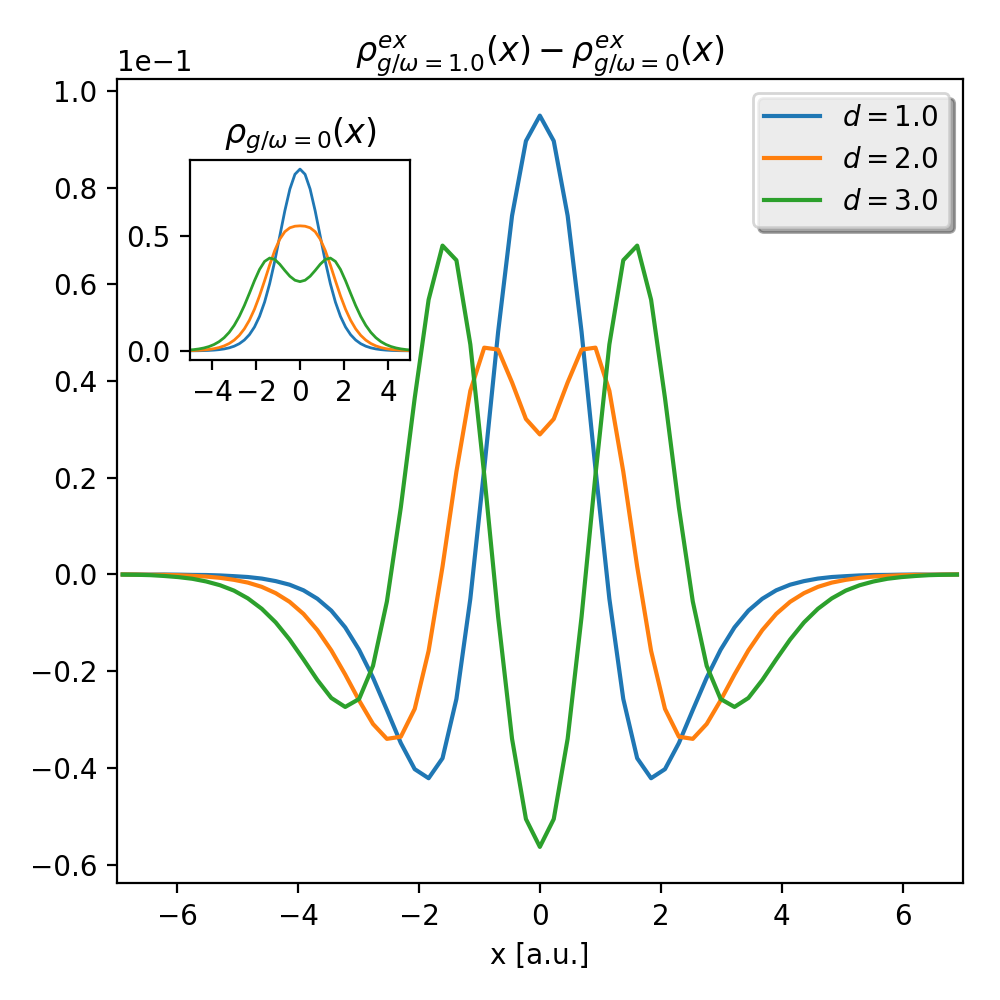}
	\includegraphics[width=0.49\columnwidth]{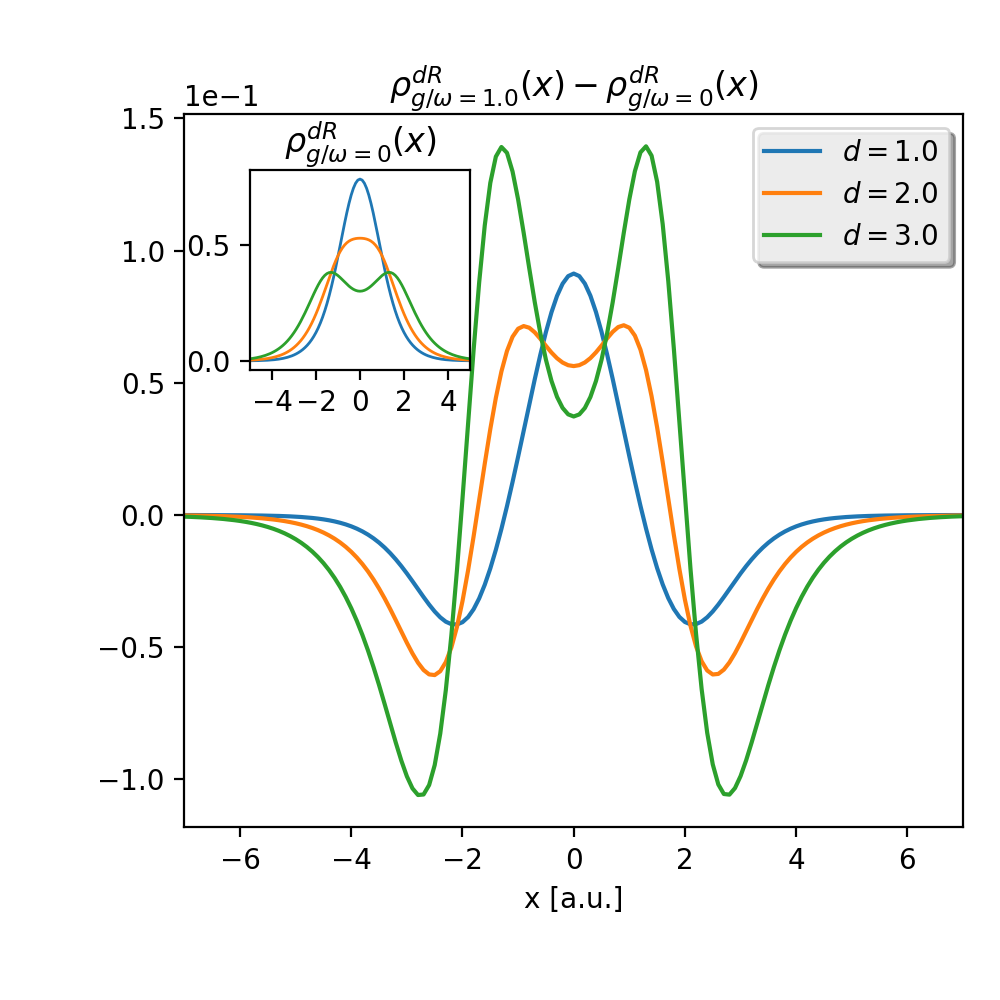}
	\caption{We show the differences in the electronic density of the $H_2$ molecule for 3 different bond lengths $d$ (as examples of the dissociation) for $g/\omega=1.0$ compared to $g/\omega=0$, calculated exactly ($\rho^{ex}_{g/\omega}(x)$, left) and with dressed RDMFT ($\rho^{dR}_{g/\omega}(x)$, right). We see that for small $d$, the cavity mode reduces the electronic repulsion and localizes the charges at the bond center ($d=1<d_{eq}=1.628$) in comparison to the free molecule (insets.) For larger $d$, the electronic repulsion is locally enhanced such that the charge deviations are separated in two peaks ($d=2$.) For very large $d$, this interplay between local suppresion and enhancement of repulsion becomes more pronounced ($d=3$.) The dressed RDMFT calculations capture the behavior very well.}
	\label{fig:H2_diss}
\end{figure}
We conclude this section with two examples, for which the light-matter interaction changes the bare systems non-trivially, depending not only on the coupling strength but also on the details of the electronic structure. We start with the dissociation of $H_2$ as an example for a chemical reaction, where we use $v_{H_2}(x)$ with different $d$. In Fig. \ref{fig:H2_diss}, we see the density of two H-atoms under variation of the distance $d$ with and without the (strong) coupling to the cavity. We see that the influence of the cavity mode strongly depends on the exact electronic structure. The interaction with the cavity mode can locally reduce or enhance the electronic repulsion due to the Coulomb interaction, where the exact interplay between both effects depends on the interatomic distance. Thus, we can observe a number of different effects like pure localization of the density towards the center of charge ($d=1$) or localization combined with a local enhancement of repulsion such that the density deviations exhibit a double peak structure ($d=2$.) The local enhancement of electronic repulsion can grow so strong that the density at the center of charge is reduced but at the same time the density maxima shift closer to each other, which is an effective suppression of electronic repulsion ($d=3$.) This interplay is reflected in the natural orbitals and occupation numbers. The coupling shifts a considerable amount of occupation from the first natural orbital to the second and third one. The contribution to the total density of the former (latter) has the character of enhanced (suppressed) electron repulsion. 
To show the potential of these effects, we present calculations in the deep-strong coupling regime with $g/\omega= 1.0$, where the effects reach the order of 10 \% of the unperturbed density, which is enormous. For smaller coupling strengths of the order of $g/\omega=0.1$, these effects are as diverse, but naturally smaller with density deformations of the order of $10^{-3}$. However, as every observable depends on the density, such deviations are significant. Remarkably, dressed RDMFT reproduces the effects accurately.
In the second example, we compare the behavior of the $He$ and $Be$ atoms under the influence of the cavity. Though the shapes of the electronic density of the two bare systems are very similar (see insets in Fig. \ref{fig:He_Be_comp},) they behave very differently under the influence of the cavity, which can be seen in Fig. \ref{fig:He_Be_comp}. The electronic density of $He$ is pushed towards its center of charge with increasing coupling strength, which can be understood as a suppression of the electronic repulsion induced by the Coulomb interaction. As $He$ can be understood very well with only one orbital this is to be expected.\footnote{For $g/\omega=0.8$ we still observe $n_1=1.85$. However, it should be noted that the good agreement of dressed RDMFT with the exact calculation in comparison to dressed HF is exactly because of the contribution of the second natural orbital, that is (still considerably) occupied with $n_2=0.14$.}
Things change for $Be$, where we have several dominant orbitals. With increasing coupling strength, we see like in the dissociation example a subtle interplay between suppression and local enhancement of the electronic repulsion, that depends on the coupling strength. Thus for the same coupling strength, we can observe opposite ($g/\omega=0.1$ and $0.4$ in the plot) but also similar effects ($g/\omega=0.8$ in the plot) in two systems that have almost the same ``bare'' density shape. Like in the dissociation example, this intricate behavior can be understood by the interplay of the different natural orbitals contributing to the electronic density. In this particular case, the main physics happens in the second and third natural orbital, where the former (with a double-peak structure) loses a considerable amount of occupation to the latter (with a triple peak structure) with increasing coupling strength.

These (seemingly simple) examples show how subtle details of the electronic structure influence the changes induced by the coupling to photons. We see a non-trivial interplay between local suppression and enhancement of the Coulomb induced repulsion between the particles. This is reflected in the natural orbitals and occupation numbers of the light-matter system and thus influences all possible observables. Such small changes have been shown to theoretically affect chemical properties and reactions strongly, which are determined by an intricate interplay between Coulomb and photon induced correlations~\cite{Schafer2018}. Whether these modifications of the underlying electronic structure are indeed a major player in the changes of chemical and physical properties still needs to be seen. However, to capture such modifications in the first place (and study their influence) clearly needs an ab-initio theory that is able to treat both types of (strong) correlations accurately and is predictive inside as well as outside of a cavity. We have shown here that dressed RDMFT is a viable option to predict and analyze these intricate structural changes.

%% file: 6summary.tex
\section*{Conclusion}
\label{sec:conclusion}
In this work we presented a RDMFT formalism for coupled matter-photon systems. This formalism is capable to account for the full quantum-mechanical degrees of freedom of the coupled fermion-boson problem. We discussed that extending the standard formulation of electronic RDMFT to systems with coupled fermionic and bosonic degrees of freedom is not straightforward. Then, we presented an alternative approach which overcomes most of the intricate representability issues by embedding the coupled matter-photon system in a higher-dimensional auxiliary space. Specifically, we introduced for a problem with $N$ electrons coupled to $M$ photon modes, $(N-1)M$ auxiliary coordinates, which allowed us to ``fermionize'' the coupled problem with respect to new polaritonic coordinates. The resulting dressed fermionic particles are governed by a Hamiltonian with only 1-body and 2-body terms and thus, one can apply any standard electronic-structure method. The extension is constructed in such a way that the auxiliary dimensions do not modify the original physical system and the physical observables are easy to recover. Notably, besides the possibility to study modifications of electronic systems due to a cavity mode, dressed RDMFT offers also the possibility to calculate purely photonic observables like the mode occupation or fluctuations of the electric and magnetic field.  We used this framework to develop and implement dressed RDMFT in the electronic-structure code Octopus~\cite{Andrade2015}, and tested it with the Hartree-Fock and Müller functional. For simple one-dimensional models of atoms and molecules the obtained approximate results were in good agreement with the exact results from the weak to the deep-strong coupling regime. 
We then used our method to show that the modifications due to strong matter-photon coupling are far from trivial and depend on the detailed electronic structure. For a molecular as well as an atomic system we showed that strong coupling can locally enhance and suppress the Coulomb induced repulsion between electrons. This behaviour does not only depend on the strength of the matter-photon coupling but also on the details of the matter subsystem (e.g. the interatomic distance of the atoms of a molecule.) We showed that our method allows to predict the structures accurately inside and outside of the cavity and furthermore extends the well-established tools of natural orbitals to analyze coupled light-matter systems.

Although the presented method is practical only for a few photon modes, since the number of photon modes determines the dimension of the involved dressed orbitals, it is exactly these cases that are the most relevant in cavity and circuit QED experiments. Since dressed RDMFT is non-perturbative and seems to be accurate over a wide range of couplings, it is a promising tool to investigate long-standing problems of quantum optics, such as the quest for a super-radiant phase in the ground state of strongly-coupled matter-photon systems~\cite{Bernardis2018}. Moreover, it is a very promising tool to investigate changes in the ground-state due to matter-photon coupling that can possibly modify chemical reactions~\cite{Thomas2016}.  Recently, it has been shown that charge-transfer processes can be considerably modified due to strong coupling to a cavity mode\cite{Schafer2018}. And although the presented results were for reduced dimensionality, an extension to three spatial dimensions is straightforward. We can rely here again on an already existing implementation for RDMFT in Octopus. Work along these lines is in progress.
Besides such interesting applications and fundamental questions of light matter-interactions, there are many open questions to answer also in the presented theory itself. For instance, how strong is the influence of the hitherto negelected $q$-exchange symmetry? First calculations for many particles indicate that it will become important to enforce this extra symmetry to stay accurate when going from the weak to the deep-strong coupling regime. Furthermore, it might become beneficial to avoid the ``fermionization'' that we employed and then very interesting mathematical questions about $N$-representability for coupled fermion-boson systems need to be addressed. Here, the understanding how to enforce the $q$-exchange symmetries in the dressed formulation could be very useful.

\begin{acknowledgement}
F.B. would like to thank Nicole Helbig, Klaas Giesbertz, Micael Oliveira, and Christian Schäfer for stimulating and useful discussions. We acknowledge financial support from the European Research Council (ERC-2015-AdG-694097).
\end{acknowledgement}

\begin{suppinfo}
	Survey on the bosonic symmetry of the photon wave function. Details about the convergence study of the numerical examples shown in the paper. Protocol for the convergence of a dressed HF/RDMFT calculation. 
\end{suppinfo}